\documentclass[12pt]{article}
\usepackage{amsmath,amsfonts,amssymb,amsthm,bm,bbm,mathtools,graphicx,color,tcolorbox,algorithm,algorithmic}
\usepackage{caption,fancyhdr,titlesec,indentfirst,booktabs,verbatim,hyperref,cases,titletoc,multirow,wasysym,empheq,setspace,natbib,chngcntr, subcaption,caption}
\usepackage[font=small,labelfont=bf]{caption}

\newcommand{\rom}[1]{\uppercase\expandafter{\romannumeral #1\relax}}
\usepackage[page,header]{appendix}
\usepackage[english]{babel}
\usepackage[utf8]{inputenc}

\newtheorem{theorem}[]{Theorem}
\newtheorem{lemma}[]{Lemma}

\newtheorem{proposition}[]{Proposition}

\def\bc{{\bf{c}}}

\def\bt{{\bf{t}}}
\def\bx{{\bf{x}}}
\def\bnu{{\boldsymbol{\nu}}}
\def\bmu{{\boldsymbol{\mu}}}

\def\by{{\bf{y}}}

\def\bV{\mathbb V}

\def\WW{\mathcal W}
\def\HH{\mathcal H}
\def\EE{\mathcal E}

\def\XX{\mathcal X}

\def\RR{\mathcal R}

\def\KK{\mathcal K}

\def\NN{\mathcal N}
\def\AA{\mathcal A}

\def\R{\mathbb R}
\def\E{\mathbb E}

\def\Z{\mathbb Z}
\def\N{\mathbb N}
\def\P{\mathbb P}

\def\E{\mathbb{E}}

\def\bPsi{\mathbf \Psi}

\newcommand{\change}[1]{{\leavevmode\color{black}{#1}}}

\usepackage{xr}
\begin{document}

\title{\Large{\textbf{Nonparametric Estimation via Partial Derivatives}}} 
\author{
\bigskip
{\sc Xiaowu Dai} \\
{\it {\normalsize University of California, Los Angeles}}\\\\
{\normalsize To appear in \it{Journal of the Royal Statistical Society: Series B.}}
}
\date{}
\maketitle

\begin{footnotetext}[1]
{\textit{Address for correspondence:} Xiaowu Dai, Department of Statistics and Data Science, UCLA, 8125 Math Sciences Bldg \#951554, CA 90095, USA. E-mail: dai@stat.ucla.edu. }
\end{footnotetext}

\begin{abstract}
\noindent
Traditional nonparametric estimation methods often lead to a slow convergence rate in large dimensions and require unrealistically enormous sizes of datasets for reliable conclusions. We develop an approach based on \change{partial derivatives}, either observed or estimated, to effectively estimate the function at near-parametric convergence rates. The novel approach and computational algorithm could lead to methods useful to practitioners in many areas of science and engineering. 
Our theoretical results reveal a behavior universal to this class of nonparametric estimation problems.
We explore a general setting involving tensor product spaces and build upon the smoothing spline analysis of variance (SS-ANOVA) framework.
For $d$-dimensional models under full interaction, the optimal rates with gradient information on $p$ covariates are identical to those for the $(d-p)$-interaction models without gradients and, therefore, the models are immune to the ``curse of interaction."  For additive models, the optimal rates using gradient information are root-$n$, thus achieving the ``parametric rate."
We demonstrate aspects of the theoretical results through synthetic and real data applications.

\end{abstract}
\bigskip

\noindent{\bf Key Words:} Gradient; Interaction;  Reproducing kernel Hilbert space; Smoothing spline ANOVA; Time series.

\newpage
\baselineskip=22pt


\section{Introduction}
\label{sec:Intro}
\noindent
Gradient information for complex systems arises in many areas of science and engineering.
Economists estimate cost functions, where data on factor demands and costs are collected together. By Shephard's Lemma, the demand functions are the first-order partial derivatives of the cost function \citep{hall2007nonparametric}. 
In actuarial science, demography provides mortality force data, which, along with samples from the survival
distribution, yield gradients for the survival distribution function  \citep{frees1998understanding}. 
In stochastic simulation, gradient estimation has been studied for a large class of problems \citep{glasserman2013monte}. 
In discrete event simulation, the gradient can be estimated with a negligible computational burden compared to the effort for obtaining a new response \citep{chen2013enhancing}. 
In meteorology,  wind speed and direction are gradient functions of barometric pressure and 
 are measured over broad geographic regions \citep{breckling2012analysis}.
In dynamical and time series applications, gradient information can be observed or estimated,  as in biological and infectious disease modeling \citep{ramsay2007parameter, dai2021kernel, dai2024post}.
 In traffic engineering, real-time motion sensors can record velocity in addition to positions \citep{solak2002derivative}. 

This paper focuses on nonparametric function estimation under smoothness constraints. 
Rates of convergence often limit the applications of traditional nonparametric estimation methods in high-dimensional settings, where the number of covariates is large \citep{stone1980optimal, stone1982optimal}. 
A considerable amount of research effort has been devoted to countering this curse of dimensionality. 
The additive model is a popular choice  \citep{stone1985additive, hastie1990generalized}. 
An additive model assumes the high-dimensional function to be a sum of one-dimensional functions and drops interactions among covariates in order to reduce the variability of an estimator. 
\cite{stone1985additive} showed that the optimal convergence rate for additive models is the same as that for univariate nonparametric estimation problems.  Thus, the additive models effectively mitigate the curse of dimensionality. 
Additive models, however,  could be too restrictive and lead to wrong conclusions in applications where interactions among the covariates may be present. As a more flexible alternative, smoothing spline analysis of variance (SS-ANOVA) models, the analogs of parametric ANOVA models, have attracted lots of attention \citep{wahba1995smoothing, huang1998projection,lin2006component, zhu2014structured}. In particular, SS-ANOVA models include additive models as special cases. \cite{lin2000tensor} proved that when the interactions among covariates are in
tensor product spaces, the optimal rates of convergence for SS-ANOVA models  \emph{exponentially} depend on the order of interaction. Thus, when SS-ANOVA models are used in problems that involve high-order interactions,
it leads to the requirement of unrealistically enormous dataset sizes for reliable conclusions. 
We call this phenomenon the \emph{curse of interaction}.

We develop a new approach based on partial derivatives to effectively compromise the curse of interaction. 
Let $\{(\bt_i^{(0)},y_i^{(0)}): i=1,\ldots, n\}$ be the function data that follow a regression model,
 \begin{equation}
\label{modelequation1}
Y^{(0)}  = f_0(\bt^{(0)}) + \epsilon^{(0)}.
\end{equation}
Here $\epsilon^{(0)}\in\R$ is a  random error, $f_0:\XX^d\mapsto\R$ is a  function of $d$ covariates  $\bt=(t_1,\ldots,t_d)$, and $\bt^{(0)}\in\XX^d\equiv[0,1]^d$ is the design point.
Write $\partial f_0(\bt)/\partial t_j$ as the $j$th partial derivative of $f_0(\bt)$. 
Let $\{(\bt_i^{(j)},y_i^{(j)}): i=1,\ldots, n; j=1,\ldots,p\}$ be the partial derivatives that follow a regression model, 
\begin{equation}
\label{modelequation}
Y^{(j)}  =\frac{\partial f_0(\bt^{(j)})}{\partial t_j}+\epsilon^{(j)},   \quad j=1,\ldots, p.
\end{equation}
Here $\epsilon^{(j)}$s  are random errors, and $\bt^{(j)}$s are  the design points in $\XX^d$. 
We allow  $Y^{(j)}$ to be directly observable or estimated from function data.
The $p\in\{1,\ldots,d\}$ denotes the number of gradient types.
Without loss of generality, we focus on the first $p$ covariates in model (\ref{modelequation}). 
In particular, when $p=d$, model (\ref{modelequation}) gives the \emph{full} gradient data.
We allow for a relaxed error structure for both function and gradient data. 
Specifically, we assume the random errors $\epsilon^{(0)}$ and $\epsilon^{(j)}$s in models  \eqref{modelequation1} and \eqref{modelequation} to satisfy,
\begin{equation}
\label{eqn:errorstruct}
\begin{aligned}
 &\E[\epsilon_i^{(j)}] = o(n^{-1/2}), \ \  \text{Var}[\epsilon_i^{(j)}] = \sigma_j^2<\infty,\\
& \text{Cov}[\epsilon_i^{(j)},\epsilon_{i'}^{(j')}] =O\left(|i-i'|^{-\Upsilon}\right) \text{ for some } \Upsilon>1,
\end{aligned}
\end{equation}
where $i\neq i'$ and $j,j'=0,1,\ldots,p$. 
We assume the short-range correlation in (\ref{eqn:errorstruct}) with some $\Upsilon>1$. 
This assumption is generally valid in practice, as gradient data are often estimated by using local function data through methods such as finite-difference techniques.
\change{We provide three concrete examples in Appendix to elaborate on the assumption  \eqref{eqn:errorstruct}.}
Moreover, random errors in \eqref{eqn:errorstruct} can be uncentered and correlated, which are typical for estimated gradients, and include the i.i.d. errors in \citet{hall2007nonparametric} as a special case.

The SS-ANOVA model \citep{wahba1995smoothing} amounts to the assumption that
\begin{equation}
\label{eqn:anovadecompfti}
f_0(\bt) = \mbox{constant} + \sum_{j=1}^d f_{0j}(t_j) + \cdots + \sum_{1\leq j_1<j_2<\cdots<j_r\leq d}f_{0j_1j_2\cdots j_r}(t_{j_1},t_{j_2},\ldots,t_{j_r}),
\end{equation}
where the component functions include main effects $f_{0j}$s, two-way interactions $f_{0j_1j_2}$s, and so on. 
Component functions are modeled nonparametrically, and we assume that they reside in certain reproducing kernel Hilbert spaces  \citep[RKHS,][]{wahba1990}.
The series on the right-hand side of \eqref{eqn:anovadecompfti} is truncated to some order $r$ of interactions to enhance interpretability.
We call $f_0(t)$ as \emph{full} or \emph{truncated} interaction SS-ANOVA model if $r=d$ or $1\leq r<d$,  respectively.
The SS-ANOVA model \eqref{eqn:anovadecompfti} can be identified with space,
\begin{equation}
\label{eqn:anovadechi}
\begin{aligned}
 \HH &  = \{1\} \oplus \sum_{j=1}^d \HH^{j} \oplus\cdots\oplus\sum_{1\leq j_1<j_2<\cdots<j_r\leq d} \left[\HH^{j_1}\otimes \HH^{j_2}\otimes\cdots\otimes\HH^{j_r}\right].
\end{aligned}
\end{equation}
The components of the SS-ANOVA model in (\ref{eqn:anovadecompfti}) are in the mutually orthogonal subspaces of $\HH$ in  (\ref{eqn:anovadechi}). 
The additive model can be viewed as a special case of the SS-ANOVA model \eqref{eqn:anovadecompfti}
 by taking $r=1$.  We assume that all component functions come from a common RKHS $(\HH_1,\|\cdot\|_{\HH_1})$ given by $\HH^{j} \equiv\HH_1$ for  $j=1,\ldots,d$.
Obviously the linear model is a special example of \eqref{eqn:anovadecompfti} by taking $r=1$ and letting $\HH_1$  be the collection of all univariate linear functions defined over $\XX$. \change{Another canonical example of $\{1\}\oplus\HH_1$ is the $m$th  order Sobolev space $\mathcal W_2^m(\XX)$; see, e.g., \cite{wahba1990} for further examples.}

We study the possibility of near-parametric rates in the general setting of SS-ANOVA models.
Suppose the eigenvalues of the kernel function decay polynomially, i.e., its $\nu$th largest eigenvalue is of the order $\nu^{-2m}$. Our results show that the minimax optimal rates for estimating $f_0$ under the \emph{full} interaction (i.e., $r=d$) are 
\begin{equation}
\label{eqn:minimaxraternnrd}
\RR(n,d,r,p)  = \begin{cases}
\left[n(\log n)^{1+p-d}\right]^{-\frac{2m}{2m+1}},  & \mbox{ if } 0\leq p< d,\\
n^{-\frac{2md}{(2m+1)d-2}}\mathbbm{1}_{d\geq 3}+n^{-1}(\log n)^{d-1}\mathbbm{1}_{d< 3},& \mbox{ if } p=d.
\end{cases}
\end{equation}
 The rates in \eqref{eqn:minimaxraternnrd} present an interesting two-regime dichotomy between the scenerios of $0\leq p<d$ and $p=d$.
When $0\leq p<d$, the  rate given by \eqref{eqn:minimaxraternnrd} matches with the  minimax optimal rate for estimating a  $(d-p)$-interaction model  without gradient information \citep{lin2000tensor}. 
\change{For example, when $p=0$ with no partial derivative data, the rate from \eqref{eqn:minimaxraternnrd} is $[n(\log n)^{1-d}]^{-2m/(2m+1)}$. This rate aligns with the known rate for estimating a $d$-interaction SS-ANOVA model \citep{lin2000tensor}. However, with a large $d$, this rate is heavily affected by the exponential term $(\log n)^{d-1}$, which makes the estimation challenging and leads to the curse of interaction. The inclusion of gradient data provides a substantial advantage in overcoming these challenges.}
For instance, when $p=d-1$,   the rate in \eqref{eqn:minimaxraternnrd} becomes $n^{-2m/(2m+1)}$, which is the same as the optimal rate for estimating additive models without gradient information and independent of $d$ \citep{stone1985additive}. 
This  indicates that  SS-ANOVA models can be immune to the curse of interaction  through the use of  partial derivative data.  

On the other hand, when $p=d\geq 3$, the rate in (\ref{eqn:minimaxraternnrd}) becomes
\begin{equation*}
\RR(n,d,r,p)  = n^{-\frac{2md}{(2m+1)d-2}}.
\end{equation*}
This rate converges \textit{faster} than the optimal rate for additive models $n^{-2m/(2m+1)}$. When $p=d=2$, the rate in  (\ref{eqn:minimaxraternnrd}) is
$\RR(n,d,r,p) = n^{-1}\log n$.
If $p=d=1$, the  rate in (\ref{eqn:minimaxraternnrd}) is the same as the \emph{parametric} convergence rate,
$\RR(n,d,r,p) = n^{-1}$.
\change{It is also worth noting that when $f_0$ has truncated interaction (i.e., $r<d$), the rates also improve by incorporating partial derivatives, which will be discussed in Section \ref{sec:randomdesign}.} In particular, the rate for additive models (i.e., $r=1$) under $p=d$ matches with the \emph{parametric}  rate,
$\RR(n,d,r,p) = n^{-1}$.

\change{In the literature, various studies have outlined the construction of linear estimators for the linear functionals of $f_0$, with the difficulty of estimation characterized by a modulus of continuity \citep{donoho1991geometrizing, donoho1994statistical, klemela2001sharp, cai2005adaptive}. These studies are relevant to our work in two ways: first, they demonstrate the feasibility of achieving a parametric rate in estimating a univariate function $f_0$ from noisy derivative data, which
 aligns with the rate in our paper as a special case in the univariate context. Second, they provide the optimal rate for estimating partial derivatives of $f_0$ from observations of $f_0$, which differs from our target of estimating $f_0$ itself. 
Our methodology and new convergence rates bridge a gap in these studies by focusing on incorporating noisy gradient data for  multivariate function estimation. 
A similar observation of accelerated rates has been noted earlier with \emph{higher-order} derivatives \citep{hall2007nonparametric, hall2010nonparametric}.  Our results suggest that such a phenomenon holds with \emph{first-order} derivatives and applies to  general SS-ANOVA models involving tensor product spaces.
While our theoretical comparison primarily involves \citet{hall2007nonparametric} due to its seminal importance and relevance to integrating noisy gradients in nonparametric regression, we recognize the continuous advancements in the field over the last decade. These developments include applications of joint models \eqref{modelequation1} and \eqref{modelequation} in areas such as stochastic simulations and Gaussian process methodologies, where gradient data enhances estimation and prediction \citep[see, e.g.,][]{riihimaki2010gaussian,  chen2013enhancing, fu2014regression, wang2016estimating, zhang2023gradient, lim2024estimating}.  
Nonetheless, a comprehensive statistical theory explaining the benefit of incorporating noisy gradient data has been lacking.  This paper develops a theoretical framework that shows how gradient data can mitigate the curse of interaction and significantly enhance the scalability of nonparametric modeling, especially for high-dimensional SS-ANOVA models.
}

\subsection{Our contributions}
\noindent
We develop an approach and computational algorithm to incorporate partial derivatives and lead to methods useful to practitioners in many areas of science and engineering.
We obtain a new theory that reveals a behavior universal to this class of nonparametric estimation problems. 
Our proposal and theoretical results considerably differ from the existing works in multiple ways, which are summarized as follows.

Firstly, our results broaden the i.i.d. error structure by allowing the random errors in function data and gradient data to be biased and correlated. This relaxed assumption is in line with applications when the gradient data are estimated \citep{chen2013enhancing}.

Secondly, we develop a new approach and computational algorithm in RKHS that can easily incorporate gradient information. The proposed estimator also enjoys interpretability by providing a direct description of interactions. We also find that partial derivatives can reduce interactions in terms of the minimax convergence rates.  


Finally, we obtain a sharper theory on the estimation with partial derivatives. We show that when $p=d-1$, the optimal rate for estimating $d$-dimensional SS-ANOVA models under full interaction is $n^{-2m/(2m+1)}$, which is independent of the interaction order $r$. Hence the SS-ANOVA models are immune to the \emph{curse of interaction} via using gradients.  In contrast, \cite{hall2007nonparametric} showed that when $p=d-1$, the convergence rate for estimating $d$-dimensional functions is $n^{-2m/(2m+d-1)}$, which has the curse of dimensionality in $d$. Therefore, our results show that partial derivatives are useful for the scalability of nonparametric estimation in high dimensions, particularly when using the SS-ANOVA models.

The rest sections are organized as follows. 
We first provide background in  Section \ref{sec:notation}, and show main results in Section \ref{sec:randomdesign}.
Section \ref{sec:realexamples} presents synthetic and real data examples.
Section \ref{sec:relatedwork} discusses related works.
We provide conclusion in Section \ref{sec:discussion}.
The results under other types of designs and their proofs, together with additional numerical examples, are relegated to the Appendix.


\section{Background}
\label{sec:notation}
\noindent
We begin with a motivating example with partial derivatives. Then we briefly review basic facts about RKHS for the setting of our interest. 

\subsection{Motivating example}
\label{sec:ssanovaanderror}

 \begin{figure}[!ht]
    \centering
    \begin{subfigure}[b]{0.33\textwidth}
        \centering
        \includegraphics[width=\textwidth]{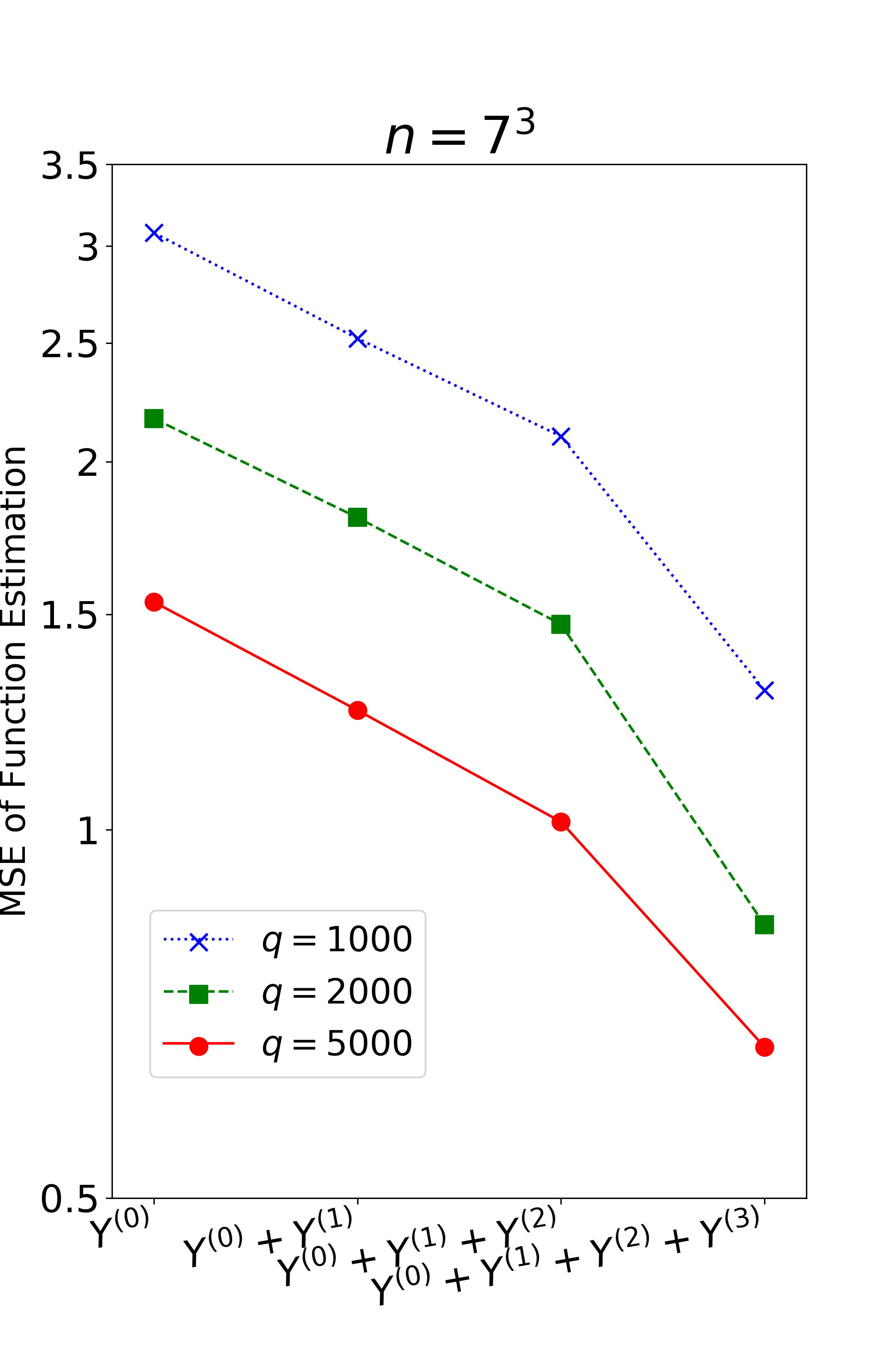}
    \end{subfigure}%
    \begin{subfigure}[b]{0.33\textwidth}
        \centering
        \includegraphics[width=\textwidth]{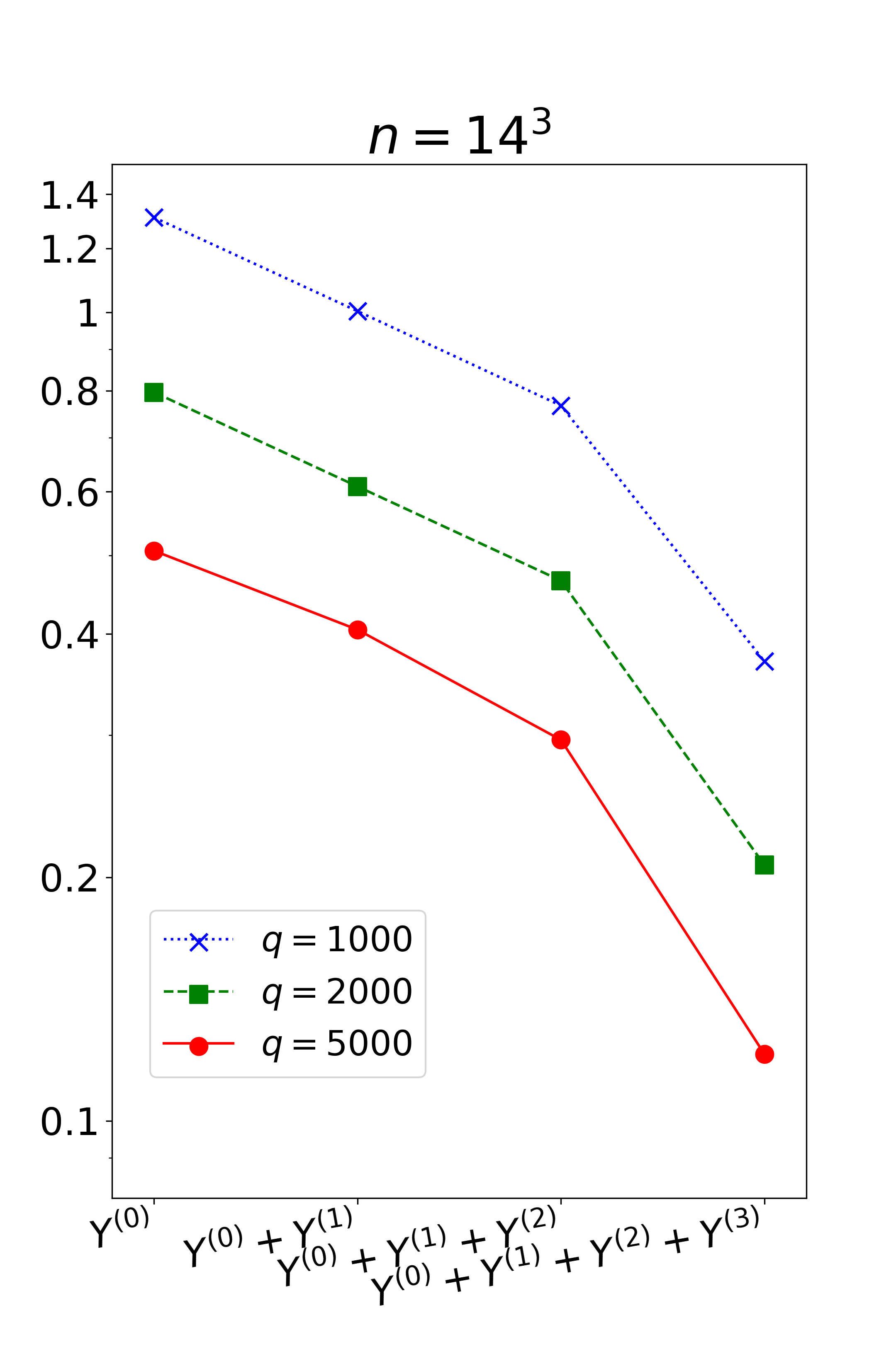}
    \end{subfigure}    
       \begin{subfigure}[b]{0.33\textwidth}
        \centering
        \includegraphics[width=\textwidth]{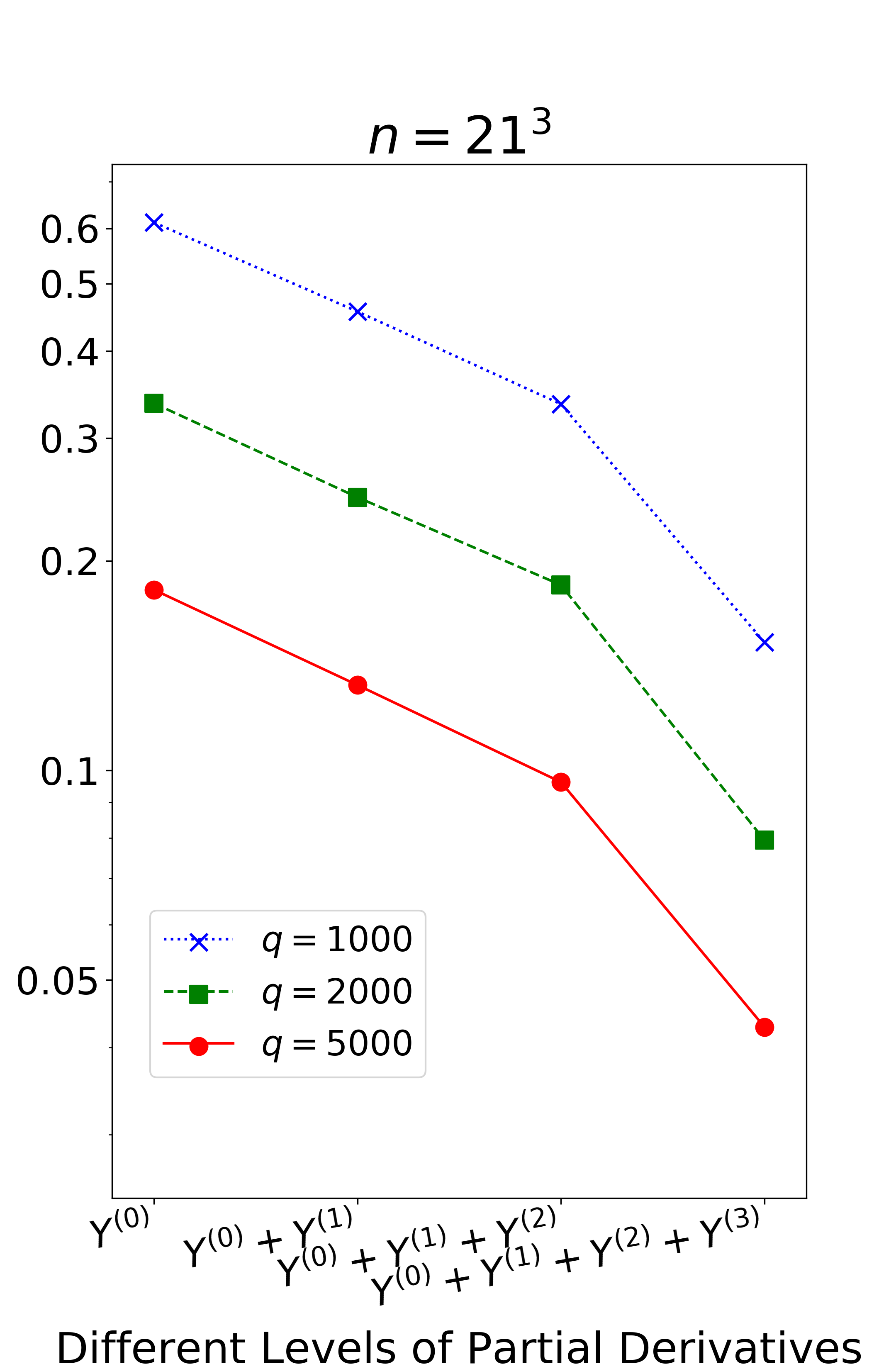}
    \end{subfigure}
    \caption{Estimation error of our estimator incorporating different levels of gradient information,  for the stochastic simulation example.   The $y$-axis is in the log scale. }
\label{fig:blacksholes}
\end{figure}

\noindent
We study a stochastic simulation application to motivate models  \eqref{modelequation1} and \eqref{modelequation}. 
Let $h(\bt,\omega)$ be the response of  a stochastic simulation, which has a design point $\bt\in\XX^d$ and a random variable $\omega$. 
It is of interest to build  fast  and accurate estimation for $f_0(\bt) =\E_\omega[h(\bt,\omega)]$ \citep{chen2013enhancing, glasserman2013monte}. 
At each replication $k=1,\ldots,q$, the stochastic simulation has a different random variable $\omega_k$.  
A user can select design $\bt^{(0)}$ and run the stochastic simulation to obtain a response
$Y_k(\bt^{(0)})  =h(\bt^{(0)},\omega_k) = f_0(\bt^{(0)}) + \epsilon^{(0)}_k$, where   $\epsilon^{(0)}_k$ is i.i.d. centered simulation noise. 
In practice, it is common to average  responses  to reduce the variance of simulation noises, i.e., let
$Y^{(0)} = [Y_1(\bt^{(0)}) +Y_2(\bt^{(0)}) + \cdots +Y_q(\bt^{0})]/q$, where $q$ is the number of simulation replications and is at the order of hundreds or thousands.
Then  the response $Y^{(0)}$ follows model (\ref{modelequation1}), where
 $ \epsilon^{(0)}$ is the averaged simulation noise.
Under regularity  conditions ensuring the interchange
of expectation and differentiation \citep{l1990unified}, the infinitesimal perturbation analysis (IPA) gives the  gradient estimator of $f_0(\bt)$ that follows model \eqref{modelequation},
\begin{equation*}
Y^{(j)}=\frac{\partial }{\partial t_j}h(\bt^{(j)},\omega),\quad \bt^{(j)}\in\XX^d, \ j=1,\ldots,p, \ 1\leq p\leq d.
\end{equation*}
Moreover, the IPA estimators are unbiased, $\E_\omega[Y^{(j)}]=\partial f_0/\partial t_j$ \citep{glasserman2013monte}. We provide details of our stochastic simulation in Section \ref{exp:blacksholes}. The results are reported in Figure  \ref{fig:blacksholes}, which shows mean-squared errors (MSEs) for varying sample size $n$, replication number $q$, and different methods. Those include stochastic kriging with function data (i.e., $p=0$), our estimator with function and one type of gradient data (i.e., $p=1$), two types of gradient data (i.e., $p=2$), the full gradient data (i.e., $p=3$). A significant decrease in MSEs is observed when incorporating partial derivatives. 
Moreover, the computational cost for obtaining the gradient estimator is relatively low, as calculating the IPA estimator $Y^{(j)}$ does not need additional replication of the simulation.  In contrast, getting a new function response $Y^{(0)}$ requires $q$ new replications of the simulation, and each replication could incur a high cost. 

\subsection{Reproducing kernel for partial derivatives}
\noindent
We briefly review some basic facts about RKHS. Interested readers are referred to \cite{aronszajn1950theory} and \cite{wahba1990} for further details. 
\change{Let $K$ be a Mercer kernel that is a symmetric positive semi-definite and square-integrable function on $\XX\times \XX$.}
\change{It can be uniquely identified with the Hilbert space $\HH_1$ that is the completion of  $\{\sum_{i=1}^Nc_iK(t_i,\cdot):t_i\in\XX, c_i\in\R, i=1,\ldots,N\}$}
under the inner product
$\left\langle \sum_ic_iK(t_i,\cdot), \sum_jc_jK(t_j,\cdot)\right\rangle_{\HH_1} = \sum_{i,j}c_ic_jK(t_i,t_j)$.
Most commonly used kernels are differentiable, which we shall assume in what follows. In particular, we assume that
\begin{equation}
\label{eqn:partial2tt'kcx1x1}
\frac{\partial^2 }{\partial t\partial t'} K(t,t')\in \mathcal C(\XX\times \XX).
\end{equation}
where $\mathcal C(\cdot)$ is the space of continuous functions.
Let the kernel
$K_d((t_1,\ldots,t_d)^\top,(t_1',\ldots,t_d')^\top) = K(t_1,t_1') \cdots  K(t_d,t_d')$.
Then $K_d(\cdot,\cdot)$ is the kernel corresponding to the RKHS $(\HH,\|\cdot\|_{\HH})$ in \eqref{eqn:anovadechi}; see, e.g., \cite{aronszajn1950theory}.
The condition \eqref{eqn:partial2tt'kcx1x1} together with the  continuity of $\langle\cdot,\cdot\rangle_{\HH}$ yield that for any $g\in\HH$, $\partial g(\bt)/\partial t_j  = \partial \langle g, K_d(\bt,\cdot)\rangle_\HH/\partial t_j =  \left\langle g, \partial K_d(\bt,\cdot)/\partial t_j\right\rangle_\HH.$
Thus, the gradient $\partial g(\bt)/\partial t_j$  is a bounded linear functional in $\HH$ and has a representer $\partial K_d(\bt,\cdot)/\partial t_j$.
By Mercer’s theorem \citep{riesz1955}, the kernel function $K$ admits an eigenvalue decomposition:
\begin{equation}
\label{eqn:spectral}
K(t,t') = \sum_{\nu\geq 1}\lambda_{\nu}\psi_{\nu}(t)\psi_{\nu}(t'),
\end{equation}
where $\lambda_{1}\geq\lambda_{2}\geq\cdots\geq0$ are eigenvalues and $\{\psi_{\nu}:\nu\geq 1\}$ are the corresponding eigenfunctions. \change{For example, $\lambda_\nu\asymp \nu^{-2m}$  for $\WW_2^m(\XX)$ under the Lebesgue measure \citep{wahba1990}, which will be also discussed in Appendix.}

\section{Main Results}
\label{sec:randomdesign}
\noindent
In this section, we present a new approach for nonparametric estimation via partial derivatives and develop a fast algorithm. We also derive a new theory and show a convergence behavior universal to this class of estimation problems. 

\subsection{Estimation via partial derivatives}
\noindent
We introduce a method that merges function and derivative information for better estimation. 
When the function $f_0$ in \eqref{eqn:anovadecompfti} is smooth in $\HH$, we add the empirical loss of partial derivatives as a penalty. Combining these information, we derive the function $\widehat{f}_n$ that meets the smoothness criteria and aligns closely with the observed data,
\begin{equation}
\label{scheme1}
\begin{aligned}
\change{\widehat{f}_{n}} = \underset{\change{\|f\|_\HH\leq R_n}}{\arg\min}& \left\{\frac{1}{n}\sum_{i=1}^{n}\left[y_i^{(0)}- f(\bt_i^{(0)})\right]^2+\sum_{j=1}^pw_j\cdot \frac{1}{n}\sum_{i=1}^{n}\left[y_i^{(j)}-\frac{\partial f}{\partial t_j}(\bt_i^{(j)})\right]^2\right\}.
\end{aligned}
\end{equation}
Here $\change{R_n}\geq 0$ is an appropriately chosen Hilbert radius, and $w_j\geq 0$ is a weight parameter, where a natural choice is $w_j = \sigma_0^2/\sigma_j^2$.
If $\sigma^2_0$ and $\sigma^2_j$ are unknown, we can replace them with consistent estimators for variances \citep{hall1990asymptotically}.
\change{The concept of derivative-based penalty has also been employed in the generalized profiling approach of  \citet{ramsay2007parameter}, which derives a penalty by comparing the derivative of the estimated function to a trajectory generated by ordinary differential equations (ODEs). However, the approach in \eqref{scheme1} is different by directly comparing the derivative of the estimated function with either observed or estimated  derivatives at discrete data points, which avoids the complexities associated with ODE computations.}  
The following theorem gives a closed-form solution to  \eqref{scheme1}.
\begin{theorem}
\label{thm:uniqueminimizer}
Assume that kernel $K$ satisfies the differentiability condition \eqref{eqn:partial2tt'kcx1x1}.
\change{Then, for any $R_n\geq 0$, there exists a  minimizer $\widehat{f}_{n}(\bt)$ of \eqref{scheme1} in a finite-dimensional space,
\begin{equation*}
\widehat{f}_n(\bt) = \sum_{i=1}^n\alpha_{i0} K_d(\bt_i^{(0)},\bt) + \sum_{j=1}^p\sum_{i=1}^n \alpha_{ij}\frac{\partial K_d}{\partial t_j}(\bt_i^{(j)},\bt),
\end{equation*}
where the coefficients ${\boldsymbol \alpha}_j = (\alpha_{1j},\ldots,\alpha_{nj})^\top\in\R^n$ for $j=0,1,\ldots,p$.}
\end{theorem}
\noindent
This theorem is a generalization of the well-known representer lemma for smoothing
splines \citep{wahba1990}. It in effect turns an infinity-dimensional optimization
problem into an optimization problem over finite number of coefficients. We will devise a fast algorithm for this optimization in Section \ref{sec:compalg} and show its scalability for large data.

The estimator (\ref{scheme1}) is different from existing methods of incorporating gradients. For example, \cite{morris1993bayesian} proposed a stationary Gaussian process to combine noiseless gradients, whereas the estimator \eqref{scheme1}  applies to noisy gradients. 
\cite{hall2007nonparametric} studied a regression-kernel estimator to incorporate noisy derivatives and required special structures on the observed derivatives.
However, the estimator \eqref{scheme1}  can incorporate all types of estimated or observed partial derivatives. 
\cite{hall2010nonparametric} used a series-type estimator but could have a curse of dimensionality problem. In contrast, \eqref{scheme1}  can scale up to a large dimension $d$.
\cite{chen2013enhancing} considered a stochastic kriging method, where the correlation coefficients between gradients and function data are required to be estimated. 
Differently, it is unnecessary to estimate such correlations for implementing \eqref{scheme1}.
Moreover, we will demonstrate that the estimator (\ref{scheme1})  outperforms competing alternatives through numerical examples in Section \ref{sec:realexamples}.

\subsection{Computational algorithm}
\label{sec:compalg}
\noindent
We now develop a  fast algorithm for computing the  minimizer $\widehat{f}_{n}(\bt)$ in Theorem \ref{thm:uniqueminimizer}.
Note that $\widehat{f}_{n}(\bt)$ can be further written as, for any $\bt\in\XX^d$, 
\begin{equation} \label{eqn:representer}
\widehat{f}_n(\bt) = \tilde{\bPsi}_d(\bt)^\top \tilde{\bc}_0 + \sum_{j=1}^p\frac{\partial \tilde{\bPsi}_d(\bt)^\top \tilde{\bc}_j}{\partial t_j},
\end{equation} 
where $\tilde{\bPsi}_d(\bt) = \left[ \tilde{\bPsi}^{\otimes_1}(t_1)^\top, \ldots, \tilde{\bPsi}^{\otimes_1}(t_d)^\top, \tilde{\bPsi}^{\otimes_2}(t_1,t_2)^\top,\ldots, \tilde{\bPsi}^{\otimes_r}(t_{d-r+1},t_{d-r+2},\ldots,t_d)^\top  \right]^\top$. \change{The column vector $\widetilde{\bPsi}^{\otimes_1}(t)$ has the $\nu$th element equal to $\sqrt{\lambda_\nu}\psi_\nu(X)$ for $\nu\geq 1$.
The vector $\bPsi^{\otimes_2}(t_i,t_j) = \bPsi^{\otimes_1}(t_i)\otimes\bPsi^{\otimes_1}(t_j)$ is generated by the Kronecker product that combines two vectors $\bPsi^{\otimes_1}(t_i)$ and $\bPsi^{\otimes_1}(t_j)$ into a single vector, where for each element in the first vector $\bPsi^{\otimes_1}(t_i)$, we multiply the entire second vector $\bPsi^{\otimes_1}(t_j)$ by that element, and the resulting vectors from each multiplication are then concatenated, forming a long vector that captures all pairwise interactions between the elements of $\bPsi^{\otimes_1}(t_i)$ and $\bPsi^{\otimes_1}(t_j)$.
Similarly, $\tilde{\bPsi}^{\otimes_r}(t_{d-r+1},t_{d-r+2},\ldots,t_d) = \bPsi^{\otimes_1}(t_{d-r+1})\otimes \bPsi^{\otimes_1}(t_{d-r+2})\otimes\cdots \otimes\bPsi^{\otimes_1}(t_d) $ is the Kronecker product of the $r$ corresponding vectors.}
 Here $\tilde{\bc}_j  = \Big[ \widetilde{\bPsi}_d(\bt^{(j)}_1),\ldots,\widetilde{\bPsi}_d(\bt^{(j)}_{n}) \Big] {\boldsymbol \alpha}_j$ is the infinite-dimensional coefficient vector, where $j=0,1,\ldots,p$.
 
The key idea is to employ the random feature mapping \citep{RR07, dai2022kernel} to approximate the kernel function, which enables us to construct a projection operator between the RKHS and the original predictor space. Specifically,   
if the kernel functions that generate $\mathcal{H}_1$ are shift-invariant, i.e., $K(t,t') = K(t-t')$, and integrate to one, i.e., $\int_\XX K(t-t') d(t-t') = 1$, then the Bochner's theorem \citep{bochner1934theorem} states that such kernel functions satisfy the Fourier expansion: 
\begin{equation*}
\begin{aligned}
K(t-t') &=\int_\R p(w)\exp\left\{\sqrt{-1}w(t-t')\right\}dw,
\end{aligned}
\end{equation*}
where $p(w)$ is a probability density defined by
\begin{equation*} \label{eqn:probdensityf}
p(w) = \int_\XX K(t)e^{-2\pi\sqrt{-1} wt}dt.
\end{equation*}
We note that many kernel functions are shift-invariant and integrate to one. Examples include the Mat\'ern  kernel, $K(t,t') = \tilde{\tau}_1(1+|t-t'|/\tau_1+|t-t'|^2/3\tau_1^2) e^{-|t-t'|/\tau_1}$,
 the Laplacian kernel, $K(X,X')=\tilde{\tau}_2e^{-|X-X'|/\tau_2}$, the Gaussian kernel, $K(X,X')=\tilde{\tau}_3e^{-\tau_3^2|X-X'|^2/2}$, and the Cauchy kernel, $K(X,X')=\tilde{\tau}_4(1+\tau_4^2|X-X'|^2)^{-1}$, where $\tilde{\tau}_1,\tilde{\tau}_2,\tilde{\tau}_3,\tilde{\tau}_4$ are the normalization constants, and $\tau_1, \tau_2, \tau_3,\tau_4$ are the scaling parameters. It is then shown that \citep{RR07} the minimizer in Theorem \ref{thm:uniqueminimizer} can be approximated by,
\begin{equation*}
\widehat{f}_n(\bt) = \bPsi_d(\bt)^\top {\bc_{0}} + \sum_{j=1}^p \frac{\partial\bPsi_d(\bt)^\top {\bc_{j}}}{\partial t_j},
\end{equation*}
where $\bPsi_d(\bt) = \left[ \bPsi^{\otimes_1}(t_1)^\top,\ldots,\bPsi^{\otimes_1}(t_d)^\top, \bPsi^{\otimes_2}(t_1,t_2)^\top,\ldots, \bPsi^{\otimes_r}(t_{d-r+1},t_{d-r+2},\ldots,t_d)^\top  \right]^\top$, and $\bPsi^{\otimes_1}(t_j) = \left[ \tilde{\psi}_1(t_j),\ldots,\tilde{\psi}_s(t_j) \right]^\top \in \R^s$ is a vector of $s$ Fourier bases with the frequencies drawn from the density $p(w) $, i.e., 
\begin{equation}
\label{eqn:fouriermc}
\begin{aligned}
& \omega_{j,\nu} \overset{\text{i.i.d.}}{\sim} p(\omega), \quad\quad  b_{j,\nu} \overset{\text{i.i.d.}}{\sim} \textrm{Uniform}[0,2\pi], \\
&\tilde{\psi}_\nu(t_j) = \sqrt{\frac{2}{s}}\cos(t_j \omega_{j,\nu} + b_{j,\nu}), \quad\quad j=1,\ldots,d, \; \nu=1,\ldots,s,
\end{aligned}
\end{equation}
and  $\bPsi^{\otimes_2}(t_i,t_j) = \bPsi^{\otimes_1}(t_i)\otimes\bPsi^{\otimes_1}(t_j) \in \R^{s^2}$, and so on. 
We write the augmented random feature vector as,
\begin{equation} \label{eqn:psi2p}
\bPsi_{(p+1)d}(\bt) = \left(\bPsi_d(\bt)^\top, \frac{\partial \bPsi_d(\bt)^\top}{\partial t_1}, \ldots,\frac{\partial \bPsi_d(\bt)^\top}{\partial t_p} \right)^\top.
\end{equation}
Then the minimizer in Theorem \ref{thm:uniqueminimizer}  can be approximated by,
\begin{equation}  \label{eqn:RF}
\widehat{f}_n(\bt)  = \bPsi_{(p+1)d}(\bt)^\top {\bc_{(p+1)d}}.
\end{equation}
We estimate the coefficient vector $\bc_{(p+1)d} = (\bc_0^\top, \bc_1^\top,\ldots, \bc_{p}^\top)^\top$ by minimizing the following convex objective function,
\begin{equation}
\label{eq: kernel_reg} 
\begin{aligned}
\change{\frac{1}{n}\sum_{i=1}^{n}\left[y_i^{(0)}- \widehat{f}_n(\bt_i^{(0)})\right]^2+\sum_{j=1}^pw_j\cdot \frac{1}{n}\sum_{i=1}^{n}\left[y_i^{(j)}-\frac{\partial \widehat{f}_n}{\partial t_j}(\bt_i^{(j)})\right]^2 + \lambda\sum_{j=0}^p\|\bc_j\|_2^2,}
\end{aligned}
\end{equation}
where $\lambda\geq 0$ is the penalty parameter.
We remark that the penalty in \eqref{eq: kernel_reg} is different from the penalty in kernel ridge regression \citep{wainwright2019high}, which takes the form $\| \Psi_{(p+1)d}(\bt)^\top {\bc_{(p+1)d}}\|_{\HH}^2$. Since the random feature mapping generally cannot form an orthogonal basis, there is no closed-form representation of the RKHS norms $\| \Psi_{(p+1)d}(\bt)^\top {\bc_{(p+1)d}}\|_{\HH}^2$ in our setting. As a result, the kernel ridge regression penalty is difficult to implement, and instead we adopt the $L_2$ penalty in \eqref{eq: kernel_reg} that is easy for computing. 
We choose the smoothing parameter $\lambda$ in (\ref{eq: kernel_reg}) by generalized cross-validation (GCV)  \citep{golub1979generalized}. Let $A(\lambda)$ be the influence matrix as $\widehat{y} = A(\lambda)y$, where $y$ is the vector of function and gradient data $y = (y_1^{(0)},\ldots,y^{(0)}_n,\ldots,y^{(p)}_1,\ldots,y^{(p)}_n)^\top$, and $\widehat{y}$ is the  estimate,  $\widehat{y} =  (\widehat{f}_{n}(\bt_1^{(0)}),\ldots,\widehat{f}_{n}(\bt_n^{(0)}),\ldots,\partial\widehat{f}_{n}/\partial t_p(\bt_1^{(p)}),\ldots,\partial\widehat{f}_{n}/\partial t_p(\bt_n^{(p)}))^\top$. 
Then GCV selects $\lambda\geq 0$ by minimizing the following risk,
\begin{equation*}
\text{GCV}(\lambda) = \frac{\|\widehat{y} - y\|^2}{[n^{-1}\text{tr}(I-A(\lambda))]^2}.
\end{equation*}

The use of random feature mapping achieves potentially substantial dimension reduction. \change{More specifically, the estimator in (\ref{eqn:RF}) only requires to learn the finite-dimensional coefficient $\bc_{(p+1)d}$, compared to the estimator in (\ref{eqn:representer}) that involves an infinite-dimensional vector $\tilde{\bc}_j$ for $j=0,1,\ldots,p$.} It is known that the random feature mapping obtains the optimal bias-variance tradeoff if $s$ scales at a certain rate and $s/n\to0$ when $n$ grows \citep{rudi2017generalization}. We note that the random feature mapping also efficiently reduces the computational complexity. Given any $(d,r,p)$, the computation complexity of the estimator in (\ref{eqn:RF}) is only $O(ns^2)$, compared to the computation complexity of the kernel estimator in Theorem \ref{thm:uniqueminimizer} that is $O(n^3)$. The saving of the computation is substantial if $s/n\to 0$ as $n$ grows.

\begin{algorithm}[t!]
\caption{Estimation via partial derivatives.} 
\begin{algorithmic}[1]
\STATE \textbf{Input}:  Function data $\{(\bt_i^{(0)},y_i^{(0)}): i=1,\ldots, n\}$, partial derivatives $\{(\bt_i^{(j)},y_i^{(j)}): i=1,\ldots, n; j=1,\ldots,p\}$, weight parameters $\{w_j:j=1,\ldots,p\}$.
\STATE \textbf{Step 1}:  Sample $d$ of i.i.d. $s$-dimensional random features $\{w_\nu,b_\nu\}_{\nu=1}^s$ by (\ref{eqn:fouriermc}), and construct the augmented random feature vector $\bPsi_{(p+1)d}(\bt)$ by (\ref{eqn:psi2p}).  
\STATE \textbf{Step 2}: Solve the coefficient vector $\bc_{(p+1)d}$ by \eqref{eq: kernel_reg}.
\STATE \textbf{Output}: Function estimate $\widehat{f}_n(\bt)$ in \eqref{eqn:RF}.
\end{algorithmic} 
\label{alg:mixedgradient}
\end{algorithm}

We summarize the above estimation procedure in Algorithm \ref{alg:mixedgradient}.

\subsection{Minimax optimality}
\label{sec:minimaxoptimality}
\noindent
We show that our proposed estimator achieves optimality. Suppose that design points $\bt^{(0)}$  in \eqref{modelequation1} and $\bt^{(j)}$s in  \eqref{modelequation} are independently drawn from $\Pi^{(0)}$ and $\Pi^{(j)}$s, respectively, where $\Pi^{(0)}$ and $\Pi^{(j)}$s have densities bounded away from zero and infinity.
We first present a minimax lower bound in the presence of partial derivatives.
\begin{theorem}
\label{theorem:lowerbdfNlambdaregrandom}
Assume that $\lambda_\nu \asymp \nu^{-2m}$ for some $m>3/2$  \change{and the kernel $K$ admits the decomposition in \eqref{eqn:spectral}}. Under the regression models \eqref{modelequation1} and \eqref{modelequation} where $f_0$ follows the SS-ANOVA model \eqref{eqn:anovadecompfti} and \change{$\|f\|_{\HH}\leq R_n$}.  Then under the error structure (\ref{eqn:errorstruct}), there exists a constant $c$ such that
\begin{align*}
\underset{n\to\infty}{\lim\inf}\inf_{\tilde{f}}\sup_{f_0\in\HH}\P& \left\{\int_{\XX^d}\left[\tilde{f}(\bt)-f_0(\bt)\right]^2d\bt \geq c\left(\left[n(\log n)^{1-(d-p)\wedge r}\right]^{-\frac{2m}{2m+1}} \mathbbm{1}_{0\leq p<d} \right.\right.\\
& \left.\left. \quad\quad + \left[n^{-\frac{2mr}{(2m+1)r-2}} \mathbbm{1}_{r\geq 3}+n^{-1}(\log n)^{r-1}  \mathbbm{1}_{r<3}\right] \mathbbm{1}_{p=d}\vphantom{\int_{\XX^d}\left\{\tilde{f}(\bt)-f_0(\bt)\right\}^2d\bt}\right)\right\}>0,
\end{align*}
where the infimum of $\tilde{f}$ is taken over all measurable functions of the data.
\end{theorem}
\noindent
This lower bound is new in the literature, and its proof is established via Fano's lemma \citep{tsybakovintroduction}. 
Next, we show that the lower bound is attainable via our estimator.
 \begin{theorem}
\label{thm:mainupperrateestf0}
Assume that $\lambda_\nu \asymp \nu^{-2m}$ for some $m>3/2$ \change{and the kernel $K$ admits the decomposition in \eqref{eqn:spectral}.} Under the regression models \eqref{modelequation1} and \eqref{modelequation} where $f_0$ follows the SS-ANOVA model \eqref{eqn:anovadecompfti} and \change{$\|f\|_{\HH}\leq R_n$}. \change{Then under the error structure (\ref{eqn:errorstruct}) and with the number of random features in \eqref{eqn:fouriermc} set to $s=O(n\log n)$, the estimator $\widehat{f}_{n}$ in  (\ref{eqn:RF}) satisfies}
\begin{align*}
\lim_{C\to\infty}\underset{n\to\infty}{\lim\sup}& \sup_{f_0\in\HH} \P \left\{\int_{\XX^d}\left[\widehat{f}_{n}(\bt)-f_0(\bt)\right]^2d\bt \right.\leq C\left(\left[n(\log n)^{1-(d-p)\wedge r}\right]^{-\frac{2m}{2m+1}} \mathbbm{1}_{0\leq p<d} \right.\\
& \left.\left.  \quad\quad \quad\quad + \left[n^{-\frac{2mr}{(2m+1)r-2}} \mathbbm{1}_{r\geq 3}+n^{-1}(\log n)^{r-1}  \mathbbm{1}_{r<3}\right]  \mathbbm{1}_{p=d}\vphantom{\left[n(\log n)^{1-(d-p)\wedge r}\right]^{-2m/(2m+1)}}\right)\vphantom{\int_{\XX^d}\left\{\tilde{f}(\bt)-f_0(\bt)\right\}^2d\bt}\right\}=1.
\end{align*}
Here the tuning parameter $\lambda$ in (\ref{eq: kernel_reg}) is chosen by $\lambda\asymp \left[n(\log n)^{1-(d-p)\wedge r}\right]^{-2m/(2m+1)} $ when $0\leq p<d$, and $\lambda\asymp n^{-(2mr-2)/[(2m+1)r-2]}$ when $p=d, r\geq 3$, and $\lambda\asymp (n\log n)^{-(2m-1)/2m}$ when $p=d, r= 2$, and $\lambda\asymp n^{-(m-1)/m}$ when $p=d$, $r=1$. 
\end{theorem}
\noindent
The proof of Theorem \ref{thm:mainupperrateestf0} relies on
several techniques from empirical process and stochastic process theory, including the linearization method and operator gradients. 
\change{In our analysis of SS-ANOVA models incorporating gradient information, unlike the approach by \citet{lin2000tensor} which lacks such data,  we have developed a method for the simultaneous diagonalization of two positive definite kernels: one including only function data, and the other incorporating both function and gradient data. We have obtained sharper results on the minimax rates of convergence than those in \citet{lin2000tensor}.  Moreover, Theorem \ref{thm:mainupperrateestf0} demonstrates that the optimal rate in \eqref{eqn:minaxhatfttprob} can be achieved with the random feature estimator  $\widehat{f}_n(\bt)$, as defined in \eqref{eqn:RF}. This represents another contribution compared to \citet{lin2000tensor}.}


Theorems \ref{theorem:lowerbdfNlambdaregrandom} and \ref{thm:mainupperrateestf0} together 
immediately imply that the minimax optimal rate for estimating $f_0\in\HH$ is
\begin{equation}
\label{eqn:minaxhatfttprob}
\begin{aligned}
&\left[n(\log n)^{1-(d-p)\wedge r}\right]^{-\frac{2m}{2m+1}} \mathbbm{1}_{0\leq p<d}\\
&\quad\quad\quad + \left[n^{-\frac{2mr}{(2m+1)r-2}} \mathbbm{1}_{r\geq 3}+n^{-1}(\log n)^{r-1}  \mathbbm{1}_{r<3}\right] \mathbbm{1}_{p=d}.
\end{aligned}
\end{equation}
This result connects with two strands of literature--estimating SS-ANOVA models without gradient information, and estimating nonparametric functions using derivatives.

\change{Firstly, in the case of estimating SS-ANOVA models without gradient information, the result in \eqref{eqn:minaxhatfttprob} recovers the rate  known in the literature (see, e.g.,  \cite{lin2000tensor}),
\begin{equation}
\label{eqn:convlin}
\left[n(\log n)^{1-r}\right]^{-\frac{2m}{2m+1}}.
\end{equation} 
For a high-order interaction $r$, the exponential term $(\log n)^{r-1}$ in \eqref{eqn:convlin} introduces the \emph{curse of interaction} and makes the SS-ANOVA models impractical. 
Surprisingly, the result in \eqref{eqn:minaxhatfttprob} shows that incorporating gradient data mitigates the curse of  interaction. 
For example, when $d-r\leq p\leq d-1$,   the rate given by \eqref{eqn:minaxhatfttprob} becomes, 
\begin{equation}
\label{eqn:nlognd-p}
\left[n(\log n)^{1-(d-p)}\right]^{-\frac{2m}{2m+1}}.
\end{equation} 
This rate is identical to the minimax optimal rate for estimating a  $(d-p)$-interaction model without gradient information  \citep{lin2000tensor}. 
When increasing $p$ types of gradient data to $(p+1)$ types, the 
rate given by \eqref{eqn:nlognd-p} accelerates at the order of $(\log n)^{-2m/(2m+1)}$, where $p\geq d-r$
and $p+1\leq d-1$. Moreover, when $p=d-1$, the rate given by \eqref{eqn:nlognd-p} is
$n^{-2m/(2m+1)}$,
which coincides with the optimal rate for estimating additive models without gradient information \citep{stone1985additive}. The result in \eqref{eqn:minaxhatfttprob} indicates a phase transition from $0\leq p<d$ to $p=d$.  Specifically, the rate with full gradient  $p=d$ is further improved compared to that with $p\leq d-1$. We also note that
when the SS-ANOVA models have full interaction with $r=d$,  the result in  \eqref{eqn:minaxhatfttprob}  yields the  special result in \eqref{eqn:minimaxraternnrd}.}

Secondly, in the case of estimating functions using derivatives, \citet{hall2007nonparametric} pioneered the proposal of a regression-kernel method for incorporating derivative data under random design and i.i.d. errors. \cite{hall2007nonparametric} proved that with first-order partial derivatives, their estimator achieves the rate $n^{-2m/(2m+d-1)}$ for general Hölder spaces (e.g., their Theorem 3). This rate converges \emph{slower} than the rate given by \eqref{eqn:minaxhatfttprob} when $d \geq 2$, and it suffers from the curse of dimensionality when $d$ is large. In contrast, our work, employing a reproducing-kernel approach within the function space of SS-ANOVA models, a subspace of Hölder spaces characterized by a tensor-product structure, achieves the \emph{improved} convergence rate in \eqref{eqn:minaxhatfttprob}. This new result shows the practical value of gradient information in enhancing the scalability of nonparametric modeling, especially in high-dimensional settings typical of SS-ANOVA models.

\subsection{Extensions of the main results}
\noindent
\change{We discuss various ways for extending the optimal rates established in Theorems \ref{theorem:lowerbdfNlambdaregrandom} and \ref{thm:mainupperrateestf0}. For instance, these rates can be extended to scenarios where the function values and partial derivatives have different sample sizes. Let \( n_j \) denote the sample size for the dataset \( \{(\bt_i^{(j)}, y_i^{(j)}): i = 1, \ldots, n_j\} \), where \( j = 0, 1, \ldots, p \). By applying the same arguments as in our proof, the rate in these theorems can be expressed as
\begin{equation*}
\begin{aligned}
&\min\Big\{\left[n_0(\log n_0)^{1-r}\right]^{-\frac{2m}{2m+1}},\quad \Big[\big(\min_{j\geq 1}n_j\big)\Big(\log \big(\min_{j\geq 1}n_j\big)\Big)^{1-(d-p)\wedge r}\Big]^{-\frac{2m}{2m+1}} \mathbbm{1}_{0\leq p<d}\\
&\quad\quad\quad\quad\quad\quad\quad + \Big[\big(\min_{j\geq 1}n_j\big)^{-\frac{2mr}{(2m+1)r-2}} \mathbbm{1}_{r\geq 3} + \big(\min_{j\geq 1}n_j\big)^{-1}\Big(\log \big(\min_{j\geq 1}n_j\big)\Big)^{r-1}  \mathbbm{1}_{r<3}\Big] \mathbbm{1}_{p=d}\Big\}.
\end{aligned}
\end{equation*}
This rate is essentially the minimum of two scenarios: the rate obtained by replacing \eqref{eqn:minaxhatfttprob} in terms of the value of \( \min_{j\geq 1}n_j \) and the conventional rate \eqref{eqn:convlin} based solely on the function data with \( n_0 \) samples. 
If the sample size \( n_0 \) for noisy function values is significantly smaller than \( \min_{j\geq 1}n_j \), the optimal rate in \eqref{eqn:minaxhatfttprob} still holds with \( n = \min_{j\geq 1}n_j \). In this case, the noisy function values contribute to anchoring the absolute level of the function, making function estimation identifiable. Conversely, if the dataset of noisy function values alone is substantially large, i.e., \( n_0 \) is much greater than \( \min_{j\geq 1}n_j \), the convergence rate  by Theorems \ref{theorem:lowerbdfNlambdaregrandom} and \ref{thm:mainupperrateestf0} aligns with the conventional rate \eqref{eqn:convlin} based solely on the noisy function values.
}

\change{The optimal rates in  Theorems \ref{theorem:lowerbdfNlambdaregrandom} and \ref{thm:mainupperrateestf0} also apply under deterministic designs, where the design points $\bt^{(0)}$  in \eqref{modelequation1} and $\bt^{(j)}$s in  \eqref{modelequation} are equally spaced in $\XX^d$, rather than independently drawn from distributions  $\Pi^{(0)}$ and $\Pi^{(j)}$s, respectively. This adaptation demonstrates the robustness of our result to variations in design point selection. The results for deterministic designs are given in Appendix S1. 
Additionally,  the optimal rates are valid under a more general error assumption than \eqref{eqn:errorstruct}. Specifically, it holds when $\text{Var}(\epsilon_i^{(j)}) = \sigma_j^2 + o(n^{-1/2})$. A rigorous proof of Theorem \ref{thm:mainupperrateestf0} under this general error assumption follows a similar argument to that of the original proof.}

\change{Finally, we discuss additive models, which can be regarded as a special case of the SS-ANOVA model \eqref{eqn:anovadecompfti}  by setting $r=1$. In this scenario, with gradient data  available where $p=d$, Theorems \ref{theorem:lowerbdfNlambdaregrandom} and \ref{thm:mainupperrateestf0} suggest that the estimation of additive models can achieve the parametric rate of $n^{-1}$, which is a significant improvement over the traditional optimal rate of $n^{-2m/(2m+1)}$ typically achieved without gradient information \citep{stone1985additive}. We provide intuition behind achieving the parametric rate in additive models to illustrate the benefits of incorporating gradient information in statistical estimations. Heuristically, for a univariate function $f_0$, the problem of estimating $f_0$ with noisy gradient data is analogous to settings where  $f_0$ is observed with noise, but the integral of $f_0$ is the estimation target, which can achieve the parametric rate $n^{-1}$ through nonlocal averaging \citep{donoho1991geometrizing, donoho1994statistical}. This analogy suggests that the availability of gradient data eliminates the need for smoothing or local averaging, typically necessary in nonparametric estimation, thus allowing for a faster parametric rate. In the case of multivariate additive models, where $f_0=f_{01}+\cdots+f_{0d}$, gradient data effectively provides observations on the derivatives of each component function, $df_{0j}(t_j)/dt_j$, enabling the estimation of each component at the parametric rate and, consequently, the entire function $f_0$.}

\section{Aplications}
\label{sec:realexamples}
\noindent
In this section, we demonstrate the aspects of our method and theory via various applications. We study a stochastic simulation example in Section \ref{exp:blacksholes},
and an economics example  in Section \ref{exp:costfunction}.
We analyze a real data experiment of ion channel in Section \ref{ex:computerexperiment}.
\subsection{Call option pricing with stochastic simulations}
\label{exp:blacksholes}
\noindent
We discussed a motivating example of stochastic simulation in Section \ref{sec:ssanovaanderror}.
Now we consider a detailed stochastic simulation of the call option pricing.
The Black-Scholes stochastic differential equation is commonly used  to model stock price $S_T$ at  time $T$ through
\begin{equation*}
dS_T = r_* S_TdT+\sigma_* S_TdW_T,  T\geq 0,
\end{equation*}
where  $W_T$ is the Wiener process, $r_*$ is the risk-free rate, and $\sigma_*$ is the volatility of the stock price. 
The equation has a closed-form solution: 
$S_T = S_0\exp\{(r_*-\frac{1}{2}\sigma_*^2)T+\sigma_* \sqrt{T}\omega\}$ with the standard normal variable $\omega$ and initial stock price $S_0$.
The European call option is the right to buy a stock at the prespecified time $T$ with a prespecified price $P_0$. 
The value of the European option is 
\begin{equation*}
h(\bt,\omega) = e^{-r_*T}(S_T-P_0)_+,
\end{equation*} where $\bt = (S_0,r_*,\sigma_*)$.
Our goal is to estimate the expected net present value of the option with fixed $T$ and $P_0$:
$f_0(\bt)  = \E_\omega[h(\bt,\omega)]$.
It can be seen that $f_0(\bt)$ follows the SS-ANOVA model \eqref{eqn:anovadecompfti}.
In the experiment, we fix $T=1$, $P_0=100$, and choose the design $\bt$ from equally spaced points from $S_0\in[80,120]$, $r_*\in[0.01,0.05]$, and $\sigma_* \in [0.2,1]$ with the sample size $n = 7^3, 14^3, 21^3$. The end points of each interval are always included. 
\change{We set the number of random feature $s=n/10$ for constructing the random feature estimator in \eqref{eqn:RF}.}
To address the impact of stochastic simulation noise, we simulate $q=1000, 2000, 5000$ i.i.d. replications of $S_T$  at each design point and then average the responses. Independent sampling is used across design points. 
It is known that IPA estimators for the gradient: $\partial f_0/\partial S_0$, $\partial f_0/\partial r_*$, $\partial f_0/\partial \sigma_*$ are given by \cite{glasserman2013monte},
\begin{equation}
\label{eqn:derivestimpricing}
\begin{aligned}
Y^{(1)}& = e^{-r_*T}\frac{S_T}{S_0}\cdot {\mathbf 1}\{S_T\geq P_0\},\\
Y^{(2)}& = -TY^{(0)} + e^{-r_*T}TS_T\cdot{\mathbf 1}\{S_T\geq P_0\},\\
Y^{(3)}& =  e^{-r_*T}\frac{1}{\sigma_*}\left[\log\left(\frac{S_T}{S_0}\right)-\left(r_*+\frac{1}{2}\sigma_*^2\right)T\right]S_T\cdot{\mathbf 1}\{S_T\geq P_0\}.
\end{aligned}
\end{equation}
The IPA estimators \eqref{eqn:derivestimpricing} are unbiased, $\E_\omega[Y^{(1)}]=\partial f_0/\partial S_0, \E_\omega[Y^{(2)}]=\partial f_0/\partial r_*$, $\E_\omega[Y^{(3)}]=\partial f_0/\partial \sigma_*$.  
\change{We show in Appendix  \ref{sec:exampleappendix} that  the error assumption \ref{eqn:errorstruct} holds for IPA estimators in \eqref{eqn:derivestimpricing}.}
In this example, obtaining function data at a new design point requires the generation of $q$ new random numbers and the computation of $S_T$ for each of these $q$ simulation replications. In contrast, the gradient estimator given by (\ref{eqn:derivestimpricing}) can be obtained at a negligible cost and without a new simulation.

\paragraph{Comparison with existing  method.}
Stochastic kriging \citep{ankenman2010stochastic, chen2013enhancing} is a popular method for the mean response estimation of a stochastic simulation.
We compare the results of our estimator (\ref{eqn:RF}) incorporating gradient information and the stochastic kriging method without gradient. 
Consider the tensor product Mat\'ern  kernel,
\begin{equation}
\label{eqn:maternkernel}
\prod_{j=1}^3\left(1+|t_j-t'_j|/\tau_j+|t_j-t'_j|^2/3\tau_j^2\right) \exp\left(-|t_j-t'_j|/\tau_j\right).
\end{equation} 
This kernel satisfies the differentiability condition (\ref{eqn:partial2tt'kcx1x1}), where lengthscale parameters  $\tau_j$s are chosen by the five-fold cross-validation. 
We use the actual output as the reference given by
$f_0(S_0,r_*,\sigma_*) = 
S_0\Phi\left(-d_1+\sigma_*\right) - 100e^{-r_*}\Phi\left(-d_1\right)$ when $T=1, P_0=100$, where $d_1 = \sigma_*^{-1}[\log 100 - \log(S_0) - (r_*-\sigma_*^2/2)]$ and $\Phi(\cdot)$ is the CDF of standard normal distribution.
We estimate the MSE$=\E(\widehat{f}_{n} - f_0)^2$  by a Monte Carlo  sample of $10^4$ test points in $[80,120]\times [0.01,0.05]\times [0.2,1]$.

Figure  \ref{fig:blacksholes}  reports the MSEs for different methods: stochastic kriging with only function data (i.e., $p=0$), our estimator with different types of gradient data. The results are averaged over $1000$ simulations in each setting. 
It is seen that our estimator with gradient data gives a substantial improvement in estimation compared to stochastic kriging without gradient.
For example,  the MSE of $n=7^3, q=1000$ with full gradient (i.e., $p=3$) is comparable to the MSE of $n=14^3, q=1000$ without gradient (i.e., $p=0$). Since it needs little additional cost to estimate gradients by (\ref{eqn:derivestimpricing}), our estimator essentially saves the computational cost of sampling at new designs.
It is also seen that a faster convergence rate is obtained when incorporating 
all gradient data (i.e., $p=3$) compared to $p\leq 2$. This confirms our theoretical finding in Section \ref{sec:minimaxoptimality}.

\begin{table}[ht]
\centering
  \caption{The ratios of MSE  with two types of gradient data (i.e., $p=2$) relative to  MSE with only function data (i.e., $p=0$), for the example in Section \ref{exp:blacksholes}.}
  \label{table:blacksholes}
   \scalebox{0.8}{
 \begin{tabular}{l   c c  c }
 \hline
  \rule{0pt}{\normalbaselineskip}
  $n$   \quad\quad\quad & $q=1000$ & $q=2000$  & $q=5000$  \\   [0.5ex]
 \hline
 \rule{0pt}{\normalbaselineskip}
$7^3 = 343$ \quad\quad\quad & $0.6818$ & $0.6789$ & $0.6612$  \\  [0.5ex]
 \rule{0pt}{\normalbaselineskip}
$14^3 = 2744$ \quad\quad\quad & $0.5850$ & $0.5848$ & $0.5835$  \\  [0.5ex] 
 \rule{0pt}{\normalbaselineskip}
$21^3 = 9261$ \quad\quad\quad & $ 0.5484$ & $ 0.5483$ & $ 0.5294$  \\  [0.5ex] 
 \hline
 \end{tabular}}
\end{table}

\noindent
Table \ref{table:blacksholes} reports  the ratios of the MSE  of our estimator  with two types of gradient data (i.e., $p=2$) relative to  the MSE of stochastic kriging with only function data (i.e., $p=0$). 
It is seen that incorporating  gradient data leads to a faster convergence rate, which also agrees with our finding in Section \ref{sec:minimaxoptimality}.


\subsection{Cost estimation in economics}
\label{exp:costfunction}
\noindent
We consider an economic problem of the cost function estimation. 
Write the cost function $f_0(\bt) = f_0(t_1,\ldots,t_d)$, where $t_d$ denotes the level of output and $(t_1,\ldots,t_{d-1})$ represent the prices of $d-1$ factor inputs. 
The Cobb-Douglas production function \citep{varian1992} yields that 
\begin{equation*}
f_0(t_1,\ldots,t_d) = c_0^{-\frac{1}{c}}\prod_{1\leq j\leq d-1}\left(\frac{c}{c_j}\right)^{\frac{c_j}{c}}\prod_{1\leq j\leq d-1}t_j^{\frac{c_j}{c}}t_d^{\frac{1}{c}}.
\end{equation*}  
Here $c_0$ is the efficiency parameter, $c_1,\ldots,c_{d-1}$ are elasticity parameters, and $c=c_1+\cdots+c_{d-1}$. 
Our goal is to estimate the cost function $f_0(\bt)$. 
The function data of $f_0(\bt)$ are observed at design $\bt^{(0)}\in \XX^d$.
The gradient data of $f_0(\bt)$ with respect to input prices are the quantities of factor inputs that are also observable   \citep{hall2007nonparametric},
\begin{equation*}
Y^{(j)} = \frac{\partial}{\partial t_j} f_0(\bt^{(j)}) +\epsilon^{(j)},  \quad \bt^{(j)}\in\XX^d, \ j=1,\ldots,d-1.
\end{equation*}
Here $\bt^{(j)} = \bt^{(0)}\in \XX^d$ for $1\leq j\leq d-1$ that typically follows a random design. Moreover, the observational errors are usually assumed to be i.i.d. \citep{hall2007nonparametric} and hence satisfy the error structure (\ref{eqn:errorstruct}).
Since the gradient data about $\partial f_0/\partial t_d$ is not usually observable, it motivates our modeling of $p\in\{1,\ldots,d\}$ in model (\ref{modelequation}). Clearly, $f_0(\bt)$ in this example follows the SS-ANOVA model \eqref{eqn:anovadecompfti}.
In the experiment, we consider $d=3$
and fix $t_3=1$ since the cost function is homogeneous of degree one in $(t_1,t_2,t_3)$, that is $f_0(t_1,t_2,t_3,t_4) = t_3f_0(t_1/t_3,t_2/t_3,1,t_4)$. The data are generated through,
\begin{equation*}
\begin{aligned}
Y^{(0)} & = f_0(t_1,t_2,1,t_4) + \epsilon^{(0)}, \quad Y^{(j)} & = \frac{\partial f_0(t_1,t_2,1,t_4)}{\partial t_j} + \epsilon^{(j)}  \mbox{ for }j=1,2,
\end{aligned}
\end{equation*}
where $c_0=1, c_1=0.8, c_2=0.7,c_3=0.6$, and the designs  $\bt^{(j)},j=0,1,2$ follow the i.i.d. uniform distribution in $[0.5,1.5]^3$. 
Suppose that $\epsilon^{(j)},j=0,1,2$ are Gaussian with zero means, standard deviations $0.35$, and correlation $\rho$. 
We consider varying sample size $n=100, 200, 500, 1000$, the correlation $\rho=0,0.4,0.9$,
\change{and set the number of random feature $s=n/10$ for constructing the random feature estimator in \eqref{eqn:RF}.}

\paragraph{Comparison with existing method.}
 \cite{hall2007nonparametric} proposed a regression-kernel method for incorporating gradient for cost function estimation. We compare the performance of our estimator (\ref{eqn:RF}) with that of Hall and Yatchew's estimator. 
 For the estimator in  \cite{hall2007nonparametric}, we follow Hall and Yatchew's Example 3 to
use the  tensor product Mat\'ern kernel (\ref{eqn:maternkernel}) to  average $(t_1, t_4)$ and $(t_2,t_4)$ directions locally, and then average the estimates.
The MSE is estimated by a Monte Carlo  sample of $10^4$ test points in $[0.5,1.5]^3$.

\begin{table}[ht]
\centering
 \caption{\change{The comparison of average MSEs and standard errors of our estimator with those of Hall and Yatchew's estimator, considering various gradient types,  for the example in Section \ref{exp:costfunction} with $1000$ simulations. The table shows metrics: ``average MSE (standard error),''  in units of $10^{-4}$.}}
   \label{table:costfunction}
  \scalebox{0.8}{
\change{  \begin{tabular}{l c c c  c }
 \hline
  \rule{0pt}{\normalbaselineskip}
  &   &   Our Estimator (\ref{eqn:RF}) &  \cite{hall2007nonparametric}     & Our Estimator (\ref{eqn:RF}) \\   [0.5ex]
    &   & with only $Y^{(0)}$ & with $Y^{(0)}+ Y^{(1)}+ Y^{(2)}$     & with $Y^{(0)}+ Y^{(1)}+ Y^{(2)}$ \\   [0.5ex]
 \hline
 \rule{0pt}{\normalbaselineskip}
& $\rho=0$ &  $127.1471~(22.8495)$& $61.4098~(17.4460)$    & $\textbf{47.4739~(13.5196)}$ \\  [0.5ex] 
 \rule{0pt}{\normalbaselineskip}
 $n=100$ & $\rho=0.4$  & $128.9210~(23.3594)$  & $63.1006~(17.9422)$  & $\textbf{49.8963~(13.6218)}$\\  [0.5ex]
 \rule{0pt}{\normalbaselineskip}
 & $\rho=0.9$   & $129.6300~(24.8577)$  & $64.5989~(19.8965)$ & $\textbf{51.9224~(13.6433)}$ \\  [0.5ex] \cline{2-5}
 \rule{0pt}{\normalbaselineskip}
 & $\rho=0$ & $76.6199~(15.9333)$  & $33.3001~(11.5872)$ & $\textbf{24.1501~(8.2730)}$  \\  [0.5ex] 
 \rule{0pt}{\normalbaselineskip}
 $n=200$& $\rho=0.4$  & $77.7602~(16.1079)$ & $35.0696~(11.7615)$  & $\textbf{25.5342~(8.3062)}$ \\  [0.5ex]
 \rule{0pt}{\normalbaselineskip}
& $\rho=0.9$  & $77.9138~(16.3593)$  & $36.2591~(11.9210)$ & $\textbf{27.0137~(8.6223)}$  \\  [0.5ex] \cline{2-5}
 \rule{0pt}{\normalbaselineskip}
 & $\rho=0$ & $36.1925~(8.0550)$ & $16.3861~(5.5399)$  & $\textbf{9.3499~(2.5570)}$  \\  [0.5ex] 
     \rule{0pt}{\normalbaselineskip}
$n=500$\quad\quad & $\rho=0.4$ & $38.0683~(8.2180)$  & $18.2355~(5.6164)$  & $\textbf{10.4708~(2.5619)}$ \\  [0.5ex]
     \rule{0pt}{\normalbaselineskip}
 & $\rho=0.9$ & $38.9311~(8.3654)$  & $18.7698~(5.6877)$   & $\textbf{11.0498~(2.6124)}$ \\  [0.5ex] \cline{2-5}
     \rule{0pt}{\normalbaselineskip}
  & $\rho=0$  & $21.8570~(5.6051)$  & $9.2788~(2.2411)$ & $\textbf{4.5364~(1.6147)}$ \\ [0.5ex]
     \rule{0pt}{\normalbaselineskip}
$n=1000$ & $\rho=0.4$  & $22.4943~(5.6312)$  & $10.4801~(2.2433)$ & $\textbf{5.1468~(1.6561)}$  \\ [0.5ex] 
     \rule{0pt}{\normalbaselineskip}
& $\rho=0.9$  & $22.9499~(5.6446)$ & $10.6193~(2.3386)$ & $\textbf{5.3288~(1.8550)}$  \\ [0.5ex]
 \hline
  \end{tabular}}
  }
\end{table}

\change{Table \ref{table:costfunction} reports the MSEs and standard errors for varying sample size $n$, correlation $\rho$, and different methods:
our estimator  with only function data (i.e., $p=0$),  Hall and Yatchew's estimator with function and gradient data (i.e., $p=2$), our estimator  with function and gradient data (i.e., $p=2$). The results are obtained over $1000$ simulations in each setting. 
It is seen that MSEs and standard errors of incorporating gradient information are significantly smaller than that without gradient. Moreover,  the performances of our estimator compare favorably with that of Hall and Yatchew's estimator. }

Table \ref{table:costfunction2}  reports the ratios of the MSE  of our estimator incorporating two types of gradient data (i.e., $p=2$) relative to the MSE of Hall and Yatchew's estimator incorporating two types of gradient data (i.e., $p=2$). 
It is seen that the ratio decreases with the sample size, which agrees with our theoretical finding in Section \ref{sec:minimaxoptimality}, since our estimator in this example converges at the rate $n^{-2m/(2m+1)}$ by Theorem \ref{thm:mainupperrateestf0}, and Hall and Yatchew's estimator converges at a slower rate $n^{-m/(m+1)}$.

\begin{table}[ht]
\centering
 \caption{The ratios of MSE of our estimator with two types of gradient data (i.e., $p=2$) relative to  MSE of Hall and Yatchew's estimator with two types of gradient data (i.e., $p=2$), for the example in Section \ref{exp:costfunction}.}
 \label{table:costfunction2}
  \scalebox{0.8}{
 \begin{tabular}{l  c c  c }
 \hline
  \rule{0pt}{\normalbaselineskip}
  $$  \quad\quad\quad   & \quad\quad$\rho=0$ & \quad\quad$\rho=0.4$  & \quad\quad$\rho=0.9$  \\   [0.5ex]
 \hline
 \rule{0pt}{\normalbaselineskip}
 $n=100$  \quad & \quad\quad $0.7731$ & \quad\quad $0.7907$ &\quad\quad  $0.8038$ \\  [0.5ex] 
 \rule{0pt}{\normalbaselineskip}
 $n=200$  \quad & \quad\quad$0.7252$ & \quad\quad$0.7281$ &\quad\quad $0.7450$  \\  [0.5ex] 
 \rule{0pt}{\normalbaselineskip}
 $n=500$  \quad &\quad \quad$0.5706$ &\quad\quad $0.5742$ & \quad\quad$0.5887$  \\  [0.5ex] 
     \rule{0pt}{\normalbaselineskip}
 $n=1000$  \quad  & \quad\quad$0.4889$ & \quad\quad$0.4911$ & \quad\quad$0.5018 $ \\ [0.5ex] 
 \hline
  \end{tabular}}
\end{table}

\change{Tables \ref{table:costfunction} and \ref{table:costfunction2} also indicate that $s=n/10$ yields sufficient accuracy for the estimations by the random feature estimator in \eqref{eqn:RF}.  Therefore,  in practical applications, an $s$  significantly smaller than the theoretical minimum of  $s=O(n\log n)$ in Theorem \ref{thm:mainupperrateestf0} might often suffice.}

\subsection{Ion channel experiment}
\label{ex:computerexperiment}
\noindent
We consider a real data example from a single voltage clamp experiment. The experiment is on the sodium ion channel of the cardiac cell membranes. The experiment output $z_k$ measures the normalized current for maintaining a fixed membrane potential of $-35mV$ and the input $x_k$ is the logarithm of time. The sample size of  the ion channel experiment is $N=19$.  
Computer model has been used to study the ion channel experiment \citep{plumlee2017bayesian}.
Let $\eta(x,\bt)$ be the computer model that approximates the physical system for the ion channel experiment, 
where $x$ is the experiment input and $\bt\in\XX^d$ is the calibration parameter whose value are unobservable. 
For analyzing the ion channel  experiment, the computer model is given by $\eta(x,\bt) = e_1^\top\exp(\exp(x)A(\bt))e_4$, where $\bt=(t_1,t_2,t_3)^\top\in\XX^d$, $d=3$, $e_1 = (1,0,0,0)^\top, e_4 = (0,0,0,1)^\top$, and 
\begin{equation*}
A(\bt) = \left( \begin{array}{cccc}
-t_2 - t_3 & t_1 & 0 & 0 \\
t_2 & - t_1 - t_2 &  t_1 & 0 \\
0 &  t_2 & - t_1 - t_2   & t_1 \\
0 & 0 & t_2 & -t_1 \end{array} \right).
\end{equation*}
Our goal is to estimate the function, $f_0(\bt)=\E_{(x,z)}[z - \eta(x,\bt)]^2$, which is useful for visualization, calibration, and understanding how well the computer model approximates the physical system in this experiment \citep{kennedy2001bayesian}.  
The function data at design  $\bt^{(0)}\in\XX^3$ is generated by,
\begin{equation*}
Y^{(0)} =\frac{1}{N} \sum_{k=1}^N[z_k - \eta(x_k,\bt^{(0)})]^2,\quad \text{where }N=19.
\end{equation*}
The  gradient  of computer model, i.e., $\nabla_{\bt}\eta(x,\bt)$,  can be obtained using the chain rule-based automatic differentiation. By the cheap gradient principle  \citep{griewank2008evaluating}, the  cost for computing $\nabla_{\bt}\eta(x,\bt)$ is at most four or five times the cost for function evaluation $\eta(x,\bt)$ and hence,  the gradient is cheap to obtain.
Then the estimator for the gradient of $f_0(\bt)$ is given by,
\begin{equation*}
Y^{(j)} = -\frac{2}{N}\sum_{k=1}^N\left[z_k-\eta(x_k,\bt^{(j)})\right]\frac{\partial}{\partial t_j}\eta(x_k,\bt^{(j)}), \quad \bt^{(j)}\in\XX^3, \ j=1,2,3.
\end{equation*}
In the experiment, we choose i.i.d. uniform designs for $\bt^{(j)}$s, $j=0,1,2,3$ from $\XX^3$  with the sample size  $n=1000, 2000, 3000, 5000$.

\begin{table}[ht]
\centering
  \caption{\change{The comparison of average MSEs and standard errors of our estimator with those of Morris et al.'s  estimator, considering various gradient types,  for the example in Section \ref{ex:computerexperiment} with $1000$ simulations. The table shows metrics: ``average MSE (standard error),''  in units of $10^{-6}$.} }
     \label{table:computerexp}
  \scalebox{0.8}{
\change{ \begin{tabular}{l  c c  c }
 \hline
  \rule{0pt}{\normalbaselineskip}
    &  Our Estimator  & \cite{morris1993bayesian}    & Our Estimator \\   [0.5ex]
       & with only $Y^{(0)}$ & with $Y^{(0)}+ \cdots + Y^{(3)}$     & with $Y^{(0)}+ \cdots+ Y^{(3)}$ \\   [0.5ex]
 \hline
 \rule{0pt}{\normalbaselineskip}
 $n=1000$ \quad\quad\quad\quad &  $10.6491~(4.9867)$& $8.8956~(4.8729)$    & $\textbf{7.7804~(3.6737)}$ \\  [0.5ex] 
 \rule{0pt}{\normalbaselineskip}
 $n=2000$   & $8.5302~(4.3339)$  & $6.5494~(4.0728)$  & $\textbf{5.1375~(2.4687)}$\\  [0.5ex]
 \rule{0pt}{\normalbaselineskip}
  $n=3000$  & $6.4296~(3.9595)$  & $4.1940~(3.2242)$ & $\textbf{3.1035~(1.7187)}$ \\  [0.5ex] 
 \rule{0pt}{\normalbaselineskip}
 $n=5000$  & $5.4143~(3.2268)$  & $3.0910~(1.9073)$ & $\textbf{2.1305~(0.9322)}$  \\  [0.5ex] 
 \hline
  \end{tabular}}}
  \end{table}

\paragraph{Comparison with existing method.}
\cite{morris1993bayesian} proposed a stationary Gaussian process method to incorporate gradient data for estimation. We compare the performance of our estimator 
 (\ref{eqn:RF}) with that of Morris et al.'s estimator. We use the  Mat\'ern  kernel (\ref{eqn:maternkernel}) for both our estimator and Morris et al.'s estimator, and estimate the  MSE  by a Monte Carlo sample of $10^4$ test points in $\XX^3$. 
 \change{We set the number of random feature $s=n/10$ for constructing the estimator \eqref{eqn:RF}.}
Since the true function $f_0(\bt)$ is unknown at each test point, we approximate it by using total $N=19$ real ion channel samples at each test point. The function and gradient training data are generated using $N'=10$ real ion channel samples, which are randomly chosen from the total $N=19$ samples.

\change{Table  \ref{table:computerexp} reports the MSEs and standard errors for  varying sample size $n$ and different methods:
our estimator  with only function data (i.e., $p=0$),  Morris et al.'s estimator with function and gradient data (i.e., $p=3$), our estimator  with function and gradient data (i.e., $p=3$). The results are obtained over $1000$ simulations in each setting.  It is evident that the gradient data can significantly improve the estimation performance, and
  our estimator outperforms Morris et al.'s estimator. }

\section{Related Work}
\label{sec:relatedwork}
\noindent 
We review related work from multiple kinds of literature, including nonparametric regression, function interpolation, and dynamical systems.

There is growing literature on nonparametric regression with derivatives.
Our work is related to the pioneering work of \cite{hall2007nonparametric, hall2010nonparametric}, which established the root-$n$ consistency for nonparametric estimation given mixed and sufficiently \emph{higher-order} derivatives. However, it is difficult to obtain higher-order derivatives in practice, such as in economics and stochastic simulation.
In contrast, we focus on gradient information that is \emph{first-order} derivatives and are easier to obtain in practice. 
We show that the minimax optimal rates for estimating SS-ANOVA models are accelerated by using gradient data. In particular, we show that SS-ANOVA models are immune to the curse of interaction given gradient information.

The function interpolation with gradients has been widely studied. 
For exact data and one-dimensional functions, \cite{karlin1969fundamental} and \cite{wahba1971regression} showed that at certain deterministic design for data without gradients, 
incorporating gradient to the dataset provides no new information for function interpolation. This result, however, cannot be extended to the case of noisy data. 
\cite{morris1993bayesian} incorporated noiseless derivatives for deterministic surface estimation in computer experiments. 
Unlike these works, we consider the noisy gradient information for nonparametric estimation.

Our work is also related to the literature on dynamical systems and stochastic simulation.
\cite{solak2002derivative} considered the identified linearization around an equilibrium point for estimating the derivatives in nonlinear dynamical systems. They used Gaussian processes for a combination of function and derivative observations. 
\cite{chen2013enhancing} used stochastic kriging to incorporate gradient estimators and improve surface estimation, where stochastic kriging \citep{ankenman2010stochastic} is a metamodeling technique for representing the mean response surface implied by a stochastic simulation. However, the rates of convergence are not studied in \cite{solak2002derivative} and \cite{chen2013enhancing}. We quantify the improved rates of convergence in nonparametric estimation by using gradient data.


\section{Conclusion}
\label{sec:discussion}
\noindent
Statistical modeling of gradient information becomes an increasingly important problem in many areas of science and engineering. 
We develop an approach based on \change{partial derivatives}, either observed or estimated, to effectively estimate the nonparametric function. The proposed approach and computational algorithm could lead to methods useful to practitioners. Our theoretical results showed that SS-ANOVA models are immune to the \emph{curse of interaction} using gradient information. Moreover, for the additive models, the rates using gradient information are root-$n$, thus achieving the \emph{parametric rate}.  
As a working model, we assume that the eigenvalues decay at the same
polynomial rate across component RKHS $\HH^j$s, 
which hold for Sobolev kernels, among other commonly used kernels.
It is of interest to consider incorporating gradient information in more general settings, for example, when eigenvalues decay at different rates, or if the eigenvalues for some components decay even exponentially. It is
conceivable that our analysis could be extended to deal with more general
settings, which will be left for future studies.
\section*{Acknowledgement}
\noindent
The author thanks the Editor, Associate Editor, and three anonymous reviewers for their invaluable feedback, and thanks Prof. Grace Wahba for valuable advice on this work. The author acknowledges the support of the California Center for Population Research as part of the Eunice Kennedy Shriver National Institute of Child Health and Human Development (NICHD) population research infrastructure grant P2C-HD041022.

\baselineskip=22pt
\bibliographystyle{apa}
\bibliography{references}

\begin{thebibliography}{}

\bibitem[\protect\astroncite{Ankenman et~al.}{2010}]{ankenman2010stochastic}
Ankenman, B.~E., Nelson, B.~L., and Staum, J. (2010).
\newblock Stochastic kriging for simulation metamodeling.
\newblock {\em Operations Research}, 58(2):371--382.

\bibitem[\protect\astroncite{Aronszajn}{1950}]{aronszajn1950theory}
Aronszajn, N. (1950).
\newblock Theory of reproducing kernels.
\newblock {\em Transactions of the American Mathematical Society},
  68(3):337--404.

\bibitem[\protect\astroncite{Bochner}{1934}]{bochner1934theorem}
Bochner, S. (1934).
\newblock A theorem on fourier-stieltjes integrals.
\newblock {\em Bulletin of the American Mathematical Society}, 40(4):271--276.

\bibitem[\protect\astroncite{Breckling}{2012}]{breckling2012analysis}
Breckling, J. (2012).
\newblock {\em The Analysis of Directional Time Series: Applications to Wind
  Speed and Direction}, volume~61.
\newblock Springer Science \& Business Media.

\bibitem[\protect\astroncite{Cai and Low}{2005}]{cai2005adaptive}
Cai, T.~T. and Low, M.~G. (2005).
\newblock On adaptive estimation of linear functionals.
\newblock {\em The Annals of Statistics}, 33(5):2311--2343.

\bibitem[\protect\astroncite{Chen et~al.}{2013}]{chen2013enhancing}
Chen, X., Ankenman, B.~E., and Nelson, B.~L. (2013).
\newblock Enhancing stochastic kriging metamodels with gradient estimators.
\newblock {\em Operations Research}, 61(2):512--528.

\bibitem[\protect\astroncite{Dai and Li}{2022}]{dai2021kernel}
Dai, X. and Li, L. (2022).
\newblock Kernel ordinary differential equations.
\newblock {\em Journal of the American Statistical Association},
  117(540):1711--1725.

\bibitem[\protect\astroncite{Dai and Li}{2024}]{dai2024post}
Dai, X. and Li, L. (2024).
\newblock Post-regularization confidence bands for ordinary differential
  equations.
\newblock {\em Journal of Machine Learning Research}, 25(23):1--51.

\bibitem[\protect\astroncite{Dai et~al.}{2023}]{dai2022kernel}
Dai, X., Lyu, X., and Li, L. (2023).
\newblock Kernel knockoffs selection for nonparametric additive models.
\newblock {\em Journal of the American Statistical Association},
  118(543):2158--2170.

\bibitem[\protect\astroncite{Donoho}{1994}]{donoho1994statistical}
Donoho, D.~L. (1994).
\newblock Statistical estimation and optimal recovery.
\newblock {\em The Annals of Statistics}, 22(1):238--270.

\bibitem[\protect\astroncite{Donoho and Liu}{1991}]{donoho1991geometrizing}
Donoho, D.~L. and Liu, R.~C. (1991).
\newblock Geometrizing rates of convergence, iii.
\newblock {\em The Annals of Statistics}, 19(2):668--701.

\bibitem[\protect\astroncite{Donoho et~al.}{1990}]{donoho1990minimax}
Donoho, D.~L., Liu, R.~C., and MacGibbon, B. (1990).
\newblock Minimax risk over hyperrectangles, and implications.
\newblock {\em The Annals of Statistics}, 18(3):1416--1437.

\bibitem[\protect\astroncite{Efron and
  Tibshirani}{1993}]{efron1993introduction}
Efron, B. and Tibshirani, R.~J. (1993).
\newblock {\em An Introduction to the Bootstrap}.
\newblock New York: Chapman and Hall.

\bibitem[\protect\astroncite{Frees and Valdez}{1998}]{frees1998understanding}
Frees, E.~W. and Valdez, E.~A. (1998).
\newblock Understanding relationships using copulas.
\newblock {\em North American Actuarial Journal}, 2(1):1--25.

\bibitem[\protect\astroncite{Fu and Qu}{2014}]{fu2014regression}
Fu, M.~C. and Qu, H. (2014).
\newblock Regression models augmented with direct stochastic gradient
  estimators.
\newblock {\em INFORMS Journal on Computing}, 26(3):484--499.

\bibitem[\protect\astroncite{Gelfand and Silverman}{2000}]{gelfand2000calculus}
Gelfand, I.~M. and Silverman, R.~A. (2000).
\newblock {\em Calculus of Variations}.
\newblock Courier Corporation.

\bibitem[\protect\astroncite{Glasserman}{2013}]{glasserman2013monte}
Glasserman, P. (2013).
\newblock {\em Monte Carlo Methods in Financial Engineering}.
\newblock New York: Springer Science \& Business Media.

\bibitem[\protect\astroncite{Golub et~al.}{1979}]{golub1979generalized}
Golub, G.~H., Heath, M., and Wahba, G. (1979).
\newblock Generalized cross-validation as a method for choosing a good ridge
  parameter.
\newblock {\em Technometrics}, 21(2):215--223.

\bibitem[\protect\astroncite{Griewank and
  Walther}{2008}]{griewank2008evaluating}
Griewank, A. and Walther, A. (2008).
\newblock {\em Evaluating Derivatives: Principles and Techniques of Algorithmic
  Differentiation}.
\newblock Philadelphia, PA: SIAM.

\bibitem[\protect\astroncite{Hall}{1992a}]{hall1992effect}
Hall, P. (1992a).
\newblock Effect of bias estimation on coverage accuracy of bootstrap
  confidence intervals for a probability density.
\newblock {\em The Annals of Statistics}, pages 675--694.

\bibitem[\protect\astroncite{Hall}{1992b}]{hall1992bootstrap}
Hall, P. (1992b).
\newblock On bootstrap confidence intervals in nonparametric regression.
\newblock {\em The Annals of Statistics}, pages 695--711.

\bibitem[\protect\astroncite{Hall et~al.}{1990}]{hall1990asymptotically}
Hall, P., Kay, J.~W., and Titterington, D.~M. (1990).
\newblock Asymptotically optimal difference-based estimation of variance in
  nonparametric regression.
\newblock {\em Biometrika}, 77(3):521--528.

\bibitem[\protect\astroncite{Hall and Yatchew}{2007}]{hall2007nonparametric}
Hall, P. and Yatchew, A. (2007).
\newblock Nonparametric estimation when data on derivatives are available.
\newblock {\em The Annals of Statistics}, 35(1):300--323.

\bibitem[\protect\astroncite{Hall and Yatchew}{2010}]{hall2010nonparametric}
Hall, P. and Yatchew, A. (2010).
\newblock Nonparametric least squares estimation in derivative families.
\newblock {\em Journal of Econometrics}, 157(2):362--374.

\bibitem[\protect\astroncite{H{\"a}rdle and
  Bowman}{1988}]{hardle1988bootstrapping}
H{\"a}rdle, W. and Bowman, A.~W. (1988).
\newblock Bootstrapping in nonparametric regression: local adaptive smoothing
  and confidence bands.
\newblock {\em Journal of the American Statistical Association},
  83(401):102--110.

\bibitem[\protect\astroncite{Hastie and
  Tibshirani}{1990}]{hastie1990generalized}
Hastie, T. and Tibshirani, R. (1990).
\newblock {\em Generalized Additive Models}.
\newblock London, UK: Chapman \& Hall/CRC.

\bibitem[\protect\astroncite{Huang}{1998}]{huang1998projection}
Huang, J.~Z. (1998).
\newblock Projection estimation in multiple regression with application to
  functional anova models.
\newblock {\em The Annals of Statistics}, 26(1):242--272.

\bibitem[\protect\astroncite{Jones and Mereu}{2002}]{jones2002critique}
Jones, B.~L. and Mereu, J.~A. (2002).
\newblock A critique of fractional age assumptions.
\newblock {\em Insurance: Mathematics and Economics}, 30(3):363--370.

\bibitem[\protect\astroncite{Karlin}{1969}]{karlin1969fundamental}
Karlin, S. (1969).
\newblock The fundamental theorem of algebra for monosplines satisfying certain
  boundary conditions and applications to optimal quadrature formulas.
\newblock {\em Approximations with Special Emphasis on Spline Functions}, pages
  467--484.

\bibitem[\protect\astroncite{Kennedy and O'Hagan}{2001}]{kennedy2001bayesian}
Kennedy, M.~C. and O'Hagan, A. (2001).
\newblock Bayesian calibration of computer models.
\newblock {\em Journal of the Royal Statistical Society: Series B (Statistical
  Methodology)}, 63(3):425--464.

\bibitem[\protect\astroncite{Klemel{\"a} and Tsybakov}{2001}]{klemela2001sharp}
Klemel{\"a}, J. and Tsybakov, A.~B. (2001).
\newblock Sharp adaptive estimation of linear functionals.
\newblock {\em The Annals of Statistics}, 29(6):1567--1600.

\bibitem[\protect\astroncite{L{'}Ecuyer}{1990}]{l1990unified}
L{'}Ecuyer, P. (1990).
\newblock A unified view of the ipa, sf, and lr gradient estimation techniques.
\newblock {\em Management Science}, 36(11):1293--1416.

\bibitem[\protect\astroncite{Lim}{2024}]{lim2024estimating}
Lim, E. (2024).
\newblock Estimating a function and its derivatives under a smoothness
  condition.
\newblock {\em Mathematics of Operations Research}.

\bibitem[\protect\astroncite{Lin}{1998}]{lin1998tensor}
Lin, Y. (1998).
\newblock Tensor product space anova models in multivariate function
  estimation.
\newblock {\em Thesis (Ph.D.)--University of Pennsylvania}.

\bibitem[\protect\astroncite{Lin}{2000}]{lin2000tensor}
Lin, Y. (2000).
\newblock Tensor product space anova models.
\newblock {\em The Annals of Statistics}, 28(3):734--755.

\bibitem[\protect\astroncite{Lin and Zhang}{2006}]{lin2006component}
Lin, Y. and Zhang, H.~H. (2006).
\newblock Component selection and smoothing in multivariate nonparametric
  regression.
\newblock {\em The Annals of Statistics}, 34(5):2272--2297.

\bibitem[\protect\astroncite{Morris et~al.}{1993}]{morris1993bayesian}
Morris, M.~D., Mitchell, T.~J., and Ylvisaker, D. (1993).
\newblock Bayesian design and analysis of computer experiments: Use of
  derivatives in surface prediction.
\newblock {\em Technometrics}, 35(3):243--255.

\bibitem[\protect\astroncite{Oden and Reddy}{2012}]{oden2012introduction}
Oden, J.~T. and Reddy, J.~N. (2012).
\newblock {\em An Introduction to the Mathematical Theory of Finite Elements}.
\newblock New York: John Wiley \& Sons.

\bibitem[\protect\astroncite{Plumlee}{2017}]{plumlee2017bayesian}
Plumlee, M. (2017).
\newblock Bayesian calibration of inexact computer models.
\newblock {\em Journal of the American Statistical Association},
  112(519):1274--1285.

\bibitem[\protect\astroncite{Rahimi and Recht}{2007}]{RR07}
Rahimi, A. and Recht, B. (2007).
\newblock Random features for large-scale kernel machines.
\newblock {\em Advances in Neural Information Processing systems},
  20:1177--1184.

\bibitem[\protect\astroncite{Ramsay et~al.}{2007}]{ramsay2007parameter}
Ramsay, J.~O., Hooker, G., Campbell, D., and Cao, J. (2007).
\newblock Parameter estimation for differential equations: a generalized
  smoothing approach.
\newblock {\em Journal of the Royal Statistical Society: Series B (Statistical
  Methodology)}, 69(5):741--796.

\bibitem[\protect\astroncite{Riesz and Sz.-Nagy}{1955}]{riesz1955}
Riesz, F. and Sz.-Nagy, B. (1955).
\newblock {\em Functional Analysis}.
\newblock New York: Dover Publications.

\bibitem[\protect\astroncite{Riihim{\"a}ki and
  Vehtari}{2010}]{riihimaki2010gaussian}
Riihim{\"a}ki, J. and Vehtari, A. (2010).
\newblock Gaussian processes with monotonicity information.
\newblock In {\em International Conference on Artificial Intelligence and
  Statistics}, pages 645--652.

\bibitem[\protect\astroncite{Rudi and Rosasco}{2017}]{rudi2017generalization}
Rudi, A. and Rosasco, L. (2017).
\newblock Generalization properties of learning with random features.
\newblock {\em Advances in Neural Information Processing Systems (NeurIPS)},
  30.

\bibitem[\protect\astroncite{Ruppert et~al.}{2003}]{ruppert2003semiparametric}
Ruppert, D., Wand, M.~P., and Carroll, R.~J. (2003).
\newblock {\em Semiparametric Regression}.
\newblock New York: Cambridge University Press.

\bibitem[\protect\astroncite{Schoenberg}{1964}]{schoenberg1964spline}
Schoenberg, I.~J. (1964).
\newblock Spline functions and the problem of graduation.
\newblock {\em Proceedings of the National Academy of Sciences},
  52(4):947--950.

\bibitem[\protect\astroncite{Solak et~al.}{2002}]{solak2002derivative}
Solak, E., Murray-Smith, R., Leithead, W., Leith, D., and Rasmussen, C. (2002).
\newblock Derivative observations in gaussian process models of dynamic
  systems.
\newblock {\em Advances in Neural Information Processing Systems}, 15.

\bibitem[\protect\astroncite{Stone}{1980}]{stone1980optimal}
Stone, C.~J. (1980).
\newblock Optimal rates of convergence for nonparametric estimators.
\newblock {\em The Annals of Statistics}, 8(6):1348--1360.

\bibitem[\protect\astroncite{Stone}{1982}]{stone1982optimal}
Stone, C.~J. (1982).
\newblock Optimal global rates of convergence for nonparametric regression.
\newblock {\em The Annals of Statistics}, 10(4):1040--1053.

\bibitem[\protect\astroncite{Stone}{1985}]{stone1985additive}
Stone, C.~J. (1985).
\newblock Additive regression and other nonparametric models.
\newblock {\em The Annals of Statistics}, 13(2):689--705.

\bibitem[\protect\astroncite{Suri and Leung}{1987}]{suri1987single}
Suri, R. and Leung, Y.~T. (1987).
\newblock Single run optimization of a siman model for closed loop flexible
  assembly systems.
\newblock {\em Proceedings of the 19th Conference on Winter Simulation}, pages
  738--748.

\bibitem[\protect\astroncite{Tsybakov}{2009}]{tsybakovintroduction}
Tsybakov, A.~B. (2009).
\newblock {\em Introduction to Nonparametric Estimation}.
\newblock New York: Springer.

\bibitem[\protect\astroncite{van~der Vaart and Wellner}{1996}]{wellner1996weak}
van~der Vaart, A. and Wellner, J. (1996).
\newblock {\em Weak Convergence and Empirical Processes}.
\newblock Springer, New York.

\bibitem[\protect\astroncite{Varian}{1992}]{varian1992}
Varian, H.~R. (1992).
\newblock {\em Microeconomic Analysis}.
\newblock New York: W. W. Norton \& Company.

\bibitem[\protect\astroncite{Wahba}{1971}]{wahba1971regression}
Wahba, G. (1971).
\newblock On the regression design problem of sacks and ylvisaker.
\newblock {\em The Annals of Mathematical Statistics}, pages 1035--1053.

\bibitem[\protect\astroncite{Wahba}{1990}]{wahba1990}
Wahba, G. (1990).
\newblock {\em Spline Models for Observational Data}.
\newblock Philadelphia, PA: SIAM.

\bibitem[\protect\astroncite{Wahba et~al.}{1995}]{wahba1995smoothing}
Wahba, G., Wang, Y., Gu, C., Klein, R., and Klein, B. (1995).
\newblock Smoothing spline anova for exponential families, with application to
  the wisconsin epidemiological study of diabetic retinopathy.
\newblock {\em The Annals of Statistics}, 23(6):1865--1895.

\bibitem[\protect\astroncite{Wainwright}{2019}]{wainwright2019high}
Wainwright, M.~J. (2019).
\newblock {\em High-Dimensional Statistics: A Non-Asymptotic Viewpoint},
  volume~48.
\newblock Cambridge University Press.

\bibitem[\protect\astroncite{Wang and Berger}{2016}]{wang2016estimating}
Wang, X. and Berger, J.~O. (2016).
\newblock Estimating shape constrained functions using gaussian processes.
\newblock {\em SIAM/ASA Journal on Uncertainty Quantification}, 4(1):1--25.

\bibitem[\protect\astroncite{Yuan and Cai}{2010}]{Yuan2010}
Yuan, M. and Cai, T.~T. (2010).
\newblock A reproducing kernel hilbert space approach to functional linear
  regression.
\newblock {\em Annals of Statistics}, 38(6):3412--3444.

\bibitem[\protect\astroncite{Zhang et~al.}{2023}]{zhang2023gradient}
Zhang, H., Zheng, Z., and Lavaei, J. (2023).
\newblock Gradient-based algorithms for convex discrete optimization via
  simulation.
\newblock {\em Operations Research}, 71(5):1815--1834.

\bibitem[\protect\astroncite{Zhu et~al.}{2014}]{zhu2014structured}
Zhu, H., Yao, F., and Zhang, H.~H. (2014).
\newblock Structured functional additive regression in reproducing kernel
  hilbert spaces.
\newblock {\em Journal of the Royal Statistical Society: Series B (Statistical
  Methodology)}, 76(3):581--603.

\end{thebibliography}

\newpage 

\begin{center}
\Large
\textbf{Supplementary Appendix for Nonparametric Estimation via Partial Derivatives}
\end{center}

\appendix

\section{Optimal Rates Under Deterministic Designs}
\label{sec:minmaxriskregularlat}
\noindent
We present the minimax optimal rates under deterministic designs. Specifically, we consider the regular lattice design, which is also called the tensor product design. A regular lattice of size $n=l_1\times \cdots \times l_d$ on $\XX^d$ is a collection of design points
$\{\bt_1,\ldots,\bt_n\}=\{(t_{i_1,1}, t_{i_2,2}, \ldots, t_{i_d,d}) \ |\ i_j = 1,\ldots, l_j, j=1,\ldots, d\}$,
where $t_{i,j} = i/l_j$, $i=1,\ldots, l_j, j=1,\ldots, d$. This design is widely used for SS-ANOVA models \citep{wahba1995smoothing, lin2000tensor}. 
Under regular lattices, it is without loss of generality to assume that $f_0: \XX^d \mapsto \R$ has a periodic boundary condition. This is because any finite  sequence $\{f(\bt_1),\ldots, f(\bt_n)\}$ can be associated with a periodic sequence,
\begin{equation*}
\begin{aligned}
&  f^{\text{per}} \left(i_1/l_1, \cdots, i_d/l_d\right)  \\
&  = \sum_{q_1=-\infty}^{\infty}\cdots\sum_{q_d=-\infty}^{\infty}f\left(i_1/l_1-q_1, \cdots, i_d/l_d-q_d\right),\quad \forall (i_1,\ldots,i_d)\in \Z^d,
\end{aligned}
\end{equation*}
where $\Z$ is the set of integers, and let $f(\cdot)\equiv 0$ outside and on the unobserved boundaries of $\XX^d$. On the other hand, any finite sequence $\{f(\bt_1),\ldots, f(\bt_n)\}$ can be recovered from  periodic sequence $f^{\text{per}}(\cdot)$. 
We now present the main results under deterministic design by first stating a minimax lower bound.

\begin{theorem}
\label{theorem:lowerbdfNlambdareg}
Assume that $\lambda_\nu \asymp \nu^{-2m}$ for some $m>3/2$. Under the regression models \eqref{modelequation1} and \eqref{modelequation} where $f_0$ follows the SS-ANOVA model \eqref{eqn:anovadecompfti}
and the designs $\bt^{(0)}$ and $\bt^{(j)}$s are from the regular lattice. Then under the error structure (\ref{eqn:errorstruct}), there exists a constant $c$ that does not depend on $n$ such that
\begin{equation*}
\begin{aligned}
& \underset{n\to\infty}{\lim\inf}\inf_{\tilde{f}}\sup_{f_0\in\HH}\E\int_{\XX^d}\left[\tilde{f}(\bt)-f_0(\bt)\right]^2d\bt \\
& \quad\quad\quad\quad \geq
\begin{cases}
c\left[n(\log n)^{1-(d-p)\wedge r}\right]^{-\frac{2m}{2m+1}},  & \mbox{ if } 0\leq p< d,\\
c\left[n^{-\frac{2mr}{(2m+1)r-2}} \mathbbm{1}_{r\geq 3}+n^{-1}(\log n)^{r-1}  \mathbbm{1}_{r<3}\right], & \mbox{ if } p=d,
\end{cases}
\end{aligned}
\end{equation*}
where the infimum of $\tilde{f}$ is taken over all measurable functions of the data.
\end{theorem}
\noindent
The lower bound is established via the analysis of a version of the hardest rectangular subproblem. See, e.g., \cite{donoho1990minimax}. We relegate its proof to Section \ref{subsec:reglattproof}. Next, we show that the
rates given in the lower bound in Theorem \ref{theorem:lowerbdfNlambdareg} is attainable by the estimator $\widehat{f}_{n}$ in  (\ref{scheme1}). Hence $\widehat{f}_{n}$ is also minimax rate optimal under deterministic design.

\begin{theorem}
\label{thm;deterdesgnupperbdgeneralpdreg}
Assume that $\lambda_\nu \asymp \nu^{-2m}$ for some $m>3/2$. Under the regression models \eqref{modelequation1} and \eqref{modelequation} where $f_0$ follows the SS-ANOVA model \eqref{eqn:anovadecompfti}
and the designs $\bt^{(0)}$ and $\bt^{(j)}$s are from the regular lattice. Then under the error structure (\ref{eqn:errorstruct}),  there exists a constant $C$ that does not depend on $n$ such that the estimator $\widehat{f}_{n}$ defined by (\ref{scheme1}) satisfies 
\begin{equation*}
\begin{aligned}
& \underset{n\to\infty}{\lim\sup} \sup_{f_0\in\HH} \E\int_{\XX^d}\left[\widehat{f}_{n}(\bt)-f_0(\bt)\right]^2d\bt\\
& \quad\quad\quad\quad \leq
\begin{cases}
C\left[n(\log n)^{1-(d-p)\wedge r}\right]^{-\frac{2m}{2m+1}},  & \mbox{ if } 0\leq p< d,\\
C\left[n^{-\frac{2mr}{(2m+1)r-2}} \mathbbm{1}_{r\geq 3}+n^{-1}(\log n)^{r-1}  \mathbbm{1}_{r<3}\right],& \mbox{ if } p=d.
\end{cases}
\end{aligned}
\end{equation*}
Here the tuning parameter $\lambda$ in (\ref{scheme1}) is chosen by $\lambda\asymp \left[n(\log n)^{1-(d-p)\wedge r}\right]^{-2m/(2m+1)} $ when $0\leq p<d$, and $\lambda\asymp n^{-(2mr-2)/[(2m+1)r-2]}$ when $p=d, r\geq 3$, and $\lambda\asymp (n\log n)^{-(2m-1)/2m}$ when $p=d, r= 2$, and $\lambda\asymp n^{-(m-1)/m}$ when $p=d$, $r=1$.
\end{theorem}
\noindent
The proof of Theorem \ref{thm;deterdesgnupperbdgeneralpdreg} is also presented in Section \ref{subsec:reglattproof}. Theorems \ref{theorem:lowerbdfNlambdareg} and \ref{thm;deterdesgnupperbdgeneralpdreg} together imply that under deterministic design,  the minimax optimal rate for estimating $f_0\in\HH$ with partial derivatives is
\begin{equation*}
\begin{aligned}
& \left[n(\log n)^{1-(d-p)\wedge r}\right]^{-\frac{2m}{2m+1}} \mathbbm{1}_{0\leq p<d}\\
&\quad\quad\quad + \left[n^{-\frac{2mr}{(2m+1)r-2}} \mathbbm{1}_{r\geq 3}+n^{-1}(\log n)^{r-1}  \mathbbm{1}_{r<3}\right] \mathbbm{1}_{p=d}.
\end{aligned}
\end{equation*}
This result coincides with the rate given by \eqref{eqn:minaxhatfttprob} under random design. 
Different from ours, \cite{hall2010nonparametric} proposed a series-type estimator for incorporating various derivative data under the regular lattice. \cite{hall2010nonparametric}  showed that their estimator achieves the $\sqrt{n}$-consistency when sufficiently \emph{high-order} derivatives are available. However, it is difficult to obtain high-order derivative data in practice, such as in economics and stochastic simulation.
In contrast, we focus on incorporating \emph{first-order} partial derivatives that are easier to obtain in practice. 
\cite{chen2013enhancing} studied a stochastic kriging method for incorporating partial derivatives, and analyzed its estimation error under certain widely spread designs, where the spatial correlations of observational errors at distinct design points approximately vanish.  However, rates of convergence are not studied in \cite{chen2013enhancing}. By contrast, we quantify the improved rates of convergence with partial derivatives, which result holds under the general error structure (\ref{eqn:errorstruct}).

\vspace{-0.3in}
\change{\section{Error structures of common gradient estimators}
\label{sec:exampleappendix}
\noindent
We give three examples to illustrate  that the random error assumption in \eqref{eqn:errorstruct} holds for gradient estimators that are commonly used in real-world settings.
\paragraph{Example 1: Infinitesimal perturbation analysis (IPA).} In Section \ref{exp:blacksholes}, we studied the example of call option pricing with stochastic simulations, where the unbiased gradient estimators are derived using IPA. 
Generally, IPA estimators are obtained under the condition \citep[see,][]{ankenman2010stochastic, chen2013enhancing} that
 common random numbers are not used across design points.
Then, correlation  exists only within the error terms  $(\epsilon_i^{(0)},\epsilon_i^{(1)},\ldots,\epsilon_i^{(p)})^\top$ for the same design point $i$ and not between those of different design points,
$\text{Cov}[\epsilon_i^{(j)},\epsilon_{i'}^{(j')}] = 0$,
where $i\neq i'$ and $j,j'=0,1,\ldots,p$.  Therefore, the errors of IPA gradient estimators satisfy the error assumption \eqref{eqn:errorstruct}.

Moreover, define the correlation between the simulation noise in the response and in the estimator of the $r$th gradient component as $\rho_i^{(0,j)} = \text{Corr}[\epsilon_i^{(0)},\epsilon_i^{(j)}], j=1,\ldots,p$. Let the correlation between the simulation noise in the estimators of a pair of distinct gradient components be $\rho_i^{(j_1,j_2)} = \text{Corr}[\epsilon_i^{(j_1)},\xi_i^{(j_2)}], j_1,j_2,=1,\ldots,p$ and $j_1\neq j_2$. Notably, our error assumption \eqref{eqn:errorstruct} accommodates the scenario where the correlations $\rho_i^{(0,j)}$ and  $\rho_i^{(j_1,j_2)}$ at different design points are not necessarily equal. This characteristic is consistent with the properties of the IPA estimators as shown in \citet{ankenman2010stochastic} and \citet{chen2013enhancing}.}

\vspace{-0.4in}
\change{\paragraph{Example 2: Observational gradients.} In Section \ref{exp:costfunction}, we considered the example of cost estimation in economics, where the gradient data are directly observable. 
More specifically,  the partial derivatives of $f_0(\bt)$ with respect to input prices correspond to observable quantities of factor inputs. 

In such observational studies where derivative data are available, the errors are commonly assumed to be i.i.d.  \citep{hall2007nonparametric}. Then, $\text{Cov}[\epsilon_i^{(j)},\epsilon_{i'}^{(j')}] = 0$,
where $i\neq i'$ and $j,j'=0,1,\ldots,p$.  Therefore, the errors of observational gradients satisfy the error assumption (\ref{eqn:errorstruct}).}

\vspace{-0.4in}
\change{\paragraph{Example 3: Finite difference method.}
We  explore the finite difference method as an alternative approach to derivative estimation, as applied in the life table estimation example in Appendix \ref{ex:survival}. 
Specifically, we consider the finite-difference gradient estimator at  $t^{(0)}_i\in\R$ for $i=1,\ldots,n-1$,
\begin{equation*}
\begin{aligned}
\widehat{\frac{d f_0}{d t}}(t^{(0)}_i) & \equiv \frac{y^{(0)}_{i+1}-y^{(0)}_i}{t^{(0)}_{i+1}-t^{(0)}_i} = \frac{f(t^{(0)}_{i+1})-f(t^{(0)}_i)}{t^{(0)}_{i+1}-t^{(0)}_i} + \frac{\epsilon^{(0)}_{i+1}-\epsilon^{(0)}_i}{t^{(0)}_{i+1}-t^{(0)}_i}\\
& = f'(t^{(0)}_i) + \underbrace{\left(\frac{f(t^{(0)}_{i+1})-f(t^{(0)}_i)}{t^{(0)}_{i+1}-t^{(0)}_i} -f'(t^{(0)}_i)\right)}_{\text{term I}} + \underbrace{\frac{\epsilon^{(0)}_{i+1}-\epsilon^{(0)}_i}{t^{(0)}_{i+1}-t^{(0)}_i}}_{\text{term II}}.
\end{aligned}
\end{equation*}
By the Taylor expansion, we have 
\begin{equation*}
\begin{aligned}
\text{term I}  & = \frac{1}{2}f''(\tilde{t})(t^{(0)}_{i+1}-t^{(0)}_i),
\end{aligned}
\end{equation*}
where $\tilde{t}$ lies between $t^{(0)}_{i}$ and $t^{(0)}_{i+1}$.
Assuming that the observation errors $\epsilon_i^{(0)}$s of function data are i.i.d. and centered, and considering the continuity of the second-order derivative of $f$ along with $|t^{(0)}_{i+1}-t^{(0)}_i|=o(n^{-1/2})$, the bias of the finite-difference gradient estimator satisfies,
\begin{equation*}
\begin{aligned}
\mathbb E[\epsilon_i^{(1)}]   = \E[\text{term I}] + \E[\text{term II}] = \frac{1}{2}f''(\tilde{t})(t^{(0)}_{i+1}-t^{(0)}_i)= o(n^{-1/2}).
\end{aligned}
\end{equation*}
Note that the assumption $|t^{(0)}_{i+1}-t^{(0)}_i|=o(n^{-1/2})$ is mild and typically satisfied in practical settings, such as when $t_i^{(0)}$'s are equally spaced in $\mathcal X=[0,1]$, where  $|t^{(0)}_{i+1}-t^{(0)}_i| = 1/n = o(n^{-1/2})$.
Moreover, for $|i-i'|>1$,  we have $\text{Cov}[\epsilon_i^{(0)},\epsilon_{i'}^{(1)}] = 0$ and $\text{Cov}[\epsilon_i^{(1)},\epsilon_{i'}^{(1)}] = 0$.  Hence, the covariance of the finite-difference gradient estimator satisfies, 
\begin{equation*}
\text{Cov}[\epsilon_i^{(j)},\epsilon_{i'}^{(j')}] = O(|i-i'|^{-2}),
\end{equation*}
where $i\neq i'$ and $j,j'=0,1$.
Therefore, the errors of  finite-difference  gradient estimators satisfy the error assumption (\ref{eqn:errorstruct}).
}
\section{Additional Numerical Examples}

\noindent
In this section, we provide additional numerical examples. \change{We study a manufacturing example in Section \ref{ex:productcost},  analyze a real dataset on an actuarial life table in Section \ref{ex:survival}, and explore a statistical inference example on cost estimation in Section \ref{ex:infcost}.}
\subsection{Flexible assembly systems in manufacturing}
\label{ex:productcost}
\noindent
We study a stochastic simulation in manufacturing that generates partial derivatives. Closed-loop flexible assembly system (CLFAS) is a useful tool to lower production costs and increase flexibility in manufacturing \citep{suri1987single, chen2013enhancing}.

\begin{figure}[!ht]
    \centering
         \includegraphics[width=0.55\textwidth]{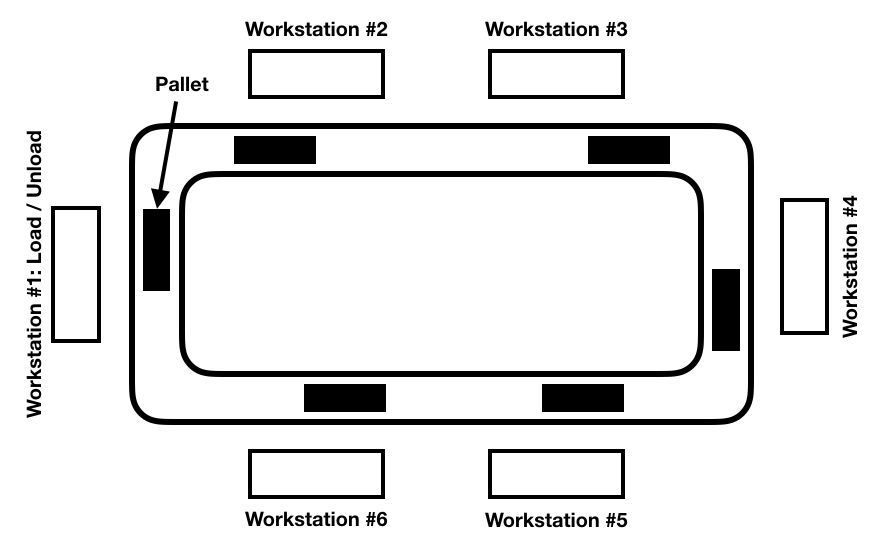}
     \caption{Diagram of CLFAS for the example in Section \ref{ex:productcost}.}
\label{fig:CLFS}
\end{figure}

Since building a CLFAS is expensive, it is important to provide a fast and accurate prediction to the CLFAS performance.
We consider a CLFAS of six automatic workstations and a conveyor with six pallets shown in  Figure \ref{fig:CLFS}.
Note that our analysis can be extended to any number of workstations or pallets. 
In this CLFAS, unfinished parts are loaded and unloaded through workstation 1 and proceed on the pallets.
The operation time at each workstation $j$, $1\leq j\leq 6$, is given by
$t_j + {\mathbf 1}\{\text{jam at station } j\} R_j$,
where $t_j$  is the fixed machine time (in minutes)  and $R_j$ is the additional random time (in minutes)  to clear the machine $j$ if it jams. 
Let $p_j$ be the probability of a part causing a jam at workstation $j$. 
Since the operation time is random, queueing may occur in the system. 
Our goal is to estimate $f_0(t_1,\ldots,t_{6})$, which denotes the expected production time of the first $5000$ parts completed by the CLFAS. Here $f_0$ can be approximated by a SS-ANOVA model in \eqref{eqn:anovadecompfti} because if  there is no queue occurs, $f_0$ has an additive structure in the covariates $(t_1,\ldots, t_{6})$. 
In the experiment, we fix $p_j=0.5\%$ and let $R_j$   i.i.d. uniformly sample from $[0.1,1.1]$.
The design points of  $(t_1,\ldots,t_{6})$  are uniformly random in $[3,9]^{6}$ with the sample size $n=100$.
To address the impact of stochastic simulation noise, we simulate $1000$ stochastic simulations of CLFAS at each design and then average the results. 

\noindent
\cite{suri1987single} proposed an IPA derivative estimators for a CLFAS as follows.
\begin{itemize}
\item[]\emph{Step 1:} Let $\AA_{j_1,j_2}$s be accumulator variables. Initialize: $\AA_{j_1,j_2} = 0$ for $j_1,j_2=1,\ldots,6$; 
\item[] \emph{Step 2:} At the end of an operation at station $j$, let $\AA_{j,j} \leftarrow \AA_{j,j} + 1$,  $j=1,\ldots,6$; 
\item[] \emph{Step 3:} If a pallet leaving station $j_1$ going to station $j_1'$ terminates an idle period of station $j_1'$, let $\AA_{j_1',j_2}\leftarrow \AA_{j_1,j_2}$, $j_2=1,\ldots,6$;  
\item[] \emph{Step 4:} If a pallet leaving station $j_1$ going to station $j_1'$ terminates a blocked period of station $j_1$, let $\AA_{j_1,j_2}\leftarrow \AA_{j_1',j_2}$, $j_2=1,\ldots,6$; 
\item[] \emph{Step 5:} At the end of the simulation, let $P$ be the total number of parts completed and $L$ be the full length of simulation in minutes. 
Output the function data $Y^{(0)}(\bt) = L/P$ and the IPA derivative estimator  $Y^{(j)}(\bt) = \AA_{6,j}/P$ for $j=1,\ldots,6$. 
\end{itemize}

\noindent
In the data generating process,  the correlation only exists for function and derivative data at the same design, not data across different design points. Hence the random errors satisfy the error structure in  (\ref{eqn:errorstruct}).
In this example, obtaining function data at a new design  requires to conduct $1000$ new simulation replications. However, it only needs to record a small matrix $\{\AA_{j_1,j_2}\}_{j_1,j_2=1}^6$ in the algorithm of \cite{suri1987single} for obtaining the IPA derivative estimators, whose computational cost is negligible compared to that of obtaining a new function data.


\begin{figure}[ht]
    \centering
         \includegraphics[width=0.6\textwidth]{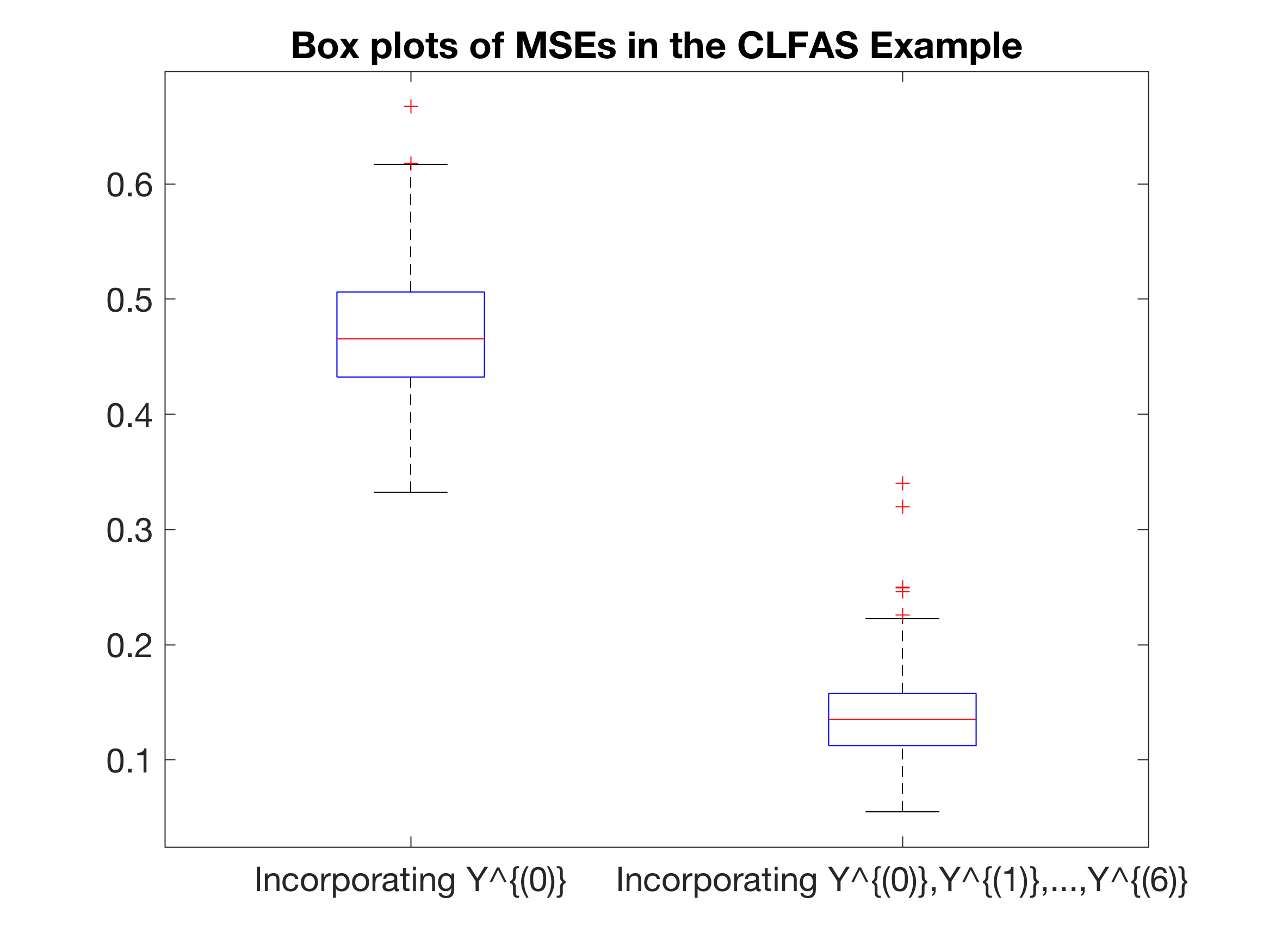}
     \caption{The box plots of  MSEs of our estimator with derivative data and the stochastic kriging without derivative data,  for the example in Section \ref{ex:productcost}.}
\label{fig:CLFSresult}
\end{figure}

\paragraph{Comparison to existing  method.}
 We compare our estimator (\ref{eqn:RF}) and the stochastic kriging method \citep{ankenman2010stochastic}.
We use the $6$-dimensional version of the tensor product Mat\'ern  kernel \eqref{eqn:maternkernel}, and choose lengthscale parameters by the five-fold cross-validation.   We estimate the MSE of estimation by a Monte Carlo sample of $10^4$ test points in $[3,9]^{6}$. Since the true production time is unknown at each test point, we approximate it by replicating $10^6$ CLFAS experiments at each test point. 

 Figure \ref{fig:CLFSresult} reports the MSEs for different methods: stochastic kriging with only function data (i.e., $p=0$), and our estimator with derivative data (i.e., $p=6$).
 The results are averaged over $1000$ simulations.
It is seen that incorporating partial derivatives leads to a significant improvement of estimation compared to without using the derivatives.

\subsection{Life table estimation}
\label{ex:survival}
\noindent
We study a real data of  U.S. 2015 period life table for the social security area (\url{www.ssa.gov/OACT/STATS/table4c6.html#fn2}), where the data separate the male and female population. 
The life table in actuarial science provides probabilities of survival and death at integer ages \citep{frees1998understanding}. 
To value payments that are not at integer ages, actuaries need to make a fractional age assumption of surviving at fractional ages.  
Our goal is to estimate the survival distribution function $f_0(t)$. Let $u(t)$ be the force of mortality function. It is known that \citep[see,][]{frees1998understanding},
\begin{equation}
\label{eqn:survivaltable}
f'_0(t) = -f_0(t)u(t).
\end{equation} 
The function data  $Y^{(0)}$ on $f_0(t)$ are generated using the death probability from life table. The force of mortality function  $u(t)$ can be estimated using the number of people that survive at age $t$, where the detail is given as follows. Denote by $l(t)$ the number of people that survive at age $t$.   Then a divided-difference estimator for $u(t)$ is \citep{jones2002critique},
\begin{equation*}
u(0) = \frac{1}{2l(0)}[3l(0) -4l(1) + l(2)], \quad u(t) = \frac{1}{2l(t)}[l(t-1) - l(t+1)] \text{ for }t>0.
\end{equation*}
The function $Y^{(0)}$, together with the estimate of $u(t)$, yield the derivative $Y^{(1)}$ according to \eqref{eqn:survivaltable}. We choose the design $t$ from equally spaced  integers from $[0,119]$ with the sample size  $n=5, 10, 15, 20$. The endpoints of $[0,119]$ are included.


\begin{table}[h!]
\centering
 \caption{\change{The comparison of average MSEs and standard errors of our estimator with those of smoothing spline estimator,  for the example in Section \ref{ex:survival} with $1000$ simulations. The table shows metrics: ``average MSE (standard error),''  in units of $10^{-4}$.}}
 \label{table:survivalfunction}
  \scalebox{0.72}{
  \change{
   \begin{tabular}{l  c c  c c c }
 \hline
  \rule{0pt}{\normalbaselineskip}
  &     & $n=5$ & $n=10$  & $n=15$ & $n=20$  \\   [0.5ex]
 \hline
 \rule{0pt}{\normalbaselineskip}
\multirow{2}{*}{M} & Smoothing spline estimator with $Y^{(0)}$  & $15.3674~(4.8815)$   & $6.7944~(2.2596)$ & $1.7687~(0.6676)$ & $0.1745~(0.0594)$ \\  [0.5ex]
 \rule{0pt}{\normalbaselineskip}
   & Our estimator   with $Y^{(0)}+ Y^{(1)}$  & $\textbf{7.4381~(2.5242)}$  & $\textbf{1.6488~(0.5009)}$ & $\textbf{0.3446~(0.1012)}$  & $\textbf{0.0227~(0.0098)}$  \\  [0.5ex]  \cline{2-6}
 \rule{0pt}{\normalbaselineskip}
\multirow{2}{*}{F}  & Smoothing spline estimator with $Y^{(0)}$  & $23.0655~(7.1699)$   & $9.9948~(3.8025)$ & $2.2299~(0.8110)$ & $0.5925~(0.1569)$  \\  [0.5ex]
 \rule{0pt}{\normalbaselineskip}
   & Our estimator   with $Y^{(0)}+ Y^{(1)}$  &$\textbf{9.4745~(3.2385)}$  & $\textbf{2.4790~(0.8654)}$ & $\textbf{0.4091~(0.1015)}$ & $\textbf{0.0755~(0.0152)}$ \\  [0.5ex]
 \hline
  \end{tabular}
}
}
\end{table}

\paragraph{Comparison to existing method.}
Smoothing spline \citep{wahba1990} is widely used for smoothing noisy data. 
We compare the results of our estimator (\ref{eqn:RF}) using the estimated derivative and the smoothing spline without using the derivative. 
We use the Mat\'ern  kernel (\ref{eqn:maternkernel}) and estimate the MSE by using the full sample at $t=0,1,\cdots, 119$.

\change{Table \ref{table:survivalfunction} reports the MSEs and standard errors  for varying sample size $n$, different population, and different methods: smoothing spline with only function data (i.e., $p=0$), and our estimator with function and derivative data (i.e., $p=1$). The results are obtained over  $1000$ simulations in each setting.  It is seen that
our estimator incorporating derivative data significantly improves the estimation results compared to the smoothing splines.}

Table \ref{table:ratiosurvivalfunction} reports the ratios of the MSE of our estimator incorporating derivative data (i.e., $p=1$) relative to the MSE of smoothing spline estimator with only  function data (i.e., $p=0$). 
It is seen that the ratio decreases with the sample size, which agrees with our theory in Section \ref{sec:minmaxriskregularlat} that incorporating derivative data accelerates the convergence rate.

\begin{table}[h!]
\centering
   \caption{The ratios of MSE of our estimator with derivative data (i.e., $p=1$) relative to MSE of spline estimator with  only function data (i.e., $p=0$), for the example in Section \ref{ex:survival}.}
 \label{table:ratiosurvivalfunction}
  \scalebox{0.8}{
 \begin{tabular}{l   c  c c c }
 \hline
  \rule{0pt}{\normalbaselineskip}
      & $n=5$ & $n=10$  & $n=15$ & $n=20$  \\   [0.5ex]
 \hline
 \rule{0pt}{\normalbaselineskip}
{Male}   & $0.4840$   & $0.2426$ & $0.1948$ & $0.1301$\\  [0.5ex]
  \rule{0pt}{\normalbaselineskip}
Female     & $0.4108$   & $0.2480$ & $ 0.1835$ & $0.1274$  \\  [0.5ex]
 \hline
  \end{tabular}}
\end{table}

\subsection{\change{Statistical inference for the cost estimation in economics}}
\label{ex:infcost}
\noindent
\change{We consider the economic problem of the cost function estimation in Section \ref{exp:costfunction}. 
We employ the bootstrap method  \citep[see, e.g.,][]{efron1993introduction} to quantify the uncertainty of our estimators (\ref{eqn:RF}) for this example. 
The process for generating a bootstrap sample includes the following steps:
(a) Produce $B$ bootstrap samples by resampling centered residuals;
(b) Re-estimate the functions to obtain $B$ bootstrap estimates of $f_0$,  denoted as 
$\widehat{f}_{b}^*$ for $b=1,\ldots,B$. From this, we can derive a bootstrap 
 confidence interval for $f_0$ at any new input $\bt_{\text{new}}$. Specifically, we determine the
$\alpha/2$ and $1-\alpha/2$ sample quantiles from $\{\widehat{f}_{1}^*(\bt_{\text{new}}),\ldots,\widehat{f}_{B}^*(\bt_{\text{new}})\}$, represented as $z_{\alpha/2}^*$ and $z_{1-\alpha/2}^*$, respectively. 
The confidence interval is thus$(z_{\alpha/2}^*,z_{1-\alpha/2}^*)$. 
Given that bias in non-parametric regression may affect the asymptotic coverage of bootstrap confidence intervals, two common correction strategies include undersmoothing and oversmoothing  \citep[see, e.g.,][]{hardle1988bootstrapping,hall1992effect, hall1992bootstrap}. Undersmoothing  is often preferred due to its simplicity and effectiveness \citep{hall1992effect}. Our estimation procedure can be easily modified to incorporate undersmoothing by selecting a smaller smoothing parameter. 
Despite the potential for a modest gain in practical performance, these strategies require another ad hoc choice of the amount of undersmoothing or oversmoothing. 
Moreover, it is quite common to ignore this bias issue, essentially
leading to the use of non-adjusted confidence intervals as suggested by \citet{efron1993introduction} and \citet{ruppert2003semiparametric}. To keep the approach simple, we use the non-adjusted confidence intervals in this example with $B=2000$.
We set the significance level at $95\%$. The empirical coverage probability is calculated as the percentage of instances in which the confidence interval covers  $f_0(\bt_{\text{new}})$ across $1000$ repetitions, with $\bt_{\text{new}}$ randomly drawn from $\XX^d$ for each repetition.

Table \ref{table:costfunctioninference} compares the coverage probability and interval length when incorporating various levels of gradients ($p=0,1,2$) using our method (\ref{eqn:RF}). The average length of the bootstrap confidence interval is computed across $1000$ repetitions. We observe in Table \ref{table:costfunctioninference} that the coverage probability of our estimator approximates $95\%$ consistently across all gradient levels ($p=0,1,2$). However, intervals without gradient information have larger lengths compared to those incorporating gradients. This observations aligns with our theoretical finding in Section \ref{sec:minimaxoptimality} that the inclusion of gradient data results in a faster decease in the MSE of the estimator compared to  excluding gradient data. }

\begin{table}[ht]
\centering
 \caption{\change{Coverage probability and length of $95\%$ bootstrap confidence intervals, incorporating various levels of gradients ($p=0,1,2$) using our method (\ref{eqn:RF}), for the example in Section \ref{ex:infcost} with $1000$ simulations.}}
 \label{table:costfunctioninference}
  \scalebox{0.8}{
\change{ \begin{tabular}{cccccccc}
\hline
\multirow{2}{*}{}  & \multirow{2}{*}{} & \multicolumn{2}{c}{with only $Y^{(0)}$} & \multicolumn{2}{c}{with $Y^{(0)}+Y^{(1)}$ } & \multicolumn{2}{c}{with $Y^{(0)}+Y^{(1)}+Y^{(2)}$ } \\[0.5ex] 
     & & Prob ($\%$)  & Length & Prob ($\%$)  & Length & Prob ($\%$) & Length  \\[0.5ex]  \hline 
 \rule{0pt}{\normalbaselineskip}
\multirow{3}{*}{$n=100$} & $\rho=0$ & 95.9722   &  14.4226   &   95.9116   &   13.4315    &   \textbf{96.8295}  &  \textbf{11.5146} \\[0.5ex] 
 \rule{0pt}{\normalbaselineskip}
        & $\rho=0.4$ &  94.4613 &  15.6566  & 96.1340 &  13.6916 & \textbf{96.9722} & \textbf{12.3477}  \\[0.5ex] 
 \rule{0pt}{\normalbaselineskip} 
        & $\rho=0.9$ & 94.1245 & 16.4833  & 94.1276 & 14.3109  & \textbf{96.1200}  & \textbf{13.4637}   \\[0.5ex]  \cline{2-8} 
 \rule{0pt}{\normalbaselineskip}
\multirow{3}{*}{$n=200$}  & $\rho=0$ & 96.3252  & 11.0673 & 96.6061 & 9.1801 & \textbf{97.3076} & \textbf{8.8906}  \\[0.5ex] 
        & $\rho=0.4$ & 95.7476 & 12.1215 & 96.5717 & 10.0875  & \textbf{96.2182}  & \textbf{9.7177}     \\[0.5ex] 
        & $\rho=0.9$ & 94.5275 & 12.5909 & 95.2201 & 11.3109  & \textbf{96.9119}  & \textbf{10.4494}   \\[0.5ex]  \cline{2-8} 
 \rule{0pt}{\normalbaselineskip}
\multirow{3}{*}{$n=500$}  & $\rho=0$ &  95.6127 & 8.4226  &  95.0207 &  6.6719 & \textbf{96.4846} & \textbf{5.7415}  \\[0.5ex] 
      & $\rho=0.4$  & 95.9650 & 8.6566  &  96.9369   &  7.4831  & \textbf{95.8447}  & \textbf{5.9061}     \\[0.5ex] 
      & $\rho=0.9$  &  95.1417 &  9.4833  & 95.4791  &  7.8287  & \textbf{95.4852} &  \textbf{6.0834}   \\[0.5ex]  \cline{2-8} 
 \rule{0pt}{\normalbaselineskip}
\multirow{3}{*}{$n=1000$} & $\rho=0$ & 95.9001 & 6.4732 &  96.2507 & 5.2168  & \textbf{97.5913} & \textbf{3.6970} \\[0.5ex] 
    & $\rho=0.4$ & 95.3200  &  6.8322 & 95.3559 & 5.7529  & \textbf{96.6146}  & \textbf{3.8210}     \\[0.5ex] 
    & $\rho=0.9$ &  95.0288 & 7.4983 &  95.9213 &  5.9815  &  \textbf{96.3667}  &  \textbf{4.1591}     \\[0.5ex]  \hline
\end{tabular}
}}
\end{table}

\subsection{\change{Additional comparisons with Hall and Yatchew's estimator}}
\noindent
\change{We present two additional examples to compare our estimator with the regression-kernel estimator in \citet{hall2007nonparametric}. 

The first example is the stochastic simulation on call option pricing in Section \ref{exp:blacksholes}. We adopt the same simulation setting, and use the actual output as the reference, which is given by
$f_0(S_0,r_*,\sigma_*) = 
S_0\Phi\left(-d_1+\sigma_*\right) - 100e^{-r_*}\Phi\left(-d_1\right)$. Here $d_1 = \sigma_*^{-1}[\log 100 - \log(S_0) - (r_*-\sigma_*^2/2)]$ and $\Phi(\cdot)$ is the CDF of standard normal distribution.
For the estimator in  \cite{hall2007nonparametric}, we follow the approach in Hall and Yatchew's Example 3 to
 average $(S_0, r_*)$ and $(S_0,\sigma_*)$ directions locally, and then average the estimates.
The MSE$=\E(\widehat{f}_{n} - f_0)^2$ is estimated using a Monte Carlo  sample of $10^4$ test points in $[80,120]\times [0.01,0.05]\times [0.2,1]$. Table \ref{table:optionhycomparison} reports the MSEs and standard errors across varying sample size $n$, replications of the simulation $q$, and levels of gradient data. The results are summarized based on $1000$ simulations for each scenario.
It is seen that our estimator significantly enhances estimation accuracy compared to Hall and Yatchew's estimator.

\begin{table}[ht]
\centering
 \caption{\change{The average MSEs and standard errors of our estimator and those of Hall and Yatchew's estimator, considering various gradient types,  for the example in Section \ref{exp:blacksholes} with $1000$ simulations. The table shows metrics: ``average MSE (standard error)," in units of $10^{-2}$.}}
   \label{table:optionhycomparison}
  \scalebox{0.8}{
\change{  \begin{tabular}{c c c c c  c }
 \hline
  \rule{0pt}{\normalbaselineskip}
  &    & \footnotesize{Hall and Yatchew with}    &   \footnotesize{Our Estimator (\ref{eqn:RF}) with}  & \footnotesize{Hall and Yatchew with}    &  \footnotesize{Our Estimator (\ref{eqn:RF}) with} \\   [0.5ex]
   $n$ &  $q$ &  \footnotesize{$Y^{(0)}+Y^{(1)}+Y^{(2)}$} &  \footnotesize{$Y^{(0)}+Y^{(1)}+Y^{(2)}$} & \footnotesize{$Y^{(0)}+Y^{(1)}+Y^{(2)}+ Y^{(3)}$}     & \footnotesize{$Y^{(0)}+Y^{(1)}+Y^{(2)}+ Y^{(3)}$} \\   [0.5ex]
 \hline
 \rule{0pt}{\normalbaselineskip}
& $1000$ &  $12.1741~(3.8190)$ &  $8.5599~(3.8415)$ & $11.4690~(3.4460)$    & $\textbf{3.9507~(1.3516)}$ \\  [0.5ex] 
 \rule{0pt}{\normalbaselineskip}
 $7^3$ & $2000$  & $11.8920~(3.3524)$   & $4.5767~(1.3534)$  & $10.8306~(3.1022)$  & $\textbf{2.2173~(0.6291)}$\\  [0.5ex]
 \rule{0pt}{\normalbaselineskip}
 & $5000$   & $10.9300~(2.8547)$ & $2.8012~(0.9527)$  & $10.1989~(2.6965)$ & $\textbf{1.8633~(0.5813)}$ \\  [0.5ex] \cline{2-6}
 \rule{0pt}{\normalbaselineskip}
 & $1000$ & $7.6601~(2.5093)$ & $2.2702~(0.8333)$  & $7.3001~(2.1872)$ & $\textbf{1.5684~(0.5730)}$  \\  [0.5ex] 
 \rule{0pt}{\normalbaselineskip}
 $14^3$& $2000$  & $7.2160~(2.4019)$ & $1.7510~(0.6079)$ & $7.0696~(2.0615)$  & $\textbf{1.2402~(0.5062)}$ \\  [0.5ex]
 \rule{0pt}{\normalbaselineskip}
& $5000$  & $6.9731~(2.3591)$ & $1.4351~(0.5593)$  & $6.2591~(1.9210)$ & $\textbf{1.1468~(0.4213)}$  \\  [0.5ex] \cline{2-6}
 \rule{0pt}{\normalbaselineskip}
 & $1000$ & $6.1625~(2.0150)$ & $1.3341~(0.5150)$ & $5.3861~(1.7399)$  & $\textbf{1.0912~(0.3570)}$  \\  [0.5ex] 
     \rule{0pt}{\normalbaselineskip}
$21^3$ & $2000$ & $6.0483~(1.9180)$   & $1.1994~(0.4180)$  & $5.0355~(1.6164)$  & $\textbf{0.8988~(0.2919)}$ \\  [0.5ex]
     \rule{0pt}{\normalbaselineskip}
 & $5000$ & $5.7112~(1.8264)$ & $0.9541~(0.3654)$  & $4.7698~(1.4877)$   & $\textbf{0.7460~(0.2124)}$ \\  [0.5ex]
 \hline
  \end{tabular}}
  }
\end{table}

The second example is the single voltage clamp experiment in Section \ref{ex:computerexperiment}. We follow the same simulation setting.
For the estimator in  \cite{hall2007nonparametric}, we again follow the approach in Hall and Yatchew's Example 3 to
 average $(t_1, t_2)$ and $(t_1,t_3)$ directions locally, and then average the estimates.
The MSE$=\E(\widehat{f}_{n} - f_0)^2$ is estimated using a Monte Carlo  sample of $10^4$ test points in $\XX^3$. 
Since the true function $f_0(\bt)$ is unknown at each test point, we approximate it by using total $N=19$ real ion channel samples at each test point. The function and gradient training data are generated using $N'=10$ real ion channel samples, which are randomly chosen from the total $N=19$ samples. Table \ref{table:voltagehycomparison} reports the MSEs and standard errors across varying sample size $n$, replications of the simulation $q$, and levels of gradient data. The results are summarized based on $1000$ simulations for each scenario.
Table \ref{table:voltagehycomparison} shows that our estimator outperforms Hall and Yatchew's estimator in terms of estimation accuracy.

\begin{table}[ht]
\centering
  \caption{\change{The average MSEs and standard errors of our estimator and those of Hall and Yatchew's estimator, considering various gradient types,  for the example in Section \ref{ex:computerexperiment} with $1000$ simulations. The table shows metrics: ``average MSE (standard error),''  in units of $10^{-6}$.} }
   \label{table:voltagehycomparison}
  \scalebox{0.8}{
\change{ \begin{tabular}{c  c c c  c }
 \hline
  \rule{0pt}{\normalbaselineskip}
    &  \footnotesize{Hall and Yatchew with}   & \footnotesize{Our Estimator (\ref{eqn:RF}) with}  &  \footnotesize{Hall and Yatchew with}   & \footnotesize{Our Estimator (\ref{eqn:RF}) with}\\   [0.5ex]
$n$    &  \footnotesize{with only $Y^{(0)}$}     &  \footnotesize{with only $Y^{(0)}$} & \footnotesize{$Y^{(0)}+Y^{(1)}+Y^{(2)}+ Y^{(3)}$}     & \footnotesize{$Y^{(0)}+Y^{(1)}+Y^{(2)}+ Y^{(3)}$}  \\   [0.5ex]
 \hline
 \rule{0pt}{\normalbaselineskip}
 $1000$ & 11.0134~(5.6061) &  $10.6491~(4.9867)$& $8.6488~(4.4921)$    & $\textbf{7.7804~(3.6737)}$ \\  [0.5ex] 
 \rule{0pt}{\normalbaselineskip}
 $2000$  & 9.0626~(5.0207) & $8.5302~(4.3339)$  & $6.3674~(3.1476)$  & $\textbf{5.1375~(2.4687)}$\\  [0.5ex]
 \rule{0pt}{\normalbaselineskip}
  $3000$  & 7.0134~(4.2182) & $6.4296~(3.9595)$  & $5.0655~(2.4226)$ & $\textbf{3.1035~(1.7187)}$ \\  [0.5ex] 
 \rule{0pt}{\normalbaselineskip}
 $5000$  & 6.2315~(3.4613) & $5.4143~(3.2268)$  & $3.1745~(1.6182)$ & $\textbf{2.1305~(0.9322)}$  \\  [0.5ex] 
 \hline
  \end{tabular}}}
  \end{table}

}

\section{Proofs of the Main Results}
\label{sec:proofsofallresults}


\subsection{Proof of Theorem \ref{thm:uniqueminimizer}}
\label{sec:pfofuniqueminimizer}
\noindent
\change{We prove a more general result in the following lemma. Let  
\begin{equation*}
l_n(f) \equiv \frac{1}{n}\sum_{i=1}^{n}\left[y_i^{(0)}- f(\bt_i^{(0)})\right]^2+\sum_{j=1}^pw_j\cdot \frac{1}{n}\sum_{i=1}^{n}\left[y_i^{(j)}-\frac{\partial f}{\partial t_j}(\bt_i^{(j)})\right]^2.
\end{equation*}
Then the optimization problem \eqref{scheme1} can be rewritten as,
\begin{equation*}
\begin{aligned}
\min_{f\in\HH} l_n(f) \text{ subject to } \|f\|_{\HH}\leq R_n.
\end{aligned}
\end{equation*}
\begin{lemma}
\label{lem:strongerlemofthm1}
Let $f_{I,n}$ is the unique solution to the problem:
$\min_{f\in\HH}\|f\|_\HH \text{ subject to } l_n(f)=0$.
Then, for $0\leq R_n<\|f_{I,n}\|_{\HH}$, there exists a unique minimizer $\widehat{f}_n(\bt)$ of \eqref{scheme1} in a finite-dimensional space. Specifically, there exist coefficients ${\boldsymbol \alpha}_j = (\alpha_{1j},\ldots,\alpha_{nj})^\top\in\R^n$ for $j=0,1,\ldots,p$ such that,
\begin{equation}
\label{eqn:solntoconsprob}
\widehat{f}_n(\bt) = \sum_{i=1}^n\alpha_{i0} K_d(\bt_i^{(0)},\bt) + \sum_{j=1}^p\sum_{i=1}^n \alpha_{ij}\frac{\partial K_d}{\partial t_j}(\bt_i^{(j)},\bt),
\end{equation}
and $\|\widehat{f}_n\|_{\HH}=R_n$. For  $R_n\geq \|f_{I,n}\|_{\HH}$, $\widehat{f}_n(\bt)$ in \eqref{eqn:solntoconsprob} is one of the minimizers of \eqref{scheme1}.
\end{lemma}
\begin{proof}
Following the proof of Lemma 1 and Proposition 3 of \citet{lim2024estimating}, there exists a unique solution to the problem:
\begin{equation*}
\min_{f\in\HH}\|f\|_\HH \text{ subject to } l_n(f)=0,
\end{equation*} 
which is denoted by $f_{I,n}$. Additionally, if $1\leq R_n<J(f_I)$, there exists a unique minimizer $\widehat{f}_n(\bt)$ of \eqref{scheme1} that satisfies $\|\widehat{f}_n\|_{\HH}=R_n$.  A similar result can be found in Theorem 3 of \citet{schoenberg1964spline}.

Since the optimization problem of  \eqref{scheme1} is convex, by Lagrangian duality, it  can be reformulated as
\begin{equation*}
\begin{aligned}
\widehat{f}_{n} = \underset{f\in \HH}{\arg\min}& \left\{l_n(f) + \lambda\|f\|_\HH^2\right\}.
\end{aligned}
\end{equation*}
Here, for a fixed set of function data and partial derivatives, the smoothing parameter $\lambda\geq 0$ is a function of the radius $R_n\geq 0$. Under the condition \eqref{eqn:partial2tt'kcx1x1}, the derivative $\partial f/\partial t_j$ is a bounded linear functional in $\HH$. Following a similar argument to that of Theorem 1.3.1 in \citet{wahba1990}, $\widehat{f}_n(\bt)$ takes the form in  \eqref{eqn:solntoconsprob}. 
For $R_n\geq \|f_{I,n}\|_{\HH}$, following the proof of Proposition 6 of \citet{lim2024estimating},  $\widehat{f}_n(\bt)$ in  \eqref{eqn:solntoconsprob} is one of the minimizers  of \eqref{scheme1}. This completes the proof of Lemma \ref{lem:strongerlemofthm1}.
\end{proof}

Next, by Lemma \ref{lem:strongerlemofthm1}, we know that for any $R_n\geq 0$,  $\widehat{f}_n(\bt)$ in  \eqref{eqn:solntoconsprob} is a minimizer  of \eqref{scheme1} and it is in a finite-dimensional space spanned by $\{K_d(\bt_i^{(0)},\cdot), \frac{\partial K_d}{\partial t_j}(\bt_i^{(j)},\bt); 1\leq i\leq n, 1\leq j\leq p\}$. 
This completes the proof of Theorem \ref{thm:uniqueminimizer}.
}

\subsection{Proof of Theorem \ref{theorem:lowerbdfNlambdaregrandom}}
\label{subsubsec:prooflowerrandom}
\noindent
We establish the lower bound under random design via Fano's lemma \citep{tsybakovintroduction}.
It suffices to consider a particular case where the random errors $\epsilon^{(0)}$ and $\epsilon^{(j)}$s are independent Gaussian with zero mean and unit variance, and $\Pi^{(0)}$ and $\Pi^{(j)}$s are uniform distributions, and $\HH_1$ is generated by periodic kernels. The lower bound established for this case is at least for the general cases \citep{tsybakovintroduction}.

Let $N$ be a natural number whose value will be clear later.  We first derive the eigenvalue decay rate for the kernel $K_d$, which generates the RKHS $\HH$. 
We introduce some additional notation. 
Define a family of the multi-index $\vec{\bnu}$ by
\begin{equation}
\label{def:Vbnu}
\begin{aligned}
\bV&  = \{{\vec{\bnu} = (\nu_1,\ldots,\nu_d)^\top \in\N^d},\mbox{ where at most $r\geq 1$ of $\nu_k$s are not } 1\}.
\end{aligned}
\end{equation}
For a given $\tau>0$, the number of multi-indices $\vec{\bnu}=(\nu_1,\ldots,\nu_r)\in\N^r$ satisfying 
\begin{equation*}
\nu_1^{-2m}\cdots\nu_r^{-2m}\geq \tau
\end{equation*} is the same as the number of multi-indices such that $\nu_1\cdots\nu_r\leq \tau^{-1/(2m)}$, which amounts to
\begin{equation}
\label{eqn:tau12mlogr1}
\begin{aligned}
\sum_{\nu_2\cdots\nu_r\leq \tau^{-1/(2m)}}\tau^{-1/(2m)}/(\nu_2\cdots\nu_r) & = \tau^{-1/(2m)}\left(\sum_{\nu\leq \tau^{-1/(2m)}}1/\nu\right)^{r-1} \\
& \asymp \tau^{-1/(2m)}(\log 1/\tau)^{r-1}.
\end{aligned}
\end{equation}
Denote by $\lambda_N(K_d)$ the $N$th eigenvalues of $K_d$.
By inverting (\ref{eqn:tau12mlogr1}), we obtain 
\begin{equation*}
\lambda_N(K_d)\asymp\left[N(\log N)^{1-r}\right]^{-2m}.
\end{equation*}
Hence, the multi-indices $\vec{\bnu}=(\nu_1,\ldots,\nu_r)\in\N^r$ satisfying $\nu_1\cdots \nu_r\leq N$
correspond to the first
\begin{equation*}
c_0N(\log N)^{r-1}
\end{equation*}
eigenvalues of $K_d$, for some constant $c_0$.
Let $b$ be a length-$\{c_0N(\log N)^{r-1}\}$ binary sequence, 
\begin{equation*}
b=\{b_{\vec{\bnu}}: \nu_1\cdots\nu_r\leq N\}\in\{0,1\}^{c_0N(\log N)^{r-1}}.
\end{equation*} 
Let $\{\tilde{\lambda}_{\vec{\bnu}}:\nu_1\cdots\nu_r\leq  N\}$ be the first $c_0N(\log N)^{r-1}$ eigenvalues of $K_d$. Denote by 
\begin{equation*}
\{\tilde{\lambda}_{\vec{\bnu}+c_0N(\log N)^{r-1}}:\nu_1\cdots\nu_r\leq  N\} 
\end{equation*}the $\{c_0N(\log N)^{r-1}+1\}$th, $\{c_0N(\log N)^{r-1}+2\}$th,\ldots, $\{2c_0N(\log N)^{r-1}\}$th eigenvalues of $K_d$.

For brevity, we only prove for the case $p=d$ and $r\geq 3$. The other cases $p=d$, $r\leq 2$ and $0\leq p<d$ can be showed similarly. 
Write
\begin{align}
 & f_b(t_1,\ldots,t_r)= N^{-\frac{1}{2}+\frac{1}{r}}\sum_{\nu_1\cdots\nu_r\leq  N}b_{\vec{\bnu}}\left(1+\nu_1^2+\cdots+\nu_r^2\right)^{-\frac{1}{2}} \nonumber\\
 & \quad\quad\quad\quad\quad\quad\quad\quad\quad\quad\quad\quad\times\tilde{\lambda}_{\vec{\bnu}+c_0N(\log N)^{r-1}}^{\frac{1}{2}}\psi_{\vec{\bnu}+c_0N(\log N)^{r-1}}(t_1,\ldots,t_r),\nonumber
\end{align}
where $\psi_{\vec{\bnu}+c_0N(\log N)^{r-1}}(t_1,\ldots,t_r)$ are the corresponding eigenfunctions of $\tilde{\lambda}_{\vec{\bnu}+c_0N(\log N)^{r-1}}$ of $K_d$. Note that
\begin{align}
\|f_b\|_\HH^2 & = N^{-1+\frac{2}{r}}\sum_{\nu_1\cdots\nu_r\leq  N}b_{\vec{\bnu}}^2(1+\nu_1^2+\cdots+\nu_r^2)^{-1}\nonumber\\
& \leq N^{-1+\frac{2}{r}}\sum_{\nu_1\cdots\nu_r\leq  N}(1+\nu_1^2+\cdots+\nu_r^2)^{-1}\asymp 1,\nonumber
\end{align}
where the last step by Lemma \ref{lemma:intx1xrzxk1z2}, and this implies $f_b(\cdot)\in\HH$.

By the Varshamov-Gilbert bound, e.g., \cite{tsybakovintroduction}, there exists a collection of binary sequences $\{b^{(1)},\ldots,b^{(M)}\}\subset\{0,1\}^{c_0N(\log N)^{r-1}}$ such that 
\begin{equation*}
M\geq 2^{c_0N(\log N)^{r-1}/8},
\end{equation*} 
and
\begin{equation*}
H(b^{(l)},b^{(q)})\geq c_0N(\log N)^{r-1}/8,\quad \forall 1\leq l<q\leq M.
\end{equation*}
Here $H(\cdot,\cdot)$ denotes the Hamming distance.
Then, for $b^{(l)},b^{(q)}\in\{0,1\}^{c_0N(\log N)^{r-1}}$,
\begin{align}
& \|f_{b^{(l)}}-f_{b^{(q)}}\|_{L_2}^2\nonumber\\
& \quad \geq N^{-1+2/r}(2N)^{-2m}\sum_{\nu_1\cdots\nu_r\leq  N}(1+\nu_1^2+\cdots+\nu_r^2)^{-1}\left[b^{(l)}_{\vec{\bnu}} - b^{(q)}_{\vec{\bnu}}\right]^2\nonumber\\
&\quad \geq N^{-1+2/r}(2N)^{-2m}\sum_{c_17N/8\leq \nu_1\cdots\nu_r\leq N}(1+\nu_1^2+\cdots+\nu_r^2)^{-1}\nonumber\\
& \quad = c_2 N^{-2m}\nonumber
\end{align}
for some constants $c_1$ and $c_2$, where the last step is by Lemma \ref{lemma:intx1xrzxk1z2}.

On the other hand, for any $b^{(l)}\in\{b^{(1)},\ldots,b^{(M)}\}$, again by Lemma \ref{lemma:intx1xrzxk1z2}, \begin{equation*}
\begin{aligned}
& \|f_{b^{(l)}}\|_{L_2}^2+ \sum_{j=1}^p\| \partial f_{b^{(l)}}/\partial t_j\|_{L_2}^2\leq N^{-1+2/r}\sum_{\nu_1\cdots\nu_r\leq  N}\nu_1^{-2m}\cdots \nu_r^{-2m}\left[b^{(l)}_{\vec{\bnu}} \right]^2\\
& \quad \leq N^{-1+2/r}\sum_{\nu_1\cdots\nu_r\leq  N}\nu_1^{-2m}\cdots \nu_r^{-2m} = c_3 N^{-2m+2/r}(\log N)^{r-1}
\end{aligned}
\end{equation*}
for some constant $c_3$.

A standard argument gives that the lower bound can be reduced to the error probability in a multi-way hypothesis test \citep{tsybakovintroduction}. Specifically, let $\Theta$ be a random variable uniformly distributed on $\{1,\ldots,M\}$. Note that
\begin{equation}
\label{eqn:inftildeffoinhh1}
\begin{aligned}
& \inf_{\tilde{f}}\sup_{f_0\in\HH}\P\left\{\|\tilde{f} - f_0\|_{L_2}^2\geq \frac{1}{4}\min_{b^{(l)}\neq b^{(q)}}\|f_{b^{(l)}} - f_{b^{(q)}}\|^2_{L_2}\right\}\geq \inf_{\widehat{\Theta}}\P\{\widehat{\Theta}\neq \Theta\}.
\end{aligned}
\end{equation}
The infimum on the right-hand side is taken over all decision rules that are measurable functions of the data. By Fano's lemma,
\begin{equation}
\label{eqn:fanolemhatthetap}
\begin{split}
& \P\left\{\widehat{\Theta}\neq \Theta|\bt_1^{(0)},\ldots,\bt_n^{(0)};\ldots;\bt_1^{(p)},\ldots,\bt_n^{(p)} \right\}\\
&\quad  \geq 1-\frac{1}{\log M}\times\left[\mathbbm{1}_{\bt_1^{(0)},\ldots,\bt_n^{(0)};\ldots;\bt_1^{(p)},\ldots,\bt_n^{(p)}}(y_1^{(0)},\ldots,y_n^{(0)},\ldots,y_1^{(p)},\ldots,y_n^{(p)};\Theta)+\log 2\right],
\end{split}
\end{equation}
where 
\begin{equation*}\mathbbm{1}_{\bt_1^{(0)},\ldots,\bt_n^{(0)};\ldots;\bt_1^{e_p},\ldots,\bt_n^{e_p}}(y_1^{(0)},\ldots,y_n^{(0)},\ldots,y_1^{(p)},\ldots,y_n^{(p)})
\end{equation*} is the mutual information between $\Theta$ and $\{y_1^{(0)},\ldots,y_n^{(0)},\ldots,y_1^{(p)},\ldots,y_n^{(p)}\}$, and we fix the design points $\{\bt_1^{(0)},\ldots,\bt_n^{(0)};$ $\ldots;\bt_1^{(p)},\ldots,\bt_n^{(p)}\}$. Thus,
\begin{equation}
\label{eqn:inftildeffoinhh3}
\begin{aligned}
& \E_{\bt_1^{(0)},\ldots,\bt_n^{(0)};\ldots;\bt_1^{(p)},\ldots,\bt_n^{(p)}}\left[\mathbbm{1}_{\bt_1^{(0)},\ldots,\bt_n^{(0)};\ldots;\bt_1^{(p)},\ldots,\bt_n^{(p)}}\left(y_1^{(0)},\ldots,y_n^{(0)},\ldots,y_1^{(p)},\ldots,y_n^{(p)};\Theta\right)\right]\\
&  \leq \binom M2^{-1} \sum_{b^{(l)}\neq b^{(q)}} \E_{\bt_1^{(0)},\ldots,\bt_n^{(0)};\ldots;\bt_1^{(p)},\ldots,\bt_n^{(p)}} \mathcal{K}\left(\mathbf{P}_{f_{b^{(l)}}}| \mathbf{P}_{f_{b^{(q)}}}\right)\\
& \leq \frac{n(p+1)}{2}\binom M2^{-1} \sum_{b^{(l)}\neq b^{(q)}} \E_{\bt_1^{(0)},\ldots,\bt_n^{(0)};\ldots;\bt_1^{(p)},\ldots,\bt_n^{(p)}}\|f_{b^{(l)}} - f_{b^{(q)}}\|_{*n}^2.
\end{aligned}
\end{equation}
Here $\mathcal{K}(\cdot|\cdot)$ is the Kullback-Leibler distance, $\mathbf{P}_{f}$ is conditional distribution of $y_i^{(0)}$ and $y_i^{(j)}$s given $\{\bt_1^{(0)},\ldots,\bt_n^{(0)};\ldots;\bt_1^{(p)},\ldots,\bt_n^{(p)}\}$, and the norm $\|\cdot\|_*$ is defined  as follows,
\begin{equation*}
\|f\|_{*n}^2 = \frac{1}{n(p+1)}\sum_{i=1}^n\left\{[f(\bt_i^{(0)})]^2+\sum_{j=1}^p[\partial f(\bt_i^{(j)})/\partial t_j]^2\right\},\quad\forall f:\XX^r\mapsto\R.
\end{equation*}
Thus,
\begin{equation}
\label{eqn:inftildeffoinhh4}
\begin{aligned}
& \E_{\bt_1^{(0)},\ldots,\bt_n^{(0)};\ldots;\bt_1^{(p)},\ldots,\bt_n^{(p)}}\left[\mathbbm{1}_{\bt_1^{(0)},\ldots,\bt_n^{(0)};\ldots;\bt_1^{(p)},\ldots,\bt_n^{(p)}}(y_1^{(0)},\ldots,y_n^{(0)},\ldots,y_1^{(p)},\ldots,y_n^{(p)};\Theta)\right]\\
& \quad\leq \frac{n(p+1)}{2}\binom M2^{-1} \left.\sum_{b^{(l)}\neq b^{(q)}} \right\{\|f_{b^{(l)}} - f_{b^{(q)}}\|_{L_2}^2 \left. + \sum_{j=1}^p\|\partial f_{b^{(l)}}/\partial t_j - \partial f_{b^{(q)}}/\partial t_j\|_{L_2}^2\right\}\\
& \quad \leq \left.\frac{n(p+1)}{2} \max_{b^{(l)}\neq b^{(q)}} \vphantom{\sum_{j=1}^p}\right\{\|f_{b^{(l)}} - f_{b^{(q)}}\|_{L_2}^2 \left.+\sum_{j=1}^p\|\partial f_{b^{(l)}}/\partial t_j - \partial f_{b^{(q)}}/\partial t_j\|_{L_2}^2\right\}\\
& \quad \leq 2n(p+1)\max_{b^{(l)}\in\{b^{(1)},\ldots,b^{(M)}\}} \left\{\|f_{b^{(l)}}\|_{L_2}^2 + \sum_{j=1}^p\| \partial f_{b^{(l)}}/\partial t_j\|_{L_2}^2\right\}\\
& \quad \leq 2c_3n(p+1)N^{-2m+\frac{2}{r}}(\log N)^{r-1}.
\end{aligned}
\end{equation}
Now, (\ref{eqn:fanolemhatthetap}) yields that
\begin{align}
& \quad\inf_{\tilde{f}}\sup_{f_0\in\HH}\P\left\{\|\tilde{f} - f_0\|_{L_2}^2\geq \frac{1}{4}c_2 N^{-2m}\right\}\nonumber\\
& \quad \geq\inf_{\widehat{\Theta}}\P\{\widehat{\Theta}\neq \Theta\}\nonumber\\
& \quad \geq 1-\frac{1}{\log M}\left[\E\mathbbm{1}_{\bt_1^{(0)},\ldots,\bt_n^{(0)};\ldots;\bt_1^{(p)},\ldots,\bt_n^{(p)}}(y_1^{(0)},\ldots,y_n^{(0)},\ldots,y_1^{(p)},\ldots,y_n^{(p)};\Theta)+\log 2\right]\nonumber\\
& \quad \geq 1-\frac{2c_3n(p+1)N^{-2m+\frac{2}{r}}(\log N)^{r-1} + \log 2}{c_0(\log 2)N(\log N)^{r-1}/8}.\nonumber
\end{align}
Taking $N=c_4n^{r/(2mr+r-2)}$ with an appropriate choice of $c_4$, we have
\begin{equation*}
\underset{n\to \infty}{\lim\sup}\inf_{\tilde{f}} \sup_{f_0\in\HH}\P\left\{\|\tilde{f} - f_0\|_{L_2}^2\geq cn^{-\frac{2mr}{(2m+1)r-2}}\right\}>0,
\end{equation*}
where $c$ does not depend on $n$. \change{In addition, $\|\tilde{f} - f_0\|_{L_2}^2\geq \min_{\|f\|_\HH\leq R_n}\|f-f_0\|^2_{L_2}$.
This completes the proof of this theorem.}


\subsection{Proof of Theorem \ref{thm:mainupperrateestf0}}
\label{subsubsec:proofupperrrandom}

\paragraph{Preliminaries.}
We consider a general quadratic penalty $J(\cdot)$ for the proposed method (\ref{scheme1}), where $J(\cdot)$ is 
any squared semi-norm on the RKHS $\HH$. For example, when $\HH_1 = \WW_2^m(\XX)$, it is common to choose $J(\cdot)$ for penalizing only the smooth component of a function. In this case, an explicit form of $J(\cdot)$ is presented in  \cite{wahba1990}. The following analysis holds for replacing $J(\cdot)$ with the squared norm $\|\cdot\|_\HH^2$.

We define a new norm for any $f\in\HH$,
\begin{equation}
\label{def:normrp+1}
\begin{aligned}
\|f\|_R^2 & = \frac{1}{p+1}\left[\frac{1}{\sigma_0^2}\int f^2(\bt)d\Pi^{(0)}(\bt)  + \sum_{j=1}^p \frac{1}{\sigma_j^2}\int\left\{\frac{\partial f(\bt)}{\partial t_j}\right\}^2d\Pi^{(j)}(\bt) \right] + J(f).
\end{aligned}
\end{equation}
Note that $\|\cdot\|_R$ is a norm since it is a quadratic form and is equal to zero if and only if $f=0$. Let $\langle\cdot,\cdot\rangle_R$ be the inner product associated with $\|\cdot\|_R$. Then by Lemma \ref{lemmanormrequiv}, the norm $\|\cdot\|_R$ is equivalent to the norm $\|\cdot\|_\HH$ in RKHS $\HH$. In particular, $\|f\|_R<\infty$ if and only if $\|f\|_\HH<\infty$. 

We introduce another norm $\|\cdot\|_0$ given by
\begin{equation}
\label{def:norm0}
\begin{aligned}
\|f\|_0^2 & = \frac{1}{p+1}\left[\frac{1}{\sigma_0^2}\int f^2(\bt)d\Pi^{(0)}(\bt)+ \sum_{j=1}^p \frac{1}{\sigma_j^2}\int\left\{\frac{\partial f(\bt)}{\partial t_j}\right\}^2d\Pi^{(j)}(\bt) \right].
\end{aligned}
\end{equation}
Let a function space $F_0$ be the direct sum of some set of the orthogonal subspaces in the decomposition of $\otimes_{j=1}^dL_2(\XX)$ as in (\ref{eqn:anovadechi}) and  equipped with the norm $\|\cdot\|_0$. Write $\langle\cdot,\cdot\rangle_0$ as the inner product associated with $\|\cdot\|_0$ in $F_0$.

Finally, we define the following norm. For $f\in\HH$,
\begin{equation}
\label{eqn:fa2v1rho1}
\|f\|_{L_2(a)}^2=\sum_{{\vec{\bnu}}\in \bV} \left(1+\frac{\rho_{\vec{\bnu}}}{\|\phi_{\vec{\bnu}}\|_{L_2}^2}\right)^a f_{\vec{\bnu}}^2\|\phi_{\vec{\bnu}}\|_{L_2}^2, \quad \mbox{ for } 0\leq a\leq 1,
\end{equation}
where $f_{\vec{\bnu}} = \langle f, \phi_{\vec{\bnu}}\rangle_0$. By direct calculations,  when $a=0$ this norm coincides with $\|\cdot\|_{L_2}$ on $F_0$, and when $a=1$ this norm is equivalent to $\|\cdot\|_R$ on $\HH$.

Denote the loss function in (\ref{scheme1}) by $l_n(f)$, that is,
\begin{equation*}
l_n(f) = \frac{1}{n(p+1)}\left[\frac{1}{\sigma_0^2}\sum_{i=1}^n\{f(\bt_i^{(0)})-y_i^{(0)}\}^2+\sum_{j=1}^p\frac{1}{\sigma_j^2}\sum_{i=1}^n\left\{\frac{\partial f(\bt_i^{(j)})}{\partial t_j}-y_i^{(j)}\right\}^2\right],
\end{equation*}
and write $l_{n\lambda}(f) = l_n(f)+\lambda J(f)$. Then the estimator $\widehat{f}_{n} = \arg\min_{f\in\HH}{l_{n\lambda}(f)}$.
Denote the expected loss by $l_\infty(f)  = \E l_n(f)$
and write $l_{\infty\lambda}(f) = l_\infty(f)+\lambda J(f)$. Since $l_{\infty\lambda}(f)$ a positive quadratic form in $f\in\HH$, it has a unique minimizer in $\HH$ given by
\begin{equation*}
\bar{f}_{\infty\lambda} = \underset{f\in\HH}{\arg\min } l_{\infty\lambda}(f).
\end{equation*}
\change{Let $f^\dagger = \arg\min_{J(f)\leq R^2_n}\|f-f_0\|_{L_2}^2$.
Thus, we decompose
\begin{equation}
\label{eqn:errordecomp}
\widehat{f}_{n} - f_0 = (\widehat{f}_{n} - \bar{f}_{\infty\lambda}) + (\bar{f}_{\infty\lambda} - f^\dagger) + (f^\dagger - f_0),
\end{equation}
where $(\widehat{f}_{n} - \bar{f}_{\infty\lambda})$ is referred to \emph{stochastic error}, $(\bar{f}_{\infty\lambda} - f^\dagger)$ is referred to \emph{deterministic error}, and $(f^\dagger - f_0)$ is referred to \emph{approximation error}; see, e.g.,  \cite{wellner1996weak}.}
We omit the subscripts of $\bar{f}_{\infty\lambda}$ and $\widehat{f}_{n}$ hereafter if no confusion occurs.

\paragraph{Outline of the proof.}
Since the distributions $\Pi^{(0)}$ and $\Pi^{(j)}$s are known,  it suffices to consider the uniform distributions by the inverse transform sampling in Lemma \ref{lem:lineartransformuniform}. \change{Moreover, since $f_0$ is a functional ANOVA  model with component function spaces supported in a compact domain $\XX^d\equiv[0,1]^d$, one can smoothly extend $f_0$ to a larger compactly supported domain $[0,1+\delta]^d$ and achieve periodicity on the new boundary. This is proven in Lemma \ref{lem:periodextend}, which also shows that the eigenvalue decay rate for the RKHS associated with the extended periodic function remains the same as that for the RKHS associated with the original function.}
\change{Moreover, the probability of selecting $\bt$ within the range $\{[0,1+\delta]^d\setminus\XX^d\}$ is $O(\delta)$, which is negligible for a sufficiently small $\delta$.
Lemma \ref{lem:esterrperiodextend} shows that the estimation error of $f_0$  can be upper bounded by that of the extended periodic function. Hence,  the upper bound of the estimation for the periodic function also applies to the original function $f_0$. 
Therefore, we consider  $f_0$ has a periodic boundary in $\XX^d$ in the proof.
 A similar technique has been used in literature; e.g., \citet{hall2010nonparametric}.}

Write the trigonometrical basis on $L_2(\XX)$ as $\psi_1(t)=1$, $\psi_{2\nu}(t) = \sqrt{2}\cos 2\pi\nu t$ and $\psi_{2\nu+1}(t) = \sqrt{2}\sin 2\pi\nu t$ for $\nu\geq 1$. Let
\begin{equation}
\label{eqn:phivecbnudef}
\phi_{\vec{\bnu}}(t_1,\ldots,t_d) =\frac{\psi_{\nu_1}(t_1)\cdots\psi_{\nu_d}(t_d)}{\|\psi_{\nu_1}(t_1)\cdots\psi_{\nu_d}(t_d)\|_0}.
\end{equation}
Since $f_0$ has a periodic boundary in $\XX^d$ and $\pi^{(j)}\equiv1$, $\{\phi_{\vec{\bnu}}(\bt):{\vec{\bnu}}\in \bV\}$, where $\bV$ in (\ref{def:Vbnu}) forms an orthogonal basis for $\HH$ in $\langle\cdot,\cdot\rangle_R$; an orthogonal system for $L_2(\XX^d)$; and an orthonormal basis for $F_0$ in $\langle\cdot,\cdot\rangle_0$, that is $\langle \phi_{\vec{\bnu}}(\bt),\phi_{\vec{\bmu}}(\bt)\rangle_0 = \delta_{\vec{\bnu}\vec{\bmu}}$, where $\delta_{\vec{\bnu}\vec{\bmu}}$ is Kronecker's delta; \change{see, e.g., Chapter 2 in \cite{wahba1990}.}
 \change{The concept of simultaneous orthogonality of a basis in multiple inner product spaces has been explored in other RKHS settings; see, e.g., Section 3 in \citet{Yuan2010}.}  Hence, any $f\in\HH$ has the decomposition
\begin{equation}
\label{eqn:fsumnuthetanuphinvb}
f(t_1,\ldots,t_d) = \sum_{\vec{\bnu}\in \bV}f_{\vec{\bnu}}\phi_{\vec{\bnu}}(t_1,\ldots,t_d),\quad  \mbox{ where } f_{\vec{\bnu}} = \langle f(\bt),\phi_{\vec{\bnu}}(\bt)\rangle_0.
\end{equation}
We denote a positive scalar series $\{\rho_{\vec{\bnu}}\}_{\bnu\in \bV}$ such that $\langle \phi_{\vec{\bnu}},\phi_{\vec{\bmu}}\rangle_R = (1+\rho_{\vec{\bnu}})\delta_{\vec{\bnu}\vec{\bmu}}$. Then,
\begin{equation}
\label{eqn:JgggR0}
J(f) = \langle f,f\rangle_R - \langle f,f\rangle_0 = \sum_{\vec{\bnu}\in \bV} \rho_{\vec{\bnu}}f_{\vec{\bnu}}^2.
\end{equation}

First, we analyze the deterministic error $(\bar{f}-f^\dagger)$. By (\ref{eqn:fsumnuthetanuphinvb}), write $f^\dagger(\bt) = \sum_{\vec{\bnu}\in \bV}f_{\vec{\bnu}}^\dagger\phi_{\vec{\bnu}}(\bt)$ and $\bar{f}(\bt) = \sum_{\vec{\bnu}\in \bV}\bar{f}_{\vec{\bnu}}\phi_{\vec{\bnu}}(\bt)$. {Note the bias satisfies } $\E[\epsilon_i^{(j)}] = o(n^{-1/2})$, we have 
\begin{equation*}
l_\infty(f) = \sum_{\vec{\bnu}\in \bV} (f_{\vec{\bnu}} - f_{\vec{\bnu}}^\dagger)^2 {+ o(n^{-1/2})\sqrt{ \sum_{\vec{\bnu}\in \bV} (f_{\vec{\bnu}} - f_{\vec{\bnu}}^\dagger)^2}} + 1,
\end{equation*} and
\begin{equation}
\label{eqn:barthetavebnubias}
\bar{f}_{\vec{\bnu}} = \frac{f_{\vec{\bnu}}^\dagger{(1+\kappa_{\vec{\bnu}})}}{1{+\kappa_{\vec{\bnu}}}+\lambda\rho_{\vec{\bnu}}},\quad {\text{where } \kappa_{\vec{\bnu}}=o(1)}, \ \forall \vec{\bnu}\in \bV.
\end{equation}
An upper bound of the deterministic error will be given in Lemma \ref{lem:barff0l2ar}.

Second, we analyze the stochastic error $(\widehat{f}_{n} - \bar{f})$. 
The existence of the following Fr\'echet derivatives  is guaranteed by Lemma \ref{lem:DlNfgDl2}:
\begin{equation}
\label{eqn:frechetdlNfg}
\begin{aligned}
Dl_n(f)g & =  \frac{2}{n(p+1)}\left[\frac{1}{\sigma_0^2}\sum_{i=1}^n\{f(\bt_i^{(0)}) - y_i^{(0)}\}g(\bt_i^{(0)}) \right.\\
&\quad\quad\quad\quad\quad\quad\ \left.+ \sum_{j=1}^p\frac{1}{\sigma_j^2}\sum_{i=1}^n\left\{\frac{\partial f(\bt_i^{(j)})}{\partial t_j} - y_i^{(j)}\right\}\frac{\partial g(\bt_i^{(j)})}{\partial t_j}\right],
\end{aligned}
\end{equation}

\begin{equation}
\label{eqn:dlinftyfg2p1}
\begin{aligned}
&Dl_\infty(f)g  = \frac{2}{p+1}\left[\frac{1}{\sigma_0^2}\int\left\{f(\bt) - f_0(\bt){+o(n^{-1/2})}\right\} g(\bt)d\Pi^{(0)}(\bt)\right.\\
&\quad\quad\quad \quad \left. +\sum_{j=1}^p\frac{1}{\sigma_j^2}\int\left\{\frac{\partial f(\bt)}{\partial t_j} - \frac{\partial f_0(\bt)}{\partial t_j}{+o(n^{-1/2})}\right\}\frac{\partial g(\bt)}{\partial t_j}d\Pi^{(j)}(\bt)\right],
\end{aligned}
\end{equation}

\begin{equation}
\label{eqn:d2lnfgh}
\begin{aligned}
D^2l_n(f)gh & =  \frac{2}{n(p+1)}\left[\frac{1}{\sigma_0^2}\sum_{i=1}^ng(\bt_i^{(0)})h(\bt_i^{(0)})\right.\\
& \quad\quad\quad\quad\quad\quad\quad\quad\quad\quad\quad\left.+\sum_{j=1}^p\frac{1}{\sigma_j^2}\sum_{i=1}^n\frac{\partial g(\bt_i^{(j)})}{\partial t_j}\frac{\partial h(\bt^{(j)}_i)}{\partial t_j}\right],
\end{aligned}
\end{equation}

\begin{equation}
\label{eqn:d2linftyfgh}
\begin{aligned}
D^2l_\infty(f)gh & = \frac{2}{p+1}\left[\frac{1}{\sigma_0^2}\int g(\bt) h(\bt)d\Pi^{(0)}(\bt)\right.\\
& \quad\quad\quad\quad\quad\quad\left. +\sum_{j=1}^p\frac{1}{\sigma_j^2}\int \frac{\partial g(\bt)}{\partial t_j}\frac{\partial h(\bt)}{\partial t_j}d\Pi^{(j)}(\bt)\right] =2\langle g,h\rangle_0,
\end{aligned}
\end{equation}
where $Dl_n(f)$, $Dl_\infty(f)$, $D^2l_n(f)g$, and $D^2l_\infty(f)g$ are bounded linear operators on $\HH$. By Riesz representation theorem, with a slight abuse of notation, write
\begin{align}
Dl_n(f)g & = \langle Dl_n(f),g\rangle_R, \quad Dl_\infty(f)g = \langle Dl_\infty(f),g\rangle_R,\nonumber\\
D^2l_n(f)gh & = \langle D^2l_n(f)g, h\rangle_R, \quad D^2l_\infty(f)gh = \langle D^2l_\infty(f)g, h\rangle_R.\nonumber
\end{align}
From \cite{oden2012introduction}, there exists a bounded linear operator $U: F_0\mapsto \HH$ such that $U\phi_{\vec{\bnu}} = (1+\rho_{\vec{\bnu}})^{-1}\phi_{\vec{\bnu}}$ and $\langle f, Ug\rangle_R = \langle f,g\rangle_0$ for any $f\in \HH$ and $g\in F_0$, and the restriction of $U$ to $\HH$ is self-adjoint and positive definite.
By (\ref{eqn:d2linftyfgh}), we further derive
\begin{align}
D^2l_{\infty\lambda}(f)\phi_{\vec{\bnu}}(\bt)  = 2(U+\lambda(I-U))\phi_{\vec{\bnu}}(\bt) = 2(1+\rho_{\vec{\bnu}})^{-1}(1+\lambda\rho_{\vec{\bnu}})\phi_{\vec{\bnu}}(\bt).\nonumber
\end{align}
Define that $G_\lambda\phi_{\vec{\bnu}}=\frac{1}{2}D^2l_{\infty\lambda}(\bar{f})\phi_{\vec{\bnu}}$. By the Lax-Milgram theorem, $G_\lambda: \HH\mapsto\HH$ has a bounded inverse $G_\lambda^{-1}$ on $\HH$, and
\begin{equation}
\label{def:ginv}
G_\lambda^{-1}\phi_{\vec{\bnu}} = (1+\rho_{\vec{\bnu}})(1+\lambda\rho_{\vec{\bnu}})^{-1}\phi_{\vec{\bnu}}.
\end{equation}
Define
\begin{equation*}
\tilde{f}^* = \bar{f} - \frac{1}{2}G_\lambda^{-1}Dl_{n\lambda}(\bar{f}).
\end{equation*}
Then the stochastic error can be decomposed as
\begin{equation*}
\widehat{f}_{n} - \bar{f}= (\tilde{f}^* - \bar{f}) + (\widehat{f}_{n}- \tilde{f}^*).
\end{equation*}
The two terms on the right-hand side will be studied separately, and their upper bounds will be given in Lemma \ref{lem:tilfbarfl2a} and Lemma \ref{lem:boundhattildef}, respectively.

\paragraph{Main proof.} Now, we give the details by following the above outline.
First, we present an upper bound of the deterministic error $(\bar{f} - f^\dagger)$ in \eqref{eqn:errordecomp}.
\begin{lemma}
\label{lem:barff0l2ar}
For any $0\leq a\leq 1$, the deterministic error in \eqref{eqn:errordecomp} satisfies
\change{\begin{equation*}
\|\bar{f} - f^\dagger\|_{L_2(a)}^2 =
\begin{cases}
O\left\{\lambda^{1-a}R^2_n\right\} \quad & \mbox{ when } 0\leq p< d,\\
O\{\lambda^{\frac{(1-a)mr}{mr-1}}R^2_n\} \quad & \mbox{ when } p=d.
\end{cases}
\end{equation*}}
\end{lemma}

\begin{proof}
We first introduce some notations. For two positive sequences $a_n$ and $b_n$, we write $a_n\lesssim b_n$ (or $a_n\gtrsim b_n$) means that there exists a constant $c>0$ (or $c'>0$) such that $a_n\leq cb_n$ (or $a_n\geq c'b_n$) for all $n$.
We write $a_{n}\asymp b_{n}$ if $a_{n}/b_{n}$ is bounded away from both zero and infinity as $n\to\infty$.

For any $0\leq a\leq1$, by (\ref{eqn:JgggR0}) and (\ref{eqn:barthetavebnubias}), we have
\begin{equation}
\label{eqn:lambda2jf0supnu}
\begin{aligned}
 \|\bar{f} - f^\dagger\|_{L_2(a)}^2 
&  = \sum_{{\vec{\bnu}}\in \bV} \left(1+\frac{\rho_{\vec{\bnu}}}{\|\phi_{\vec{\bnu}}\|_{L_2}^2}\right)^a\left(\frac{\lambda\rho_{\vec{\bnu}}}{1{+\kappa_{\vec{\bnu}}}+\lambda\rho_{\vec{\bnu}}}\right)^2(f_{\vec{\bnu}}^\dagger)^2\|\phi_{\vec{\bnu}}\|_{L_2}^2 \\
& \lesssim \lambda^2\sup_{{\vec{\bnu}}\in \bV}\frac{(1+\rho_{\vec{\bnu}}/\|\phi_{\vec{\bnu}}\|_{L_2}^2)^a\rho_{\vec{\bnu}}\|\phi_{\vec{\bnu}}\|_{L_2}^2}{(1+\lambda\rho_{\vec{\bnu}})^2}\sum_{{\vec{\bnu}}\in \bV}\rho_{\vec{\bnu}}(f_{\vec{\bnu}}^\dagger)^2 \\
&\lesssim \lambda^2\change{R_n^2}\sup_{{\vec{\bnu}}\in \bV}\frac{(\prod_{k=1}^d\nu_k^{2m})^{1+a}}{(1+\sum_{j=1}^p\nu_j^2+\lambda\prod_{k=1}^d\nu_k^{2m})^2}.
\end{aligned}
\end{equation}
Write
\begin{equation*}
B_{\lambda}({\vec{\bnu}}) = \frac{(\prod_{k=1}^d\nu_k^{2m})^{1+a}}{(1+\sum_{j=1}^p\nu_j^2+\lambda\prod_{k=1}^d\nu_k^{2m})^2},\quad {\vec{\bnu}}\in \bV.
\end{equation*}
We discuss $B_{\lambda}({\vec{\bnu}})$ for $0\leq p\leq d-1$ and $p=d$ separately.

For $0\leq p\leq d-1$, since ${\vec{\bnu}}\in \bV$, there are at most $r$ of $\nu_1,\ldots,\nu_d$ not equal to 1. Suppose for any $x=\prod_{k=1}^d\nu_k^{-2m}>0$ fixed. Then $B_\lambda({\vec{\bnu}})$ is maximized by letting $\sum_{j=1}^p\nu_j^2$ be as small as possible, which implies $\nu_1=\nu_2=\cdots=\nu_p=1$. Then,
\begin{equation}
\label{eqn:supnublambdanu-a}
\begin{aligned}
\underset{{\vec{\bnu}}\in \bV}{\sup}B_{\lambda}({\vec{\bnu}}) & \asymp \sup_{(\nu_{p+1},\ldots,\nu_{(p+r)\wedge d})^\top\in\N^{r\wedge (d-p)}} \frac{\prod_{k=p+1}^{(p+r)\wedge d}\nu_k^{2m(1+a)}}{(1+\lambda\prod_{k=p+1}^{(p+r)\wedge d}\nu_k^{2m})^2}\\
& \asymp \sup_{x>0}\frac{x^{-(1+a)}}{(1+\lambda x^{-1})^2} \asymp \lambda^{-(a+1)},
\end{aligned}
\end{equation}
where the last step is achieved when $x \asymp \lambda$.

For $p=d$, since ${\vec{\bnu}}\in \bV$ and by the symmetry of coordinates $v_1,\ldots,v_d$, assume
that all indices except $v_1,\ldots,v_r$ being 1. Letting $z=\prod_{j=1}^r\nu_j^{-2m}>0$, we have
\begin{align}
\underset{{\vec{\bnu}}\in \bV}{\sup}B_{\lambda}({\vec{\bnu}})  \asymp \sup_{z>0}\frac{z^{-(1+a)}}{(z^{-1/mr}+\lambda z^{-1})^2} \asymp \lambda^{\frac{2-(1+a)mr}{mr-1}},\label{eqn:lamdba2a1amr}
\end{align}
where the last step is achieved when $z\asymp \lambda^{mr/(mr-1)}$. Combining (\ref{eqn:lambda2jf0supnu}), (\ref{eqn:supnublambdanu-a}) and (\ref{eqn:lamdba2a1amr}) we complete the proof.
\end{proof}


Before we establish an upper bound for the stochastic error, we present the Fr\'echet derivative of the operator that will be used in the proof. 
Let $X$ and $Y$ be the normed linear spaces.  The Fr\'echet derivative of an operator $F: X\mapsto Y$ is a bounded linear operator $DF(a):X\mapsto Y$ with 
\begin{equation*}
\lim_{h\to 0,h\in X}\frac{\|F(a+h)-F(a)-DF(a)h\|_Y}{\|h\|_X} = 0.
\end{equation*}
For example, if $F(a+h) - F(a) = Lh+R(a,h)$ with a linear operator $L$ and 
\begin{equation*}
\frac{\|R(a,h)\|_Y}{\|h\|_X}\to0,\quad \text{as } h\to 0,
\end{equation*} 
by definition then $L=DF(a)$ is the Fr\'echet derivative of $F(\cdot)$. The reader is referred to   \cite{gelfand2000calculus}  for a thorough investigation of the Fr\'echet derivative. 
We give the Fr\'echet derivative of the operator in our setting.

\begin{lemma}
\label{lem:DlNfgDl2}
Denote the loss function in (\ref{scheme1}) by $l_n(f)$.
With the norm $\|\cdot\|_R$ in (\ref{def:normrp+1}), the first-order Fr\'echet derivative of the functional $l_n(\cdot)$ for any $f,g\in\HH$ is
\begin{align}
Dl_n(f)g & =  \frac{2}{n(p+1)}\left[\frac{1}{\sigma_0^2}\sum_{i=1}^n\{f(\bt_i^{(0)}) - y_i^{(0)}\}g(\bt_i^{(0)}) \right.\nonumber\\
&\quad\quad\quad\quad\quad\quad\quad\quad \left.+ \sum_{j=1}^p\frac{1}{\sigma_j^2}\sum_{i=1}^n\left\{\frac{\partial f(\bt_i^{(j)})}{\partial t_j} - y_i^{(j)}\right\}\frac{\partial g(\bt_i^{(j)})}{\partial t_j}\right].\nonumber
\end{align}
The second-order Fr\'echet derivative of $l_n(\cdot)$ for any $f,g,h\in\HH$ is
\begin{equation*}
\begin{aligned}
D^2l_n(f)gh  & =  \frac{2}{n(p+1)}\left[\frac{1}{\sigma_0^2}\sum_{i=1}^ng(\bt_i^{(0)})h(\bt_i^{(0)})\right.\\
& \quad\quad\quad\quad\quad\quad\quad\quad\left.+\sum_{j=1}^p\frac{1}{\sigma_j^2}\sum_{i=1}^n\frac{\partial g(\bt_i^{(j)})}{\partial t_j}\frac{\partial h(\bt^{(j)}_i)}{\partial t_j}\right].
\end{aligned}
\end{equation*}
\end{lemma}
\begin{proof}
By direct calculations, we have
\begin{align}
& l_{n}(f+g) - l_n(f) = \frac{2}{n(p+1)}\left[\frac{1}{\sigma_0^2}\sum_{i=1}^n\{f(\bt_i^{(0)}) - y_i^{(0)}\}g(\bt_i^{(0)}) \right.\nonumber\\
&\quad\quad\quad\quad\quad\quad\quad \left.+ \sum_{j=1}^p\frac{1}{\sigma_j^2}\sum_{i=1}^n\left\{\frac{\partial f(\bt_i^{(j)})}{\partial t_j} - y_i^{(j)}\right\}\frac{\partial g(\bt_i^{(j)})}{\partial t_j}\right] + \RR_n(f,g),\nonumber
\end{align}
where
\begin{equation*}
\begin{aligned}
\RR_n(f,g) & = \frac{1}{n(p+1)}\left[\frac{1}{\sigma_0^2}\sum_{i=1}^n g^2(\bt_i^{(0)}) + \sum_{j=1}^p\frac{1}{\sigma_j^2}\sum_{i=1}^n \left\{\frac{\partial g(\bt_i^{(j)})}{\partial t_j}\right\}^2\right] \\
& = \|g\|_0^2 + O(n^{-1/2}),
\end{aligned}
\end{equation*}
and the $\|\cdot\|_0$ norm is given in (\ref{def:norm0}).
Note that $|\RR_n(f,g)|/\|g\|_R \to 0$ as $\|g\|_R\to 0$ and $n^{1/2}\|g\|_R\to \infty$. This proves the first part of the lemma.
For the second-order Fr\'echet derivative, note that
\begin{align}
& Dl_n(f+h)g- Dl_n(f)g \nonumber\\
& \quad\quad =\frac{2}{n(p+1)}\left[\frac{1}{\sigma_0^2}\sum_{i=1}^ng(\bt_i^{(0)})h(\bt_i^{(0)})+\sum_{j=1}^p\frac{1}{\sigma_j^2}\sum_{i=1}^n\frac{\partial g(\bt_i^{(j)})}{\partial t_j}\frac{\partial h(\bt^{(j)}_i)}{\partial t_j}\right],\nonumber
\end{align}
which is linear in $h$. By definition, the $D^2l_n(f)gh$ in the lemma is the valid second-order Fr\'echet derivative of $l_n(\cdot)$.
\end{proof} 
By following a similar derivation for Lemma \ref{lem:DlNfgDl2}, it is easy to obtain the first and the second-order Fr\'echet derivatives of the functional $l_\infty(\cdot)$ in (\ref{eqn:dlinftyfg2p1}) and (\ref{eqn:d2linftyfgh}), respectively.

We now establish an upper bound for the term $(\tilde{f}^*-\bar{f})$, which is a part of the stochastic error.

\begin{lemma}
\label{lem:tilfbarfl2a}
When $0\leq p<d$, we have for any $0\leq a< 1 - 1/2m$,
\begin{equation*}
\|\tilde{f}^* - \bar{f}\|_{L_2(a)}^2 =
O_\P\left\{n^{-1}\lambda^{-(a+1/2m)}[\log(1/\lambda)]^{(d-p)\wedge r-1}\right\}.
\end{equation*}
When $p=d$, we have for any $0\leq a\leq 1$,
\begin{align}
& \|\tilde{f}^* - \bar{f}\|_{L_2(a)}^2 \nonumber\\
= &
\begin{cases}
O_\P\left\{n^{-1}\change{R_n^2}\lambda^{\frac{mr}{1-mr}\left(a+\frac{r-2}{2mr}\right)}\right\},   \mbox{ if } r\geq 3;\\
O_\P\left\{n^{-1}\change{R_n^2}\log(1/\lambda)\right\},  \mbox{ if } r=2, a=0;\quad O_\P\left\{n^{-1}\change{R_n^2}\right\}, \mbox{ if } r=2, 0<a\leq 1;\\
O_\P\left\{n^{-1}\change{R_n^2}\right\},  \mbox{ if } r=1,a<\frac{1}{2m};  \quad O_\P\left\{n^{-1}\log(1/\lambda)\change{R_n^2}\right\},  \mbox{ if } r=1,a=\frac{1}{2m}; \\
O_\P\left\{n^{-1}\lambda^{\frac{1-2ma}{2m-2}}\change{R_n^2}\right\},  \mbox{ if } r=1, a>\frac{1}{2m}.
\end{cases}\nonumber
\end{align}
\end{lemma}
\begin{proof}
Notice that $Dl_{n,\lambda}(\bar{f}) = Dl_{n,\lambda}(\bar{f}) - Dl_{\infty,\lambda}(\bar{f}) = Dl_n(\bar{f}) - Dl_\infty(\bar{f})$. Hence, for any $g\in\HH$,
\begin{equation}
\label{eqn:newintermgti}
\begin{aligned}
& \E\left[\frac{1}{2}Dl_{n,\lambda}(\bar{f})g\right]^2 = \E\left[\frac{1}{2}Dl_n(\bar{f})g - \frac{1}{2}Dl_\infty(\bar{f})g\right]^2\\
&  \lesssim \frac{1}{n(p+1)^2}\sum_{j=0}^p\text{Var}\left[\frac{1}{\sigma_{j}^2}\left\{\frac{\partial\bar{f}(\bt^{(j)})}{\partial t_j} - Y^{(j)}\right\}\frac{\partial g(\bt^{(j)})}{\partial t_j}\right]\\
&  {+\sum_{j=0}^p\frac{\sigma_j^{-4}}{n^2(p+1)^2}\sum_{i\neq i'}}\text{Cov}\left[\left(\frac{\partial\bar{f}(\bt_i^{(j)})}{\partial t_j} - y_i^{(j)}\right)\frac{\partial g(\bt_i^{(j)})}{\partial t_j},\left(\frac{\partial\bar{f}(\bt_{i'}^{(j)})}{\partial t_j} - y_{i'}^{(j)}\right)\frac{\partial g(\bt_{i'}^{(j)})}{\partial t_j} \right]\\
&   {+\sum_{j\neq k}\frac{\sigma_j^{-2}\sigma_k^{-2}}{n^2(p+1)^2}\sum_{i, i'=1}^n}\text{Cov}\left[\left(\frac{\partial\bar{f}(\bt_i^{(j)})}{\partial t_j} - y_i^{(j)}\right)\frac{\partial g(\bt_i^{(j)})}{\partial t_j},\left(\frac{\partial\bar{f}(\bt_{i'}^{(k)})}{\partial t_k} - y_{i'}^{(j)}\right)\frac{\partial g(\bt_{i'}^{(k)})}{\partial t_k} \right],
\end{aligned}
\end{equation}
where the second step is due to $\sum_{i\neq i'}\text{Cov}[\epsilon_i^{(j)},\epsilon_{i'}^{(k)}] = \sum_{i\neq i'}o(|i-i'|^{{-\Upsilon}}) = o(n)$.
Note that \eqref{eqn:newintermgti} can be further bounded up to some constant  by,
\begin{equation}
\label{eqn:newintermgti2}
\begin{aligned}
& \frac{1}{n(p+1)}\left[\frac{1}{\sigma_0^4}\E\left\{\bar{f}(\bt^{(0)})-f_0(\bt^{(0)})\right\}^2\{g(\bt^{(0)})\}^2+\frac{1}{\sigma_{0}^2}\E\{g(\bt^{(0)})\}^2\right.\\
& \left.+\sum_{j=1}^p\frac{1}{\sigma_j^4}\E\left\{\frac{\partial \bar{f}(\bt^{(j)})}{\partial t_j}-\frac{\partial f_0(\bt^{(j)})}{\partial t_j}\right\}^2\left\{\frac{\partial g(\bt^{(j)})}{\partial t_j}\right\}^2+\sum_{j=1}^p\frac{1}{\sigma_{j}^2}\E\left\{\frac{\partial g(\bt^{(j)})}{\partial t_j}\right\}^2\right]\\
&  +  o(n^{-1})\frac{1}{(p+1)^2}\sum_{j,k=0}^p\E\left[\frac{\partial g(\bt^{(j)})}{\partial t_j}\right]\E\left[\frac{\partial g(\bt^{(k)})}{\partial t_k}\right],
\end{aligned}
\end{equation}
By  Lemma \ref{lemmanormrequiv}, Lemma \ref{lem:bounddejft}, and Cauchy-Schwarz inequality, we have that (\ref{eqn:newintermgti2}) is bounded up to some constant by
\begin{equation}
\label{eqn:dlnlambdang0}
\begin{aligned}
& \frac{1}{n(p+1)}\left[ \frac{1}{\sigma_0^4}c_{K}^{2d}\|\bar{f}-f_0\|_R^2\E\left\{g(\bt^{(0)})\right\}^2+\frac{1}{\sigma_{0}^2}\E\left\{g(\bt^{(0)})\right\}^2\right.\\
& \left.+\sum_{j=1}^p\frac{1}{\sigma_j^4}c_{K}^{2d}\|\bar{f}-f_0\|_R^2\E\left\{\frac{\partial g(\bt^{(j)})}{\partial t_j}\right\}^2+\sum_{j=0}^p\frac{1}{\sigma_{j}^2}\E\left\{\frac{\partial g(\bt^{(j)})}{\partial t_j}\right\}^2\right] \\
&  \lesssim n^{-1}\change{R_n^2}\|g\|_0^2,
\end{aligned}
\end{equation}
where the last step above is by Lemma \ref{lem:barff0l2ar} and the definition of the norm $\|\cdot\|_0$.
From the definition of $G_\lambda^{-1}$ in (\ref{def:ginv}), we have that $\forall g\in\HH$,
\begin{equation*}
\left\|G_\lambda^{-1}g\right\|_{L_2(a)}^2 = \sum_{{\vec{\bnu}}\in \bV}\left(1+\frac{\rho_{\vec{\bnu}}}{\|\phi_{\vec{\bnu}}\|_{L_2}^2}\right)^a\left(1+\lambda\rho_{\vec{\bnu}}\right)^{-2}\|\phi_{\vec{\bnu}}\|^2_{L_2}\langle g,\phi_{\vec{\bnu}}\rangle_R^2.
\end{equation*}
Then by the definition of $\tilde{f}^*$,
\begin{align}
& \E\|\tilde{f}^* - \bar{f}\|_{L_2(a)}^2  = \E \left\|\frac{1}{2}G_\lambda^{-1}Dl_{n\lambda}(\bar{f})\right\|_{L_2(a)}^2\nonumber\\
& = \frac{1}{4}\E\left[\sum_{{\vec{\bnu}}\in \bV}\left(1+\frac{\rho_{\vec{\bnu}}}{\|\phi_{\vec{\bnu}}\|_{L_2}^2}\right)^a(1+\lambda \rho_{\vec{\bnu}})^{-2}\|\phi_{\vec{\bnu}}\|^2_{L_2}\langle Dl_{n\lambda}(\bar{f}), \phi_{\vec{\bnu}}\rangle_R^2\right]\nonumber\\
& \leq \sum_{{\vec{\bnu}}\in \bV}\left(1+\frac{\rho_{\vec{\bnu}}}{\|\phi_{\vec{\bnu}}\|_{L_2}^2}\right)^a(1+\lambda\rho_{\vec{\bnu}})^{-2}\|\phi_{\vec{\bnu}}\|^2_{L_2}\E\left[\frac{1}{2}Dl_{n\lambda}(\bar{f})\phi_{\vec{\bnu}}\right]^2\nonumber\\
& \lesssim n^{-1}\change{R_n^2}\sum_{{\vec{\bnu}}\in \bV}\left(1+\frac{\rho_{\vec{\bnu}}}{\|\phi_{\vec{\bnu}}\|_{L_2}^2}\right)^a\left(1+\lambda\rho_{\vec{\bnu}}\right)^{-2}\|\phi_{\vec{\bnu}}\|^2_{L_2}\|\phi_{\vec{\bnu}}\|_{0}^2\nonumber\\
& \asymp n^{-1}\change{R_n^2}N_a(\lambda), \nonumber
\end{align}
where the fourth step is by (\ref{eqn:dlnlambdang0}) and the last step is because of $\|\phi_{\vec{\bnu}}\|_0=1$, $\|\phi_{\vec{\bnu}}\|^2_{L_2} \asymp (1+\sum_{j=1}^p\nu_j^2)^{-1}$, $\rho_{\vec{\bnu}}\asymp (1+\sum_{j=1}^p\nu_j^2)^{-1}\prod_{k=1}^d\nu_k^{2m}$, and $N_a(\lambda)$ is defined in Lemma \ref{lemma:varthetamultivariatetensorrank}. Hence, by Lemma \ref{lemma:varthetamultivariatetensorrank}, we complete the proof.
\end{proof}

We now give an upper bound of $(\widehat{f}_{n} - \tilde{f}^*)$, which is another part of the stochastic error.
Since $l_{n\lambda}(f)$ is a quadratic form of $f$, the Taylor expansion of $Dl_{n\lambda}(\widehat{f}_{n}) = 0$ at $\bar{f}$ gives
\begin{equation*}
Dl_{n\lambda}(\bar{f}) + D^2l_{n\lambda}(\bar{f})(\widehat{f}_{n}-\bar{f}) = 0,
\end{equation*}
and by the definition of $\tilde{f}^*$ and $G_\lambda$, we have
\begin{equation*}
Dl_{n\lambda}(\bar{f}) + D^2l_{\infty\lambda}(\bar{f})(\tilde{f}^* - \bar{f}) = 0.
\end{equation*}
Thus, $G_\lambda(\widehat{f}_{n} - \tilde{f}^*) = \frac{1}{2}D^2l_\infty(\bar{f})(\widehat{f}_{n} - \bar{f}) - \frac{1}{2}D^2l_{n}(\bar{f})(\widehat{f}_{n}-\bar{f})$, and
\begin{equation}
\label{eqn:hatfnlambdatildef}
\widehat{f}_{n} - \tilde{f}^* = G_\lambda^{-1}\left[\frac{1}{2}D^2l_\infty(\bar{f})(\widehat{f}_{n}- \bar{f}) - \frac{1}{2}D^2l_{n}(\bar{f})(\widehat{f}_{n} - \bar{f})\right].
\end{equation}

\begin{lemma}
\label{lem:boundhattildef}
If $n^{-1}\lambda^{-(2a+3/2m)}[\log(1/\lambda)]^{r-1}\to 0$ and $1/2m< a<(2m-3)/4m$, we have for any $0\leq c\leq a+1/m$,
\begin{equation*}
\|\widehat{f}_{n} - \tilde{f}^*\|_{L_2(c)}^2 = o_\P\left\{\|\tilde{f}^* - \bar{f}\|_{L_2(c)}^2\right\}.
\end{equation*}
\end{lemma}
\begin{proof}
A  sufficient condition for this lemma is that for any $1/(2m)< a<(2m-3)/(4m)$ and $0\leq c\leq a+1/m$,
\begin{equation}
 \label{eqn:anoformhatftildefc}
\begin{aligned}
& \|\widehat{f}_{n} - \tilde{f}^*\|_{L_2(c)}^2\\
&  =
\begin{cases}
O_\P\left\{n^{-1}\lambda^{-(c+a+1/2m)}[\log(1/\lambda)]^{r\wedge (d-p)-1}\right\}\\
\quad\quad\quad\quad\quad\quad\quad\quad\quad\quad\quad\cdot\|\widehat{f}_{n} - \bar{f}\|_{L_2(a+1/m)}^2, &\mbox{if } 0\leq p<d,\\
O_\P\left\{n^{-1}\lambda^{\frac{mr}{1-mr}\left(a+c+\frac{r-2}{2mr}\right)}\right\}\|\widehat{f}_{n} - \bar{f}\|_{L_2(a+1/m)}^2,  & \mbox{if } p=d, r\geq 3,\\
O_\P\left\{n^{-1}\right\}\|\widehat{f}_{n} - \bar{f}\|_{L_2(a+1/m)}, &\mbox{if } p=d, r=2,\\
O_\P\left\{n^{-1}\lambda^{\frac{1-2m(a+c)}{2m-2}}\right\}\|\widehat{f}_{n} - \bar{f}\|_{L_2(a+1/m)},   &\mbox{if } p=d, r=1.
 \end{cases}
\end{aligned}
\end{equation}
This is because once (\ref{eqn:anoformhatftildefc}) established, by letting $c=a+1/m$ and under the assumption that $n^{-1}\lambda^{-(2a+3/2m)}[\log(1/\lambda)]^{r-1}\to 0$, we have
\begin{equation*}
\|\widehat{f}_{n} - \tilde{f}^*\|_{L_2(a+1/m)}^2 = o_\P(1)\|\widehat{f}_{n} - \bar{f}\|_{L_2(a+1/m)}^2.
\end{equation*}
By the triangle inequality, we have $\|\tilde{f}^* - \bar{f}\|_{L_2(a+1/m)}  \geq \|\widehat{f}_{n} - \bar{f}\|_{L_2(a+1/m)} - \|\widehat{f}_{n} - \tilde{f}^*\|_{L_2(a+1/m)} = [1-o_\P(1)]\|\widehat{f}_{n} - \bar{f}\|_{L_2(a+1/m)}$, which implies $\|\widehat{f}_{n}  - \bar{f}\|_{L_2(a+1/m)}^2 = O_\P\{\|\tilde{f}^* - \bar{f}\|_{L_2(a+1/m)}^2\}$. Thus, by  (\ref{eqn:anoformhatftildefc}) and Lemma \ref{lem:tilfbarfl2a}, we complete the proof.

We now are in the position to prove (\ref{eqn:anoformhatftildefc}).
For any $0\leq c\leq a+1/m$, by (\ref{eqn:hatfnlambdatildef}), we have
\begin{equation}
\label{eqn:1p1ndejf}
\allowdisplaybreaks
\begin{aligned}
& \|\widehat{f}_{n} - \tilde{f}^*\|_{L_2(c)}^2  \\
& \leq \sum_{{\vec{\bnu}}\in \bV}\left(1+\frac{\rho_{\vec{\bnu}}}{\|\phi_{\vec{\bnu}}\|_{L_2}^2}\right)^c(1+\lambda\rho_{\vec{\bnu}})^{-2}\|\phi_{\vec{\bnu}}\|_{L_2}^2\cdot\frac{1}{p+1}\cdot\\
& \quad \left\{\left[\frac{\sum_{i=1}^n(\widehat{f}_{n}-\bar{f})(\bt_i^{(0)}) \phi_{\vec{\bnu}}(\bt_i^{(0)})}{n\sigma_0^2} - \frac{\int(\widehat{f}_{n}-\bar{f})(\bt)\phi_{\vec{\bnu}}(\bt)d\Pi^{(0)}(\bt)}{\sigma_0^2}\right]^2 \right.\\
& \quad\quad\left.+\sum_{j=1}^p\left[\frac{\sum_{i=1}^n\frac{\partial (\widehat{f}_{n}-\bar{f})}{\partial t_j}(\bt_i^{(j)})\frac{\partial \phi_{\vec{\bnu}}}{\partial t_j}(\bt_i^{(j)})}{n\sigma_j^2}- \frac{\int \frac{\partial (\widehat{f}_{n} - \bar{f})(\bt)}{\partial t_j}\frac{\partial \phi_{\vec{\bnu}}(\bt)}{\partial t_j}d\Pi^{(j)}(\bt)}{\sigma_j^2}\right]^2\right\}.
\end{aligned}
\end{equation}
Let $g_{j}(\bt) = \frac{1}{\sigma_j^2}\frac{\partial (\widehat{f}_{n} - \bar{f})}{\partial t_j} \frac{\partial \phi_{\vec{\bnu}}}{\partial t_j}$ and $g_0(\bt) = \frac{1}{\sigma_0^2}(\widehat{f}_{n} - \bar{f})\phi_{\vec{\bnu}}$.
Hence, we can do the expansion on the basis $\{\phi_{\vec{\bmu}}\}_{{\vec{\bmu}}\in \N^d}$,
\begin{equation}
\label{eqn:gjbtsumbmund}
g_{j}(\bt)=\sum_{{\vec{\bmu}}\in \N^d} Q^{j}_{{\vec{\bmu}}}\phi_{\vec{\bmu}}(\bt),\quad  \mbox{ where } Q^{j}_{\vec{\bmu}} = \langle g_j(\bt),\phi_{\vec{\bmu}}(\bt)\rangle_0.
\end{equation}
Unlike (\ref{eqn:fsumnuthetanuphinvb}) with the multi-index $\vec{\bnu}\in \bV$, we require ${\vec{\bmu}}\in \N^d$ in (\ref{eqn:gjbtsumbmund}) since now $g_{j}(\bt)$is a product function.
By Cauchy-Schwarz inequality,
\begin{equation}
\label{eqn:cgammnua1mf35}
\begin{aligned}
& \left[\frac{1}{n\sigma_j^2}\sum_{i=1}^n\frac{\partial (\widehat{f}_{n} - \bar{f})}{\partial t_j}(\bt_i^{(j)})\frac{\partial \phi_{\vec{\bnu}}}{\partial t_j}(\bt_i^{(j)})  - \frac{1}{\sigma_j^2}\int \frac{\partial (\widehat{f}_{n} - \bar{f})(\bt)}{\partial t_j}\frac{\partial \phi_{\vec{\bnu}}(\bt)}{\partial t_j}\right]^2\\
&  =  \left[\sum_{{\vec{\bmu}}\in \N^d} Q^{j}_{{\vec{\bmu}}}\left(\frac{1}{n}\sum_{i=1}^n\phi_{\vec{\bmu}}(\bt^{(j)}_i)-\int\phi_{\vec{\bmu}}(\bt)\right)\right]^2\\
& \leq \left[\sum_{{\vec{\bmu}}\in \N^d} (Q^{j}_{{\vec{\bmu}}})^2\left(1+\frac{\rho_{\vec{\bmu}}}{\|\phi_{\vec{\bmu}}\|_{L_2}^2}\right)^a\|\phi_{\vec{\bmu}}\|_{L_2}^2\right]\\
& \cdot \left[\sum_{{\vec{\bmu}}\in \N^d} \left(1+\frac{\rho_{\vec{\bmu}}}{\|\phi_{\vec{\bmu}}\|_{L_2}^2}\right)^{-a}\|\phi_{\vec{\bmu}}\|_{L_2}^{-2}\left(\frac{1}{n}\sum_{i=1}^n\phi_{\vec{\bmu}}(\bt^{(j)}_i)-\int\phi_{\vec{\bmu}}(\bt)\right)^2\right].
\end{aligned}
\end{equation}
By Lemma \ref{lemma:inequanund1rhonua}, if $a> 1/2m$, then the sum of the first part in the right-hand side of (\ref{eqn:cgammnua1mf35}) over $j=0,1,\ldots,p$ is bounded by
\begin{equation}
\label{eqn:boundgjl2a1m}
\begin{aligned}
&  \sum_{j=0}^p\sum_{{\vec{\bmu}}\in \N^d} \left(1+\frac{\rho_{\vec{\bmu}}}{\|\phi_{\vec{\bmu}}\|_{L_2}^2}\right)^a\|\phi_{\vec{\bmu}}\|_{L_2}^2\left\langle\frac{\partial (\widehat{f}_{n} - \bar{f})}{\partial t_j}\frac{\partial \phi_{\vec{\bnu}}}{\partial t_j},\phi_{\vec{\bmu}}\right\rangle_0^2\\
&  \lesssim  \|\widehat{f}_{n} - \bar{f}\|_{L_2(a+1/m)}^2\sum_{j=0}^p\sum_{{\vec{\bmu}}\in \N^d}\left(1+\frac{\rho_{\vec{\bmu}}}{\|\phi_{\vec{\bmu}}\|_{L_2}^2}\right)^a\|\phi_{\vec{\bmu}}\|_{L_2}^2\left\langle \frac{\partial \phi_{\vec{\bnu}}}{\partial t_j},\phi_{\vec{\bmu}}\right\rangle_0^2\\
& \lesssim \|\widehat{f}_{n} - \bar{f}\|_{L_2(a+1/m)}^2\left(1+\frac{\rho_{\vec{\bnu}}}{\|\phi_{\vec{\bnu}}\|_{L_2}^2}\right)^a\|\phi_{\vec{\bnu}}\|_{L_2}^2\left(1+\sum_{j=1}^p\nu_j^2\right)\\
& \asymp \|\widehat{f}_{n} - \bar{f}\|_{L_2(a+1/m)}^2 \left(1+\frac{\rho_{\vec{\bnu}}}{\|\phi_{\vec{\bnu}}\|_{L_2}^2}\right)^a.
\end{aligned}
\end{equation}
The second part on the right-hand side of (\ref{eqn:cgammnua1mf35}) can be  bounded by
\begin{equation}
\label{eqn:enuganmmacnf}
\begin{aligned}
& \E\left[\sum_{{\vec{\bmu}}\in \N^d}\left(1+\frac{\rho_{\vec{\bmu}}}{\|\phi_{\vec{\bmu}}\|_{L_2}^2}\right)^{-a}\|\phi_{\vec{\bmu}}\|_{L_2}^{-2}\left(\frac{1}{n}\sum_{i=1}^n\phi_{\vec{\bmu}}(\bt_i^{(j)}) - \int\phi_{\vec{\bmu}}(\bt)\right)^2\right]\\
&\quad  \leq n^{-1}\sum_{{\vec{\bmu}}\in \N^d}\left(1+\frac{\rho_{\vec{\bmu}}}{\|\phi_{\vec{\bmu}}\|_{L_2}^2}\right)^{-a}\|\phi_{\vec{\bmu}}\|_{L_2}^{-2}\int\phi_{\vec{\bmu}}^2(\bt)\\
& \quad \asymp n^{-1}\sum_{{\vec{\bmu}}\in \N^d}\left(1+\frac{\rho_{\vec{\bmu}}}{\|\phi_{\vec{\bmu}}\|_{L_2}^2}\right)^{-a} \lesssim n^{-1}\sum_{{\vec{\bmu}}\in \N^d}\mu_1^{-2ma}\cdots\mu_d^{-2ma}\\
& \quad \leq n^{-1}\left(\sum_{\mu_1=1}^\infty\mu_1^{-2ma}\right)^d\asymp n^{-1},
\end{aligned}
\end{equation}
where the third step uses $\rho_{\vec{\bmu}}/\|\phi_{\vec{\bmu}}\|_{L_2}^2\asymp \mu_1^{2m}\cdots\mu_d^{2m}$, and the fourth step holds for $a> 1/2m$. Combing (\ref{eqn:boundgjl2a1m}) and (\ref{eqn:enuganmmacnf}),  we have that for
$a>1/2m$,
\begin{equation}
\label{eqn:e1p1n-1hatfbarfl2a1m}
\begin{aligned}
&\sum_{j=0}^p\E\left[\sum_{{\vec{\bmu}}\in \N^d} Q^{j}_{{\vec{\bmu}}}\left(\frac{1}{n}\sum_{i=1}^n\phi_{\vec{\bmu}}(\bt^{(j)}_i)-\int\phi_{\vec{\bmu}}(\bt)\right)\right]^2\\
& \lesssim \frac{1}{n} \|\widehat{f}_{n} - \bar{f}\|_{L_2(a+1/m)}^2\left(1+\frac{\rho_{\vec{\bnu}}}{\|\phi_{\vec{\bnu}}\|_{L_2}^2}\right)^a.
\end{aligned}
\end{equation}

\paragraph{Putting all together.}
Therefore, if $1/2m< a<(2m-3)/4m$ and $0\leq c\leq a+1/m$,  (\ref{eqn:1p1ndejf})  and (\ref{eqn:e1p1n-1hatfbarfl2a1m}) imply that
\begin{align}
 \E\|\widehat{f}_{n} -\tilde{f}^*\|_{L_2(c)}^2  \lesssim n^{-1}\|\widehat{f}_{n} - \bar{f}\|_{L_2(a+1/m)}^2N_{a+c}(\lambda).\nonumber
 \end{align}
 By Lemma \ref{lemma:varthetamultivariatetensorrank} we complete the proof for  (\ref{eqn:anoformhatftildefc}) and this lemma.
\end{proof}

Finally, we combine \eqref{eqn:errordecomp} and Lemmas \ref{lem:barff0l2ar}--\ref{lem:boundhattildef} to obtain the following proposition.

\begin{proposition}
\label{thm:convergenceforhatfnlambda}
Under the conditions of Theorem \ref{theorem:lowerbdfNlambdaregrandom} and assuming the distributions $\Pi^{(0)}$ and $\Pi^{(j)}$s are known.
If $1/2m< a<(2m-3)/4m$, and $n^{-1}\lambda^{-(2a+3/2m)}[\log(1/\lambda)]^{r-1}\to 0$, then for any $c\in[0,a+1/m]$, the $\widehat{f}_{n}$ given by (\ref{scheme1}) satisfies, when  $0\leq p<d$,
\begin{align}
 \|\widehat{f}_{n}- f_0\|_{L_2(c)}^2  = O\left\{\change{\min_{J(f)\leq R_n^2}\|f-f_0\|_{L_2(c)}^2} + \lambda^{1-c}\change{R_n^2}\right\}  + O_\P\left\{n^{-1}\change{R_n^2}\lambda^{-(c+1/2m)}[\log(1/\lambda)]^{r\wedge (d-p)-1}\right\},\nonumber
\end{align}
and when $p=d$,
\begin{align}
& \|\widehat{f}_{n} - f_0\|_{L_2(c)}^2 \nonumber\\
& =
\begin{cases}
O\left\{\change{\min_{J(f)\leq R_n^2}\|f-f_0\|_{L_2(c)}^2} + \lambda^{\frac{(1-c)mr}{mr-1}}\change{R_n^2}\right\}  + O_\P\left\{n^{-1}\change{R_n^2}\lambda^{\frac{mr}{1-mr}\left(c+\frac{r-2}{2mr}\right)}\right\}\quad   \mbox{ if } r\geq 3,\\
O\left\{\change{\min_{J(f)\leq R_n^2}\|f-f_0\|_{L_2(c)}^2} +\lambda^{\frac{2m}{2m-1}}\change{R_n^2}\right\}  + O_\P\left\{n^{-1}\change{R_n^2}\log(1/\lambda)\right\}  \mbox{ if } r=2, c=0,\\
O\left\{\change{\min_{J(f)\leq R_n^2}\|f-f_0\|_{L_2(c)}^2} +\lambda^{\frac{2(1-c)m}{2m-1}}\change{R_n^2}\right\}  + O_\P\left\{n^{-1}\change{R_n^2}\lambda^{\frac{2mc}{1-2m}}\right\}  \mbox{ if } r=2, c>0,\\
O\left\{\change{\min_{J(f)\leq R_n^2}\|f-f_0\|_{L_2(c)}^2} +\lambda^{\frac{(1-c)m}{m-1}}\change{R_n^2}\right\}  + O_\P\left\{n^{-1}\change{R_n^2}\right\}  \mbox{ if } r=1, c<\frac{1}{2m},\\
O\left\{\change{\min_{J(f)\leq R_n^2}\|f-f_0\|_{L_2(c)}^2} +\lambda^{\frac{2m-1}{2(m-1)}}\change{R_n^2}\right\}  + O_\P\left\{n^{-1}\change{R_n^2}\log(1/\lambda)\right\}  \mbox{ if } r=1, c=\frac{1}{2m},\\
O\left\{\change{\min_{J(f)\leq R_n^2}\|f-f_0\|_{L_2(c)}^2} +\lambda^{\frac{(1-c)m}{m-1}}\change{R_n^2}\right\}  + O_\P\left\{n^{-1}\change{R_n^2}\lambda^{\frac{1-2mc}{2m-2}}\right\}  \mbox{ if } r=1, c>\frac{1}{2m}.
 \end{cases}\nonumber
\end{align}
\end{proposition}

By Proposition \ref{thm:convergenceforhatfnlambda}, we can derive the convergence rates by  the estimator $\widehat{f}_{n}$ defined by  (\ref{scheme1}). In fact, for $p=d$ and $r\geq 3$, by letting $\lambda \asymp n^{-\frac{2mr-2}{(2m+1)r-2}}$, $a=1/2m+\epsilon$ for some $\epsilon>0$ and $c=0$, we have that
$ n^{-1}\lambda^{-(2a+3/2m)}[\log(1/\lambda)]^{r-1}\to 0$
is equivalent to
\begin{equation*}
-1 + \frac{5(mr-1)}{2m^2r+mr-2m}<0.
\end{equation*}
Thus, the conditions for Proposition \ref{thm:convergenceforhatfnlambda} are satisfied. Similarly, we can verify that when $p=d$ and $r=2$,  $\lambda\asymp [n(\log n)]^{-(2m-1)/2m}$ satisfies the conditions for Proposition \ref{thm:convergenceforhatfnlambda}. When $p=d$ and $r=1$, $\lambda\asymp n^{-(m-1)/m}$ satisfies the conditions for the above proposition. When $0\leq p\leq d-r$, $\lambda\asymp [n(\log n)^{1-r}]^{-2m/(2m+1)}$  satisfies the conditions for the above Proposition,  as well as
when $d-r< p< d$ by letting $\lambda\asymp [n(\log n)^{1+p-d}]^{-2m/(2m+1)}$.  This observation leads to the following theorem for $\widehat{f}_{n}$ in (\ref{scheme1}).
 \begin{theorem}
 \label{eqn:esterrororgest}
Assume that $\lambda_\nu \asymp \nu^{-2m}$ for some $m>3/2$. Under the regression models \eqref{modelequation1} and \eqref{modelequation} where $f_0$ follows the SS-ANOVA model \eqref{eqn:anovadecompfti} and \change{$\|f\|_{\HH}\leq R_n$}. Then under the general error structure (\ref{eqn:errorstruct}), the estimator $\widehat{f}_{n}$ defined by  (\ref{scheme1}) satisfies
\begin{align*}
\lim_{C\to\infty}\underset{n\to\infty}{\lim\sup}& \sup_{f_0\in\HH} \P \left\{\int_{\XX^d}\left[\widehat{f}_{n}(\bt)-f_0(\bt)\right]^2d\bt \right.\leq C\left(\left[n(\log n)^{1-(d-p)\wedge r}\right]^{-\frac{2m}{2m+1}} \mathbbm{1}_{0\leq p<d} \right.\\
& \left.\left.  \quad\quad \quad\quad + \left[n^{-\frac{2mr}{(2m+1)r-2}} \mathbbm{1}_{r\geq 3}+n^{-1}(\log n)^{r-1}  \mathbbm{1}_{r<3}\right]  \mathbbm{1}_{p=d}\vphantom{\left[n(\log n)^{1-(d-p)\wedge r}\right]^{-2m/(2m+1)}}\right)\vphantom{\int_{\XX^d}\left\{\tilde{f}(\bt)-f_0(\bt)\right\}^2d\bt}\right\}=1.
\end{align*}
Here the tuning parameter $\lambda$ in (\ref{eq: kernel_reg}) is chosen by $\lambda\asymp \left[n(\log n)^{1-(d-p)\wedge r}\right]^{-2m/(2m+1)} $ when $0\leq p<d$, and $\lambda\asymp n^{-(2mr-2)/[(2m+1)r-2]}$ when $p=d, r\geq 3$, and $\lambda\asymp (n\log n)^{-(2m-1)/2m}$ when $p=d, r= 2$, and $\lambda\asymp n^{-(m-1)/m}$ when $p=d$, $r=1$. 
\end{theorem}

\change{Finally, we approximate the estimator $\widehat{f}_n$ in \eqref{scheme1} with the random feature estimator defined by \eqref{eqn:RF},
\begin{equation*}  
\widehat{f}_n^{\text{RF}}  =  \bPsi_{(p+1)d}(\bt)^\top {\bc_{(p+1)d}}.
\end{equation*}
We have the following decomposition,
\begin{equation}
\label{eqn:approxerrordecomp}
\begin{aligned}
\widehat{f}_n^{\text{RF}}-\widehat{f}_{n} & = \underbrace{(S_s\widehat{C}_{s,\lambda}^{-1}\widehat{S}_s^*y - L_sL_{s,\lambda}^{-1}y)}_{\text{Error I}} + \underbrace{(L_sL_{s,\lambda}^{-1}y - LL_{\lambda}^{-1}y)}_{\text{Error II}}.
\end{aligned}
\end{equation}
Here the notations are similar to those in the Definition 2 of \citet{rudi2017generalization}. Specifically, 
let $y$ be the vector of data, $y = (y_1^{(0)},\ldots,y^{(0)}_n,\ldots,y^{(p)}_1,\ldots,y^{(p)}_n)^\top$. Moreover,
\begin{itemize}
\item The approximated kernel $K_s=\bPsi_{(p+1)d}(\bt)^\top(\bt)\bPsi_{(p+1)d}(\bt')$. 
\item $S_s$: $(S_s\beta)(\cdot)=\bPsi_{(p+1)d}(\cdot)^\top\beta$.
\item $S_s^*$: $S_s^*g = \frac{1}{\sqrt{s}}\int\bPsi_{(p+1)d}(\bt)g(\bt)d\bt$.
\item $L_s$: $(L_sg)(\cdot) = \int K_s(\cdot,\bt)g(\bt)d\bt$.
\item $C_s$: $C_s = \int\bPsi_{(p+1)d}(\bt)\bPsi_{(p+1)d}(\bt)^\top d\bt$.
\item $\widehat{C}_s$: $\widehat{C}_s=\frac{1}{n}\sum_{i=1}^n\bPsi_{(p+1)d}(\bt_i)\bPsi_{(p+1)d}(\bt_i)^\top$.
\item The random feature mapping estimator $\widehat{f}_n^{\text{RF}} = S_s\widehat{C}_{s,\lambda}^{-1}\widehat{S}_s^* y$.
\end{itemize}
We analyze the two error terms  in \eqref{eqn:approxerrordecomp} separately. For the Error I, note that, $L_sL_{s,\lambda}^{-1} = S_sC_{s,\lambda}^{-1}S_s^*$. Then,
\begin{equation*}
\begin{aligned}
\text{Error I}&= S_s\widehat{C}_{s,\lambda}^{-1}\widehat{S}_s^*y - L_sL_{s,\lambda}^{-1}y\\
& = S_s\widehat{C}_{s,\lambda}^{-1}(\widehat{S}_s^*-S_s^*)y +S_s(\widehat{C}_{s,\lambda}^{-1} - C_{s,\lambda}^{-1})S_s^*y\\
& = S_s\widehat{C}_{s,\lambda}^{-1}(\widehat{S}_s^*-S_s^*)y +S_s\widehat{C}_{s,\lambda}^{-1}(C_{s,\lambda}- \widehat{C}_{s,\lambda})C_{s,\lambda}^{-1}S_s^*y\\
& =S_s\widehat{C}_{s,\lambda}^{-1}(\widehat{S}_s^*-S_s^*)y + (S_s\widehat{C}_{s,\lambda}^{-1}C_{s,\lambda}^{1/2})\left[C_{s,\lambda}^{-1/2}(C_{s,\lambda}- \widehat{C}_{s,\lambda})\right]C_{s,\lambda}^{-1}S_s^*y.
\end{aligned}
\end{equation*}
By Lemma 7 in \citet{rudi2017generalization}, we obtain that,
\begin{equation*}
\begin{aligned}
\|\text{Error I}\|_{L_2}&\leq O\left(\lambda^{-1/2}n^{-1}+n^{-1/2}\lambda^{-1/4m}\right).
\end{aligned}
\end{equation*}
By Lemma \ref{lem:tilfbarfl2a}, both the term $\lambda^{-1/2}n^{-1} $ and the term $n^{-1/2}\lambda^{-1/4m}$ are dominated by $\|\tilde{f}^* - \bar{f}\|_{L_2}$. By Proposition \ref{thm:convergenceforhatfnlambda},
\begin{equation}
\label{eqn:bdonerrori}
\|\text{Error I}\|_{L_2}= O(\|\widehat{f}_{n}-f_0\|_{L_2}^2).
\end{equation}
For the Error II, by Lemma 8 and Equation (14) in \citet{rudi2017generalization}, we have
\begin{equation*}
\begin{aligned}
\|\text{Error II}\|_{L_2}&= O\left( \sqrt{\frac{\log(1/\lambda)}{s}} \right),
\end{aligned}
\end{equation*}
By Proposition \ref{thm:convergenceforhatfnlambda}, and letting $s = O(n\log n)$, we have $\|\text{Error II}\|_{L_2} = O(n^{-1/2})$. Hence
\begin{equation}
\label{eqn:bdonerrorii}
\|\text{Error II}\|_{L_2} = O(\|\widehat{f}_{n}-f_0\|_{L_2}^2).
\end{equation}
By combining \eqref{eqn:bdonerrori} and \eqref{eqn:bdonerrorii}, we have $\|\widehat{f}_n^{\text{RF}} - \widehat{f}_{n}\|_{L_2} = O(\|\widehat{f}_{n}-f_0\|_{L_2})$. By triangle inequality,
\begin{equation}
\label{eqn:bdfrff0}
\begin{aligned}
\|\widehat{f}_n^{\text{RF}}-f_0\|_{L_2} & = \|\{\widehat{f}_n^{\text{RF}}-\widehat{f}_{n}\} + \{\widehat{f}_{n} - f_0\}\|_{L_2}\\
& \leq \|\widehat{f}_n^{\text{RF}}-\widehat{f}_{n}\|_{L_2} + \|\widehat{f}_{n} - f_0\|_{L_2} \\
& = O(\|\widehat{f}_{n} - f_0\|_{L_2}).
\end{aligned}
\end{equation}
Using Theorem \ref{eqn:esterrororgest} and \eqref{eqn:bdfrff0}, we complete the proof of Theorem \ref{thm:mainupperrateestf0}.}


\subsection{Auxiliary Lemmas for Theorems \ref{theorem:lowerbdfNlambdaregrandom} and  \ref{thm:mainupperrateestf0}}
\label{subsubsec:lemma1}

\begin{lemma} 
\label{lemma:intx1xrzxk1z2}
Suppose that $\beta\geq 0$ and $0<\alpha\leq 2$. Then, as $\Xi\to \infty$,
\begin{align*}
& \int_{x_1\cdots x_r\leq \Xi, x_k\geq 1}\prod_{k=1}^rx_k^\beta (x_1^\alpha+x_2^{\alpha}+\cdots +x_r^\alpha)^{-1}dx_1\cdots dx_r\\
& \asymp 
\begin{cases}
\Xi^{\beta+1-\alpha/r},   \mbox{ if } r\geq 3;\\
\log(\Xi),  \mbox{ if } r=2, \beta=\alpha/2-1;\quad \Xi^{\beta+1-\alpha/2} \mbox{ if } r=2, \beta>\alpha/2-1;\\
1,  \mbox{ if } r=1,\beta<\alpha-1;  \quad \log(\Xi)  \mbox{ if } r=1, \beta=\alpha-1; \\
\Xi^{\beta-\alpha+1}  \mbox{ if } r=1, \beta>\alpha-1.
\end{cases}
\end{align*}
\end{lemma}
\begin{proof}
By the symmetry of covariates, 
\begin{align}
& \int_{x_1\cdots x_r\leq \Xi, x_k\geq 1} \prod_{k=1}^rx_k^\beta(x_1^\alpha+x_2^\alpha+\cdots +x_r^\alpha)^{-1}dx_1\cdots dx_r\nonumber\\
& \asymp \int_{x_1\cdots x_r\leq \Xi, x_1\geq x_2\geq\cdots\geq x_r\geq 1}\prod_{k=1}^rx_k^\beta (x_1^\alpha+x_2^\alpha+\cdots +x_r^\alpha)^{-1}dx_r\cdots dx_1\nonumber\\
& :=\EE.\nonumber
\end{align}

First, we prove when $r\geq 3$,  as $\Xi\to \infty$, we have
\begin{align}
\label{eqn:r3intx1xraa12r}
\EE\lesssim \Xi^{\beta+1-\alpha/r}.
\end{align}
For this, define the set
$\KK=\left\{0\leq k\leq r-2: \left(\frac{\Xi}{x_1\cdots x_{r-k-1}}\right)^{1/(k+1)}\leq x_{r-k-1}\right\}.$
If $\KK$ is not empty, we denote the smallest element in $\KK$ by $k^*$. Then $0\leq k^*\leq r-2$. For any $(x_1,\ldots, x_r)\in\{(x_1,\ldots,x_r):x_1\cdots x_r\leq \Xi, x_1\geq x_2\geq\cdots\geq x_r\geq 1,x_r\leq x_{r-1}\leq \frac{\Xi}{x_1\cdots x_{r-1}}\}$, we have
\begin{equation}
\label{eqn:condxrk1xkstar}
\begin{cases}
1\leq x_{r-k}\leq x_{r-k-1}& \quad\mbox{ for } 0\leq k\leq k^*-1,\\
1\leq x_{r-k^*}\leq \left(\frac{\Xi}{x_1\cdots x_{r-k^*-1}}\right)^{1/(k^*+1)}& \quad \mbox{ for }  k=k^*,\\
x_{r-k}\geq \left(\frac{\Xi}{x_1\cdots x_{r-k-1}}\right)^{1/(k+1)} & \quad\mbox{ for } k^*+1\leq k\leq r-2,\\
x_{1}\geq \Xi^{1/r}& \quad\mbox{ for }  k= r-1.
\end{cases}
\end{equation}
Thus, as $\Xi\to \infty$, 
\begin{equation}
\label{eqn:a1-2rkknotempty}
\begin{aligned}
\EE& \lesssim \int_{x_1\cdots x_r\leq \Xi, x_1\geq x_2\geq\cdots\geq x_r\geq 1}\\
& \quad \quad\quad\quad \left\{(x_1)^{\beta-\alpha/(r-1)}\cdots (x_{r-k^*-1})^{\beta-\alpha/(r-1)}\right\} x_{r-k^*}^\beta\\
& \quad\quad\quad\quad\quad\quad\cdot \left\{(x_{r-k^*+1})^{\beta-\alpha/(r-1)}\cdots (x_{r})^{\beta-\alpha/(r-1)}\right\}d\bx\\
& \asymp \int_{x_1\cdots x_r\leq \Xi, x_1\geq x_2\geq\cdots\geq x_r\geq 1}\\
& \quad \quad\quad\quad  \left\{(x_1)^{\beta-\alpha/(r-1)}\cdots (x_{r-k^*-1})^{\beta-\alpha/(r-1)}\right\} \\
& \quad\quad\quad \quad\quad\quad
\cdot (x_{r-k^*})^{[\beta+1-\alpha/(r-1)]k^*+\beta} dx_{r-k^*}dx_{r-k^*-1}\cdots dx_1\\
& \asymp \int_{x_1\cdots x_r\leq \Xi, x_1\geq x_2\geq\cdots\geq x_r\geq 1}\\
& \quad \quad\quad\quad\left\{(x_1)^{-1-\alpha/[(r-1)(k^*+1)]}\cdots (x_{r-k^*-1})^{-1-\alpha/[(r-1)(k^*+1)]}\right\}\\
& \quad\quad\quad \quad\quad\quad
\cdot \Xi^{\beta+1-\alpha k^*/[(r-1)(k^*+1)]}  dx_{r-k^*-1}\cdots dx_1\\
& = \Xi^{\beta+1-\alpha/r},
\end{aligned}
\end{equation}
where the first step uses $x_{r-k^*}\geq 1$ and Lemma \ref{lem:aversioofyoungsineq},  the second step uses $x_{r-k}\leq x_{r-k-1}$ for all $k\leq k^*-1$ in (\ref{eqn:condxrk1xkstar}), the third step uses the upper bound on $x_{r-k^*}$ in (\ref{eqn:condxrk1xkstar}), the fourth step uses the lowers bounds on $x_{r-k}$ for all $k^*+1\leq k\leq r-2$ in  (\ref{eqn:condxrk1xkstar}).
If $\KK$ is empty, then for any $(x_1,\ldots, x_r)\in\{(x_1,\ldots,x_r):x_1\cdots x_r\leq \Xi, x_1\geq x_2\geq\cdots\geq x_r\geq 1,x_r\leq x_{r-1}\leq \Xi/(x_1\cdots x_{r-1})\}$, it satisfies 
\begin{align*}
1\leq x_k\leq x_{k-1} \mbox{ for any } 2\leq k\leq r, \quad \mbox{ and }\quad 1\leq x_1\leq \Xi^{1/r}.
\end{align*}
Thus,  as $\Xi\to \infty$,
\begin{equation}
\label{eqn:a1-2rkkempty}
\begin{aligned}
\EE& = \int_1^{\Xi^{1/r}}\cdots\int_1^{x_{r-2}}\int_1^{x_{r-1}}\\
& \quad\quad\quad\quad\quad\prod_{k=1}^rx_k^\beta(x_1^\alpha+x_2^\alpha+\cdots+x_{r-1}^\alpha+x_r^\alpha)^{-1}dx_rdx_{r-1}\cdots dx_1\\
& \lesssim  \int_1^{\Xi^{1/r}}\cdots\int_1^{x_{r-2}}\int_1^{x_{r-1}}\\
& \quad\quad\quad\quad\quad
x_1^{\beta-\alpha/r}\cdots x_{r-1}^{\beta-\alpha/r}x_r^{\beta-\alpha/r}dx_rdx_{r-1}\cdots  dx_1\asymp \Xi^{\beta+1-\alpha/r}.
\end{aligned}
\end{equation}
Combining (\ref{eqn:a1-2rkknotempty}) and (\ref{eqn:a1-2rkkempty}) completes the proof for (\ref{eqn:r3intx1xraa12r}).

On the other hand, when $r\geq 3$ and as $\Xi\to \infty$,
\begin{equation}
\label{eqn:r3intx1xraa12rlower}
\begin{aligned}
\EE& \geq \int_1^{\Xi^{1/r}}\cdots\int_1^{x_{r-2}}\int_1^{x_{r-1}}\\
& \quad\quad\quad\quad\quad\prod_{k=1}^rx_k^\beta(x_1^\alpha+\cdots+x_{r-1}^\alpha+x_r^\alpha)^{-1}dx_rdx_{r-1}\cdots dx_1\\
& \geq \int_{1}^{\Xi^{1/r}}\cdots\int_1^{x_{r-2}}\int_{1}^{x_{r-1}}\\
& \quad\quad\quad\quad\quad\prod_{k=1}^{r}x_k^\beta\cdot r^{-1}x_{1}^{-\alpha}dx_rdx_{r-1}\cdots dx_1 \asymp \Xi^{\beta+1-\alpha/r}.
\end{aligned}
\end{equation}
Therefore, combining (\ref{eqn:r3intx1xraa12r}) and (\ref{eqn:r3intx1xraa12rlower}) completes the proof of the lemma for $r\geq 3$. 

Then we consider for $r=2$. For  $0<\alpha\leq 2$,
\begin{align}
\EE& \leq  2\int_{1}^{\sqrt{\Xi}}\int_1^{x_1}x_1^{\beta-\alpha}x_2^\beta dx_2dx_1 + 2\int_{\sqrt{\Xi}}^{\Xi}\int_1^{\Xi/x_1}x_1^{\beta-\alpha}x_2^\beta dx_2dx_1\nonumber\\
& \asymp 
\begin{cases}
\log(\Xi)  \quad\mbox{ when }2\beta+2-\alpha=0\\
\Xi^{\beta+1-\alpha/2} \quad \mbox{ when } 2\beta+2-\alpha>0
\end{cases}\quad \mbox{ as } \Xi\to \infty.
\label{eqn:r=2equivx1xrleqxix1gt}
\end{align}
On the other hand, we have
\begin{equation}
\label{eqn:r=2r3intx1xraa12}
\begin{aligned}
\EE& \geq \int_1^{\sqrt{\Xi}}\int_1^{x_1}x_1^\beta x_2^\beta(x_1^\alpha+x_2^\alpha)^{-1}dx_2dx_1\\
 & \geq 2^{-1}\int_1^{\sqrt{\Xi}}\int_1^{x_1}x_1^{\beta-2}x_2^\beta dx_2dx_1 \\
 & \asymp 
\begin{cases}
\log(\Xi)  \quad\mbox{ when }2\beta+2-\alpha=0\\
\Xi^m \quad \mbox{ when } 2\beta+2-\alpha>0
\end{cases} \mbox{ as } \Xi\to \infty.
\end{aligned}
\end{equation}
Combining (\ref{eqn:r=2equivx1xrleqxix1gt}) and (\ref{eqn:r=2r3intx1xraa12}) completes the proof of the lemma for $r=2$. 

Finally, we consider for $r=1$. Note that $\int_{1}^\Xi x_1^\beta x_1^{-\alpha}dx_1 \asymp 1$ when $0\leq \beta<\alpha-1$, and $\int_{1}^\Xi x_1^\beta x_1^{-\alpha}dx_1 \asymp \log(\Xi)$ when $\beta=\alpha-1$, and $\int_{1}^\Xi x_1^\beta x_1^{-\alpha}dx_1 \asymp \Xi^{\beta-\alpha+1}$ when $\beta>\alpha-1$. This completes the proof.
\end{proof}

\begin{lemma}
\label{lemmanormrequiv}
The norm $\|\cdot\|_R$ is equivalent to $\|\cdot\|_{\HH}$ in $\HH$.
\end{lemma}
\begin{proof}
Observe that for any $g\in \HH$, by the assumption that $\Pi^{(0)}$ and $\Pi^{(j)}$s are bounded away from  0 and infinity, we have
\begin{align}
& \frac{1}{p+1}\left[\frac{1}{\sigma_0^2}\int g^2(\bt)\Pi^{(0)}(\bt) + \sum_{j=1}^p \frac{1}{\sigma_j^2}\int\left\{\frac{\partial g(\bt)}{\partial t_j}\right\}^2\Pi^{(j)}(\bt) \right]  \nonumber\\
&\leq c_1\left[\int g^2(\bt) + \sum_{j=1}^p \int\left\{\frac{\partial g(\bt)}{\partial t_j}\right\}^2\right] \leq c_2\cdot c_{K}^{2d}\|g\|_{\HH}^2,\nonumber
\end{align}
for some constant $c_1$ and $c_2$, where the last step is by Lemma \ref{lem:bounddejft}. Hence 
\begin{equation}
\label{eqn:grnormlessc2ckghhnorm}
\|g\|_R^2 \leq (c_2c_{K}^{2d}+1)\|g\|_\HH^2.
\end{equation}

On the other hand, for any $g\in\HH$ we can do the orthogonal decomposition $g=g^0+g^1$ where $\langle g^0,g^1\rangle_\HH=0$, $g^0$ is in the null space of $J(\cdot)$ and $g^1$ is in the orthogonal space of the null space of $J(\cdot)$ in $\HH$.  Since the null space of $J(\cdot)$ has a finite basis that forms a positive definite kernel matrix, we assume the minimal eigenvalue of the kernel matrix is $\mu_{\min}'>0$. Then there exists a constant $c_3>0$ such that
\begin{equation}
\label{eqn:f0r2c2f0l2}
\|g^0\|_R^2\geq c_3\|g^0\|_{L_2}^2 \geq c_3 \mu_{\min}' \|g^0\|_\HH^2.
\end{equation}
For $g^1$, we have $\|g^1\|_R^2\geq J(g^1) = \|g^1\|_\HH^2$. Thus, for any $g\in\HH$,
\begin{align}
\|g\|_R^2 & \geq c_3\int \left(g^0+g^1\right)^2 + \|g^1\|_\HH^2\nonumber\\
& \geq c_3 \left\{\|g^0\|_{L_2}^2 + \frac{1+c_3}{c_3}\|g^1\|_{L_2}^2 - 2\|g^0\|_{L_2}\|g^1\|_{L_2}\right\}\nonumber\\
& \geq \frac{c_3}{1+c_3}\|g^0\|^2_{L_2},\nonumber
\end{align}
where the second inequality is by $\|g^1\|_\HH^2\geq \|g^1\|_{L_2}^2$. By (\ref{eqn:f0r2c2f0l2}), we obtain $\|g\|_R^2 \geq (1+c_3)^{-1}c_3\mu_{\min}'\|g^0\|_\HH^2$. Together with $\|g\|_R^2\geq J(g^1) = \|g^1\|_\HH^2$, we have
\begin{equation}
\label{eqn:grnormgtrc2ckghhnorm}
\|g\|_R^2 \geq \left(1+ \frac{1+c_3}{c_3\mu_{\min}'}\right)^{-1}\|g\|_\HH^2.
\end{equation}
Combining (\ref{eqn:grnormlessc2ckghhnorm}) and (\ref{eqn:grnormgtrc2ckghhnorm}) completes the proof.
\end{proof}

\begin{lemma}[Inverse transformation]
\label{lem:lineartransformuniform}
Suppose that designs $\bt^{(j)}$, $j=0,\ldots,p$ are independently drawn from known distributions $\Pi^{(j)}$ supported in $\XX^d$. Then, there exists a linear transformation to data $(\bt^{(j)},Y^{(j)})$ such that transformed design points $\bx^{(j)}$s are independently uniformly distributed on $\XX^d$.
\end{lemma}
\begin{proof}
First, we consider function and derivative data sharing a common design, i.e., $\bt_i^{(j)}=\bt_i^{(k)}$, $\forall 1\leq i\leq n, 0\leq j<k\leq p$. Write $\bt^{(j)} = (t^{(j)}_1,\ldots,t^{(j)}_d)\in\XX^d$. We allow covariates of $\bt^{(j)}$ can be correlated; that is, the density of $\bt^{(j)}$ is decomposed as:
\begin{equation*}
d\Pi^{(j)}(t_1,\ldots,t_d)  = d\Pi^{(j)}_d(t_d)d\Pi^{(j)}_{d-1}(t_{d-1}|t_d)\cdots d\Pi^{(j)}_1(t_1|t_d,t_{d-1}, \ldots,t_{2}).
\end{equation*}
Now let
\begin{equation*}
\begin{aligned}
x^{(j)}_d=\Pi^{(j)}_d(t^{(j)}_d),   \quad x^{(j)}_{d-1}=\Pi^{(j)}_{d-1}(t^{(j)}_{d-1}|t^{(j)}_d), \ldots,  \quad x^{(j)}_1=\Pi^{(j)}_1(t^{(j)}_1|t^{(j)}_d,t^{(j)}_{d-1}\ldots,t^{(j)}_{2}).
\end{aligned}
\end{equation*} 
Then, $\bx^{(j)} = (x_1^{(j)},x_2^{(j)},\ldots,x_d^{(j)})$ is uniformly distributed on $\XX^d$. 
Define that
\begin{equation*}
\begin{aligned}
& h(x_1, x_2,\ldots,x_d) \\
&  = f(\{\Pi_1^{(j)}\}^{-1}(x_1|x_d,\ldots,x_2),\{\Pi_2^{(j)}\}^{-1}(x_2|x_d,\ldots,x_3),\ldots, \{\Pi_d^{(j)}\}^{-1}(x_d)).
\end{aligned}
\end{equation*} 
Thus,
\begin{align}
\frac{\partial h(\bx)}{\partial x_j}  = \sum_{k=1}^j\frac{\partial f(\bt)}{\partial t_k}\cdot\frac{\partial t_k}{\partial x_j}  = \sum_{k=1}^{j-1}\frac{\partial f}{\partial t_k}\cdot\frac{\partial t_k}{\partial x_j} + \frac{\partial f}{\partial t_j}\cdot\frac{1}{d\Pi^{(j)}_j(t_j|t_d,\ldots,t_{j+1})}.\nonumber
\end{align}
With the design $\bx^{(j)}$ defined, we transform the responses $Y^{(j)}$s to $Z^{(j)}$s  by letting $Z^{(0)} = Y^{(0)}$ and  for any $j=1,\ldots,p$,
\begin{equation*}
Z^{(j)} = \sum_{k=1}^{j-1} Y^{(k)}\frac{\partial t_k^{(j)}(x_d^{(j)}, x_{d-1}^{(j)}\ldots,x_k^{(j)})}{\partial x_j}+ \frac{Y^{(j)}}{d\Pi^{(j)}_j(t_j^{(j)}|t_d^{(j)},\ldots,t_{j+1}^{(j)})}.
\end{equation*}
Write
\begin{equation*}
\tilde{\sigma}_{j}^2 = \sum_{k=1}^{j-1}\sigma_k^2\left[\frac{\partial t_k^{(j)}}{\partial x_j}(x_{d}^{(j)},x_{d-1}^{(j)},\ldots,x_{k}^{(j)})\right]^2+\frac{\sigma_j^2}{[d\Pi_j^{(j)}(t^{(j)}_{j}|t^{(j)}_{d}, \ldots,t^{(j)}_{j+1})]^2}.
\end{equation*}
Then it is clear that  $Z^{(j)} = \partial h/\partial x_j(\bx^{(j)}) + \widetilde{\epsilon^{(j)}}$, where the errors $\widetilde{\epsilon^{(j)}}$s are independent and centered noises with variance $\tilde{\sigma}_j^2$s.

Second, we consider that not all types of function observations and partial derivatives data  share a common design, i.e., $\exists 0\leq j\neq k\leq p$ and $1\leq i\leq n$ such that $\bt^{(j)}_i\neq \bt^{(k)}_i$. We require the covariates of each $\bt^{(j)}$ are independent; that is, the density of $\bt^{(j)}$ can be decomposed as:
\begin{equation*}
d\Pi^{(j)}(t_1,\ldots,t_d) = d\Pi_1^{(j)}(t_1)d\Pi_2^{(j)}(t_2)\cdots d\Pi_d^{(j)}(t_d)
\end{equation*} 
Now let 
\begin{equation*}
x_1^{(j)}=\Pi_1^{(j)}(t_1^{(j)}), \quad x_2^{(j)} = \Pi_2^{(j)}(t_2^{(j)}),\quad\ldots,\quad x_d^{(j)} = \Pi_d^{(j)}(t_d^{(j)}).
\end{equation*} 
Then $\bx^{(j)} = (x_1^{(j)}, x_2^{(j)},\ldots, x_d^{(j)})$ is uniformly distributed on $\XX^d$. Define the function
\begin{equation*}
h(x_1,\ldots,x_d) = f(\{\Pi_1^{(j)}\}^{-1}(x_1),\{\Pi_2^{(j)}\}^{-1}(x_2),\ldots,\{\Pi_d^{(j)}\}^{-1}(x_d)).
\end{equation*} 
Thus, we have 
\begin{equation*}
\frac{\partial h(\bx)}{\partial x_j} = \frac{\partial f(\bt)}{\partial t_j}\cdot\frac{\partial t_j(x_j)}{\partial x_j} =  \frac{\partial f(\bt)}{\partial t_j}\cdot\frac{1}{d\Pi_j^{(j)}(t_j)}.
\end{equation*} 
Correspondingly, the responses $Y^{(j)}$ is transformed to $Z^{(j)}$, $0\leq j\leq p$, by letting $Z^{(0)} =Y^{(0)}$ and $Z^{(j)} = Y^{(j)}/d\Pi_j^{(j)}(t_j^{(j)})$ for $1\leq j\leq d$, and write the transformed variance $\tilde{\sigma}_{j}^2  = \sigma_j^2/[d\Pi_j^{(j)}(t_{j}^{(j)})]^2$.
\end{proof}

\change{
\begin{lemma}
\label{lem:periodextend}
Suppose that $f_0$ follows the SS-ANOVA model in (4),  defined on $\XX^d\equiv[0,1]^d$.
Then, there exists a periodic  function $\tilde{f}_0$ on the expanded domain $[0,1+\delta]^d$ for any $\delta>0$ such that  $\tilde{f}_0(\bt)\equiv f_0(\bt)$ for $\bt\in\XX^d$, and $\tilde{f}_0$ maintains the same order of smoothness as $f_0$, in the sense that  $\tilde{f}_0$ follows the same RKHS in (5),  defined on $[0,1+\delta]^d$.
\end{lemma}

\begin{proof} The construction of the periodic function  consists of four main steps.

\emph{Step 1:} We show that when $\lambda_\nu \asymp \nu^{-2m}$, the $m$-th order Sobolev space on $\XX$ can be embedded into the RKHS $\HH_1$.  Specifically, let  $\mathcal{W}_2^m(\mathcal{X})$ denote the Sobolev space of order $m$, consisting of functions whose derivatives up to order $m-1$ are absolutely continuous and whose $m$-th derivative is square-integrable:
\begin{equation*}
\begin{aligned}
\mathcal{W}_2^m(\mathcal{X}) = \Big\{ g : \mathcal{X} \to \mathbb{R} \, \, g, dg/dt, \ldots, &d^{m-1}g/dt^{m-1}  \text{ are absolutely}\\
& \text{ continuous, and } d^mg/dt^{(m)} \in L_2 \Big\}.
\end{aligned}
\end{equation*}
There are many possible norms that can be quipped with $\mathcal W_2^m$ to make it a RKHS. For example, it can be endowed with the norm,
\begin{equation*}
\|g\|_{\mathcal{W}_2^m}^2 = \sum_{q=0}^{m-1}\left( \int g^{(q)} \right)^2 + \int \left(g^{(m)}\right)^2.
\end{equation*}
Following the results in Chapter 2 of \citet{wahba1990}, the eigenvalues of the associated kernel decay at a rate of $\lambda_\nu \asymp \nu^{-2m}$ for $\nu\geq 1$.
\vspace{0.1in}

\emph{Step 2:} For any $f_{0j}\in\HH_1$ on $\XX$, $j=1,\ldots,d$, we construct the function $g_j$ as,
\begin{equation}
\label{eqn:constperidg}
g_j(t_j) = \sum_{k=0}^{2m+1}c_{jk}t_j^k, \text{ for } t_j\in[1,1+\delta],
\end{equation}
where the coefficients $\{c_{jk}\}_{k=0}^{2m+1}$ are computed by satisfying the linear system:
\begin{equation}
\label{eqn:linearsysg}
 g_j^{(q)}(1) = f_{0j}^{(q)}(1) \ \text{ and } \ g_j^{(q)}(1+\delta) = f_{0j}^{(q)}(0), \quad \forall q=0,1,\ldots,m.
\end{equation}
Since the linear system \eqref{eqn:linearsysg} has $2m+2$ equations and the function $g_j$ in \eqref{eqn:constperidg} has $2m+2$ free coefficients $\{c_{jk}\}_{k=0}^{2m+1}$, there is a unique solution.
We define the extended function as,
\begin{equation*}
\tilde{f}_{0j}(t_j) = 
\begin{cases}
f_{0j}(t_j),\quad t_j\in[0,1],\\
g_j(t_j),\quad t_j\in[1,1+\delta],
\end{cases}
\end{equation*}
where $g_j$ is the $(2m+1)$-th order polynomial defined in \eqref{eqn:constperidg}. Since $g_j$ is continuous and has $m-1$ absolutely continuous derivatives, together with the property that the $m$-th derivative of $g_j$ is in $L_2$, we know that $\tilde{f}_{0j}(t_j)\in \WW_2^m([0,1+\delta])$.
By the result in Step 1,  the $m$-th order Sobolev space  $\mathcal{W}_2^m(\mathcal{X})$ can be embedded to the RKHS $\HH_1$. Hence $\tilde{f}_{0j}(t_j)$ follows the same RKHS as $f_{0j}(t_j)$ with the expanded domain on $[0,1+\delta]$.
\vspace{0.1in}

\emph{Step 3:}  For any $f_{0j_1j_2\cdots j_r}\in\HH_1\otimes\HH_1\otimes\cdots\otimes \HH_1$, $1\leq j_1<j_2<\cdots<j_r\leq d$ and $1\leq r\leq d$,  there exists a finite integer $s$ and functions $f_{0j_1\nu},f_{0j_2\nu},\ldots,f_{0j_r\nu}\in\HH_1$ for $\nu=1,\ldots,s$, such that
\begin{equation*}
f_{0j_1j_2\cdots j_r}(t_{j_1},t_{j_2},\ldots,t_{j_r})=\sum_{\nu=1}^sf_{0j_1\nu}(t_{j_1})f_{0j_2\nu}(t_{j_2})\cdots f_{0j_r\nu}(t_{j_r}).
\end{equation*}
By the construction in Step 2, we can find $g_{j\nu}(t_{j_1}) = \sum_{k=0}^{2m+1}c_{j\nu k}t_{j}^k$ for $t_{j}\in[1,1+\delta]$ and $j=j_1,j_2,\ldots,j_r$, such that,
\begin{equation}
\label{eqn:indgjbnd}
\begin{aligned}
 g_{j\nu}^{(q)}(1) = f_{0j\nu}^{(q)}(1) \ \text{ and } \ g_{j\nu}^{(q)}(1+\delta) = f_{0j\nu}^{(q)}(0), \quad \forall q=0,1,\ldots,m,
\end{aligned}
\end{equation}
Since the linear system \eqref{eqn:indgjbnd} has $(2m+2)$ equations and the function $g_{j\nu}(t_{j_1}) = \sum_{k=0}^{2m+1}c_{j\nu k}t_{j}^k$ has $(2m+2)$ free coefficients $\{c_{j\nu k}\}_{k=0}^{2m+1}$, there is a unique solution.
We define the extended function as,
\begin{equation*}
\tilde{f}_{0j_1j_2\cdots j_r}(t_{j_1},t_{j_2},\ldots,t_{j_r})=  \sum_{\nu=1}^sh_{j_1\nu}(t_{j_1})h_{j_2\nu}(t_{j_2})\cdots h_{j_r\nu}(t_{j_r}),
\end{equation*}
for any $(t_{j_1},t_{j_2},\ldots,t_{j_r})\in[0,1+\delta]^r$, where  for any $j=j_1,j_2,\ldots,j_r$, the function $h_{j\nu}$ is defined as,
\begin{equation*}
h_{j\nu}(t_j) = \begin{cases}
f_{0j\nu}(t_j)\quad t_j\in[0,1],\\
g_{j\nu}(t_j)\quad t_j\in[1,1+\delta],
\end{cases}
\end{equation*}
and $g_{j\nu}(t_{j_1}) = \sum_{k=0}^{2m+1}c_{j\nu k}t_{j}^k$ is the $(2m+1)$-th order polynomial. Since $g_{j\nu}$ is continuous and has $(m-1)$ absolutely continuous derivatives, together with the property that the $m$-th derivative of $g_{j\nu}$ is in $L_2$, we know that $\tilde{f}_{0j_1j_2\cdots j_r}(t_{j_1},t_{j_2},\ldots,t_{j_r})\in \WW_2^m([0,1+\delta])\otimes  \WW_2^m([0,1+\delta])\otimes\cdots\otimes \WW_2^m([0,1+\delta])$. By the result in Step 1,  the $m$-th order Sobolev space  $\mathcal{W}_2^m(\mathcal{X})$ can be embedded to the RKHS $\HH_1$. Hence $\tilde{f}_{0j_1j_2\cdots j_r}(t_{j_1},t_{j_2},\ldots,t_{j_r})$ follows the same RKHS as $f_{0j_1j_2\cdots j_r}$ with the expanded domain on $[0,1+\delta]^r$.
\vspace{0.1in}

\emph{Step 4:} For $f_0$ follows the SS-ANOVA model (4) on $\XX^d$,
we can define the function $\tilde{f}_0(\bt)$ that extends $f_0$ from $\XX^d$ to $[0,1+\delta]^d$ for any $\delta>0$. Specifically, let
\begin{equation*}
\tilde{f}_0(\bt) = \mbox{constant} + \sum_{j=1}^d \tilde{f}_{0j}(t_j) + \cdots + \sum_{1\leq j_1<j_2<\cdots<j_r\leq d}\tilde{f}_{0j_1j_2\cdots j_r}(t_{j_1},t_{j_2},\ldots,t_{j_r}).
\end{equation*}
By the construction in Steps 2 and 3, we have that $\tilde{f}_0(\bt) = f_0(\bt)$ for $\bt\in\XX^d$, which implies that $\tilde{f}_0(\bt)$ coincides with $f_0(\bt)$ on the original domain $\XX^d$. 
Moreover, $\tilde{f}_0(\bt)$  the same order of smoothness as $f_0(\bt)$ in the sense that  $\tilde{f}_0(\bt)$ follows the same RKHS in (5)  defined on $[0,1+\delta]^d$.
Hence, the eigenvalue decay rate of the RKHS for  $\tilde{f}_0(\bt)$ is the same as that of the RKHS for  $f_0(\bt)$.
Finally, by \eqref{eqn:linearsysg} and \eqref{eqn:indgjbnd}, we have that for any $j=1,\ldots,d$ and $(t_1,\ldots,t_{j-1},t_{j+1},\ldots,t_d) \in[0,1+\delta]^{d-1}$,
\begin{equation*}
\tilde{f}_0(t_1,\ldots,t_{j-1}, 0,t_{j+1},\ldots,t_d) = \tilde{f}_0(t_1,\ldots,t_{j-1}, 1+\delta,t_{j+1},\ldots,t_d),
\end{equation*}
which shows that the extended function $\tilde{f}_0$ has a periodic boundary on the expanded domain $[0,1+\delta]^d$ for any $\delta>0$. This completes the proof.
\end{proof}
}

\change{
\begin{lemma}
\label{lem:esterrperiodextend}
Suppose that $f_0$ follows the SS-ANOVA model in (4),  defined on $\XX^d\equiv[0,1]^d$, and 
the periodic  function $\tilde{f}_0$ is constructed in Lemma \ref{lem:periodextend}, defined on $[0,1+\delta]^d$. Then, if $\delta>0$ and for  any estimator $\widehat{f}_n$ on $[0,1+\delta]^d$, we have that
$\int_{\XX^d}[\widehat{f}_{n}(\bt)-f_0(\bt)]^2d\bt  \leq \int_{[0,1+\delta]^d}[\widehat{f}_{n}(\bt)-\tilde{f}_0(\bt)]^2d\bt$.
\end{lemma}
\begin{proof}
We decompose the estimation error of $\tilde{f}_0$ as follows:
\begin{equation*}
\begin{aligned}
&  \int_{[0,1+\delta]^d}\left[\widehat{f}_{n}(\bt)-\tilde{f}_0(\bt)\right]^2d\bt \\
& =  \int_{\XX^d}\left[\widehat{f}_{n}(\bt)-\tilde{f}_0(\bt)\right]^2d\bt +  \int_{[0,1+\delta]^d\setminus\XX^d}\left[\widehat{f}_{n}(\bt)-\tilde{f}_0(\bt)\right]^2d\bt\\
& =  \int_{\XX^d}\left[\widehat{f}_{n}(\bt)-f_0(\bt)\right]^2d\bt +  \int_{[0,1+\delta]^d\setminus\XX^d}\left[\widehat{f}_{n}(\bt)-\tilde{f}_0(\bt)\right]^2d\bt\\
& \geq  \int_{\XX^d}\left[\widehat{f}_{n}(\bt)-f_0(\bt)\right]^2d\bt,
\end{aligned}
\end{equation*}
where the second step uses the property that $\tilde{f}_0(\bt)\equiv f_0(\bt)$ for $\bt\in\XX^d$.
\end{proof}
}

\begin{lemma} 
\label{lem:bounddejft}
For any $g\in\HH$, there exists a constant $c_K$ which is independent of $g$ such that
\begin{equation*}
\sup_{\bt\in\XX^d}|g(\bt)|\leq c^d_K\|g\|_{\HH},
\end{equation*}
and
\begin{equation*}
\sup_{\bt\in\XX^d}\left|\frac{\partial g(\bt)}{\partial t_j}\right|\leq c^d_K\|g\|_{\HH}, \quad \forall 1\leq j\leq d.
\end{equation*} 
\end{lemma}
\begin{proof}
Since we assume that $K$ is continuous in the compact domain $\XX$ and satisfies (\ref{eqn:partial2tt'kcx1x1}), there exists some constant $c_K$ such that 
\begin{equation*}
\sup_{t\in\XX} |K(t,t)|\leq c_K \quad \mbox{ and }\quad\sup_{t\in\XX} \left\vert\frac{\partial^2K(t,t)}{\partial t\partial t'}\right\vert\leq c_K.
\end{equation*}
This implies for any $\bt\in\XX^d$,
\begin{equation*}
\left\|\frac{\partial K_d(\bt,\cdot)}{\partial t_j}\right\|_\HH^2 = \left\vert\frac{\partial^2 K(t_j,t_j)}{\partial t_j\partial t_j'}\right\vert \prod_{l\neq j}|K(t_l,t_l)| \leq c_K^d.
\end{equation*}
Thus, for any $g\in\HH$, by the Cauchy-Schwarz inequality,
\begin{equation*}
\sup_{\bt\in\XX^d} \left\vert\frac{\partial g(\bt)}{\partial t_j}\right\vert  \leq \sup_{\bt\in\XX^d}\left\|\frac{\partial K_d(\bt,\cdot)}{\partial t_j}\right\|_\HH\|g\|_\HH \leq c_K^d\|g\|_\HH,\quad\forall 1\leq j\leq d.
\end{equation*}
Similarly, we can show that $\sup_{\bt}|g(\bt)|\leq c_K^d\|g\|_\HH$.

\end{proof}

\begin{lemma}
\label{lemma:varthetamultivariatetensorrank}
Recall that $\bV$ as a family of multi-index $\vec{\bnu}$ is defined in (\ref{def:Vbnu}). We let
\begin{equation}
\label{eqn:defNalambdanuV}
N_a(\lambda) = \sum_{{\vec{\bnu}}\in \bV}\frac{\left(\prod_{k=1}^d\nu_k^{2m}\right)^a\left(1+\sum_{j=1}^p\nu_j^2\right)}{\left(1+\sum_{j=1}^p\nu_j^2+\lambda\prod_{k=1}^d\nu_k^{2m}\right)^2}.
\end{equation}
Then, when $0\leq p<d$, we have for any $0\leq a<1-1/2m$,
\begin{equation*}
N_a(\lambda) = O\left\{\lambda^{-a-1/2m}\left[\log (1/\lambda)\right]^{(d-p)\wedge r-1}\right\},
\end{equation*}
and when $p=d$, we have for any $0\leq a\leq 1$,
\begin{align*}
N_a(\lambda)=
\begin{cases}
O\left\{\lambda^{\frac{mr}{1-mr}\left(a+\frac{r-2}{2mr}\right)}\right\},   \mbox{ if } r\geq 3;\\
O\left\{\log(1/\lambda)\right\},  \mbox{ if } r=2, a=0;\quad O\left\{1\right\}, \mbox{ if } r=2, 0<a\leq 1;\\
O\left\{1\right\},  \mbox{ if } r=1,a<\frac{1}{2m};  \quad O\left\{\log(1/\lambda)\right\},  \mbox{ if } r=1,a=\frac{1}{2m}; \\
 O\left\{\lambda^{\frac{1-2ma}{2m-2}}\right\},  \mbox{ if } r=1, a>\frac{1}{2m}.
\end{cases}
\end{align*}
\end{lemma}

\begin{proof}
We will discuss three separate cases for $0\leq p\leq d-r$, $d-r<p<d$ and $p=d$.

First, consider  $0\leq p\leq d-r$. Since ${\vec{\bnu}}\in \bV$, there are at most $r$ of $\nu_1,\ldots,\nu_d$ not equal to 1, which implies that the number of combinations of non-1 indices being summed in (\ref{eqn:defNalambdanuV}) is no greater than $C_d^1+C_d^2+\cdots+C_d^r<\infty$. Due to the appearance of $(1+\sum_{j=1}^p\nu_j^2)$ in the denominator of (\ref{eqn:defNalambdanuV}), the largest terms of the summation (\ref{eqn:defNalambdanuV}) over ${\vec{\bnu}}\in \bV$ correspond to the combinations of $r$ indices whereas few $\nu_1,\ldots,\nu_p$ being summed as possible, which is the indices ${\vec{\bnu}}=(\nu_{k_1},\nu_{k_2},\ldots,\nu_{k_r})^\top\in \N^r$ with $k_1,k_2,\ldots,k_r>p$. Thus, by the integral approximation,
\begin{align}
& N_a(\lambda)  \asymp \sum_{\nu_{p+1}=1}^{\infty}\cdots\sum_{\nu_{p+r-1}=1}^{\infty} \sum_{\nu_{p+r}=1}^{\infty} \frac{\prod_{k=p+1}^{p+r}\nu_k^{2ma}}{\left(1+ \lambda \prod_{k=p+1}^{p+r}\nu_k^{2m}\right)^{2}}\nonumber\\
& \asymp \int_1^\infty\int_1^\infty\cdots\int_1^\infty \left(1+\lambda x_{p+1}^{b}\cdots x_{p+r-1}^b x_{p+r}^{b}\right)^{-2}dx_{p+1}\cdots dx_{p+r-1}dx_{p+r}, \nonumber
\end{align}
where $b= 2m/(2ma+1)$.
Let $z_k =x_{p+1}x_{p+2}\cdots x_{k}$ for $k=p+1,\ldots,p+r$. By using the change of variables to replace $(x_{p+1},\ldots,x_{p+r})$ by $(z_{p+1},\ldots,z_{p+r})$ and $z_{p+r}$ by $x=\lambda^{1/b}z_{p+r}$,
\begin{align}
 N_a(\lambda) &\asymp \int_1^{\infty}\int_1^{z_{p+r}}\cdots\int_1^{z_{p+2}}\left(1+\lambda z_{p+r}^{b}\right)^{-2}z_{p+1}^{-1}\cdots z_{p+r-1}^{-1}dz_{p+1}\cdots dz_{p+r-1}dz_{p+r}\nonumber\\
& \asymp \int_1^{\infty}(1+\lambda z_{p+r}^{b})^{-2}(\log z_{p+r})^{r-1}dz_{p+r}\nonumber\\
& \asymp \lambda^{-1/b} \int^\infty_{\lambda^{1/b}} (1+ x^{b})^{-2}\left(\log x - b^{-1}\log \lambda\right)^{r-1}dx\nonumber \asymp  \lambda^{-a-1/2m}\left[\log (1/\lambda)\right]^{r-1},\nonumber
\end{align}
where the last step follows from the fact that $2b>1$ for any $0\leq a <(2m-1)/(2m)$.

Second, we consider $d-r< p<d$. As discussed in the previous case, the number of combinations of non-1 indices being summed is finite, and the largest terms of the summation (\ref{eqn:defNalambdanuV}) over ${\vec{\bnu}}\in \bV$ correspond to the indices ${\vec{\bnu}} = (\nu_{k_1},\ldots,\nu_{k_{r+p-d}},\nu_{p+1},\ldots,\nu_d)^\top\in \N^r$, where the indices $k_1,\ldots,k_{r+p-d}\leq p$. Thus, by the integral approximation,
\begin{align}
 N_a(\lambda)
 & \asymp \sum_{v_{d-r+1}=1}^{\infty}\cdots\sum_{v_d=1}^{\infty}\frac{\prod_{k=d-r+1}^d\nu_k^{2ma}\left(1+\sum_{k=d-r+1}^p\nu_k^2\right)}{\left(1+\sum_{k=d-r+1}^p\nu_k^2+  \lambda \prod_{k=d-r+1}^d\nu_k^{2m}\right)^2}\nonumber\\
&  \asymp\int_1^{\infty}\cdots\int_1^{\infty} \frac{1+x_{d-r+1}^{b/m}+\cdots+x_p^{b/m}}{\left(1+x^{b/m}_{d-r+1}+\cdots+x^{b/m}_p+\lambda x_{d-r+1}^{b}\cdots x_d^{b}\right)^{2}}dx_{d-r+1}\cdots dx_d,\nonumber
\end{align}
where $b=2m/(2ma+1)$. Set $z_k=x_{p+1}x_{p+2}\cdots x_k$ for $k=p+1,\ldots,d$. By using the change the variables to replace $(x_{p+1},\ldots,x_d)$ by $(z_{p+1},\ldots,z_d)$, and $z_d$ by $x=\lambda^{1/b}z_d$, and   $x$ by $u= x_{d-r+1}\cdots x_p \cdot x$.
We have
\begin{align}
 N_a(\lambda) & \asymp \int_1^{\infty}\cdots\int_1^{\infty}\left[\int_1^{\infty}\int_1^{z_d}\cdots\int_1^{z_{p+2}}x_{d-r+1}^{b/m}\left(1+x_{d-r+1}^{b/m}+\cdots x_p^{b/m}+\lambda x_{d-r+1}^{b}\cdots x_p^{b}z_d^{b}\right)^{-2}\right.\nonumber\\
&\quad\quad\quad\quad\quad\quad\quad\quad\quad\quad\quad \left.\cdot z_{p+1}^{-1}\cdots z_{d-1}^{-1}dz_{p+1}\cdots dz_{d-1}dz_d\vphantom{\int_1^{\infty}}\right]dx_{d-r+1}\cdots dx_p\nonumber\\
& \asymp \lambda^{-1/b} \int_1^{\infty}\cdots\int_1^{\infty}\left[\int^\infty_{\lambda^{1/b}} x_{d-r+1}^{b/m}(1+x^{b/m}_{d-r+1}+\cdots x_p^{b/m}+x_{d-r+1}^{b}\cdots x_p^{b} x^{b})^{-2} \right.\nonumber\\
&\quad\quad\quad\quad\quad\quad\quad\quad\quad\quad\quad\quad 
\left.\cdot\left(\log x -b^{-1}\log \lambda\right)^{d-p-1}dx\vphantom{\int_1^{\infty}}\right]dx_{d-r+1}\cdots dx_p  \nonumber\\
& \lesssim\lambda^{-1/b} \int^\infty_{\lambda^{1/b}}\left[ \int_1^{\infty}\cdots\int_1^{\infty}
x_{d-r+1}^{b/m}\left(1+x_{d-r+1}^{b/m}+\cdots+x_p^{b/m}+ u^{b}\right)^{-2} x_{d-r+1}^{-1}\cdots x_p^{-1}\right. \nonumber\\
& \quad \cdot\left.\left(\log u - \log x_{d-r+1}-\cdots -\log x_p  -b^{-1}\log\lambda\right)^{d-p-1}dx_{d-r+1}\cdots dx_p\vphantom{\int_1^{\infty}}\right]du.\nonumber
\end{align}
By  Lemma \ref{lem:aversioofyoungsineq}, then for any $0<\tau<1$,
\begin{equation*}
\begin{aligned}
& \left(1+x_{d-r+1}^{b/m}+x_{d-r+2}^{b/m}+\cdots+x_p^{b/m}+ u^{b}\right)^{-2} \\
& \quad\quad \lesssim\left(1+x_{d-r+2}^{b/m}+\cdots+x_p^{b/m}+ u^{b}\right)^{-1+\tau} \cdot \left(x_{d-r+1}^{b/m}\right)^{-(1+\tau)}.
\end{aligned}
\end{equation*}
Together with the fact $\int_1^\infty t^{-1-\tau}(\log t)^k dt<\infty$ for any $k<\infty$, we have
\begin{align}
 N_a(\lambda) \lesssim &\lambda^{-1/b} \int^\infty_{\lambda^{1/b}} \left[ \int_1^\infty\cdots\int_1^\infty \left(1+x_{d-r+2}^{b/m}+\cdots+x_p^{b/m}+ u^{b}\right)^{-1+\tau}x_{d-r+2}^{-1}\cdots x_p^{-1}\right. \nonumber\\
&\quad\quad\quad\times\left.\left(\log u - \log x_{d-r+2}-\cdots -\log x_p  -b^{-1}\log\lambda\right)^{d-p-1}dx_{d-r+2}\cdots dx_p\vphantom{\int_1^{\infty}}\vphantom{\int_1^\infty}\right]du.\nonumber
\end{align}
Continuing this procedure gives
\begin{align*}
N_a(\lambda)\lesssim\lambda^{-1/b} \int^\infty_{\lambda^{1/b}} \left(1+ u^{b}\right)^{-(1-\tau)^{p-d+r}} \left(\log u - b^{-1}\log\lambda\right)^{d-p-1}du.
\end{align*}
Since for any $\epsilon>0$ and $d-r<p<d$, we know if $\tau<\epsilon/d$,
\begin{equation*}
(1-\tau)^{p-d+r}\geq 1- \tau(p-d+r)\geq 1-\tau(d-1)>1-\epsilon.
\end{equation*} 
 Hence, for any $0\leq a< (2m-1)/(2m)$, there exists $\tau$ such that $(1-\tau)^{p-d+r}>a+1/(2m)=1/b$.
Therefore, 
\begin{align*}
N_a(\lambda)\lesssim\lambda^{-1/b}\left[\log (1/\lambda)\right]^{d-p-1} = \lambda^{-a-1/2m}\left[\log (1/{\lambda})\right]^{d-p-1}.
\end{align*}

Finally, we consider $p=d$. As argued in the previous two cases, the number of combinations of non-1 indices being summed is finite. Now since $p=d$, by the symmetry of indices, the largest terms of the summation (\ref{eqn:defNalambdanuV}) over ${\vec{\bnu}}\in \bV$ correspond to any combinations of $r$ non-1 indices, for example, the first $r$ indices.
Thus, by the integral approximation,
\begin{equation*}
\begin{aligned}
 N_a(\lambda)
& \asymp   \sum_{\nu_{1}=1}^{\infty}\cdots\sum_{\nu_{r-1}=1}^{\infty}\sum_{\nu_r=1}^{\infty}
\frac{\prod_{k=1}^r\nu_k^{2ma}\left(1+\sum_{k=1}^r\nu_k^2\right)}{\left(1+\sum_{k=1}^r\nu_k^2+\lambda\prod_{k=1}^r\nu_k^{2m}\right)^2}\\
& \asymp \int_1^{\infty}\int_1^\infty\cdots\int_1^{\infty} \frac{1+x_1^{b/m}+\cdots+x_{r-1}^{b/m}+x_r^{b/m}}{\left(1 +x_1^{b/m}+\cdots + x_r^{b/m}+\lambda x_1^{b}\cdots x_{r-1}^bx_r^{b}\right)^{2}} dx_1\cdots dx_{r-1}dx_r,
\end{aligned}
\end{equation*}
where $b=2m/(2ma+1)$.
Observe that if $x_1\cdots x_{r-1}x_r\lesssim \lambda^{mr/[b(1-mr)]}$, then
\begin{equation*}
\lambda x_1^{b}\cdots x_{r-1}^{b}x_r^b \lesssim x_1^{b/m}+\cdots + x_{r-1}^{b/m}+x_r^{b/m}.
\end{equation*}
By  Lemma \ref{lemma:intx1xrzxk1z2} with $\beta=0$ and $\alpha=b/m\leq 2$, we have
\begin{equation}
\label{eqn:intx1xrlesslambdamrb-1}
\begin{aligned}
 N_a(\lambda)& \asymp\int_{x_1\cdots x_{r-1}x_r\lesssim\lambda^{mr/[b(1-mr)]}}   \left(1+x_1^{b/m}+\cdots+x_{r-1}^{b/m}+x_r^{b/m}\right)^{-1} dx_1\cdots dx_{r-1}dx_r\\
& \asymp
\begin{cases}
\lambda^{\frac{mr}{1-mr}\left(a+\frac{r-2}{2mr}\right)},   \mbox{ if } r\geq 3;\\
\log(1/\lambda),  \mbox{ if } r=2, a=0;\quad \lambda^{\frac{2ma}{1-2m}}, \mbox{ if } r=2, 0<a\leq 1;\\
1,  \mbox{ if } r=1,a<\frac{1}{2m};  \quad \log(1/\lambda),  \mbox{ if } r=1,a=\frac{1}{2m};\\\lambda^{\frac{1-2ma}{2m-2}},  \mbox{ if } r=1, a>\frac{1}{2m}.
\end{cases}
\end{aligned}
\end{equation}
On the other hand, if $\lambda^{mr/[b(1-mr)]}(x_1\cdots x_{r-1}x_r)^{-1}=o(1)$, then without loss of generality
 we assume $x_r=\min\{x_1,\cdots, x_r\}$. Let $z=\lambda^{1/b}x_1\cdots x_{r-1}x_r$.  By changing $x_r$ to $z$, we have
 \begin{equation}
 \label{eqn:intx1xrgtrmrlmabdalog}
\begin{aligned}
 N_a(\lambda) & \asymp\int_{\lambda^{mr/[b(1-mr)]}(x_1\cdots x_{r-1}x_r)^{-1}=o(1)}\\
& \quad\quad\left(1 +x_1^{b/m}+\cdots + x_r^{b/m}+\lambda x_1^{b}\cdots x_{r-1}^bx_r^{b}\right)^{-1}dx_1\cdots dx_{r-1}dx_r\\
&\lesssim \lambda^{-1/b}\int_{\lambda^{1/[b(1-mr)]}z^{-1}=o(1), \lambda^{-(r-1)/(br)}z^{(r-1)/r}\leq x_1\cdots x_{r-1}\leq \lambda^{-1/b}z}\\
& \quad\quad\left(1 +x_1^{b/m}+\cdots + x_{r-1}^{b/m}+z^{b}\right)^{-1} x_1^{-1}\cdots x_{r-1}^{-1}dx_{1}\cdots dx_{r-1}dz \\
&\lesssim \lambda^{-1/b}\int_{\lambda^{1/[b(1-mr)]}z^{-1}=o(1)}\left[\int_{ \lambda^{-(r-1)/(br)}z^{(r-1)/r}\leq x_1\cdots x_{r-1}\leq \lambda^{-1/b}z} \right.\\
& \quad\quad\left.\left(x_1^{b/m}+\cdots + x_{r-1}^{b/m}\right)^{-\tau}x_1^{-1}\cdots x_{r-1}^{-1}dx_{1}\cdots dx_{r-1} \vphantom{\int_1^\infty}
\right] z^{b(-1+\tau)}dz\\
&\lesssim \lambda^{-1/b}\int_{\lambda^{1/[b(1-mr)]}z^{-1}=o(1)} \lambda^{\tau/(mr)}z^{-\tau b/(mr)}\cdot z^{b(-1+\tau)}dz = o\left[\lambda^{\frac{mr}{1-mr}\left(a+\frac{r-2}{2mr}\right)}\right],
\end{aligned}
\end{equation}
where the third step follows from the Lemma \ref{lem:x1xrgtrxi} for $\beta=-1$ and $\alpha=\tau b/m$.
Combining (\ref{eqn:intx1xrlesslambdamrb-1}) and (\ref{eqn:intx1xrgtrmrlmabdalog}), we complete the proof for $p=d$ and this lemma.
\end{proof}

\begin{lemma}[Bounding the norm of the product of functions]
\label{lemma:inequanund1rhonua}
For any $f,g\in\otimes^d\HH_1$, $ a>1/2m$, and $1\leq p\leq d$, we have that
\begin{align}
&\sum_{{\vec{\bnu}}\in\N^d}\left(1+\frac{\rho_{\vec{\bnu}}}{\|\phi_{\vec{\bnu}}\|_{L_2}^2}\right)^a\|\phi_{\vec{\bnu}}\|_{L_2}^2\left\langle \frac{\partial f(\bt)}{\partial t_j}\frac{\partial g(\bt)}{\partial t_j}, \phi_{\vec{\bnu}}(\bt)\right\rangle_0^2\nonumber\\
& \quad\quad\lesssim \|f\|^2_{L_2(a+1/m)} \left[\sum_{{\vec{\bnu}}\in\N^d}\left(1+\frac{\rho_{\vec{\bnu}}}{\|\phi_{\vec{\bnu}}\|_{L_2}^2}\right)^a\|\phi_{\vec{\bnu}}\|_{L_2}^2\left\langle\frac{\partial g(\bt)}{\partial t_j}, \phi_{\vec{\bnu}}(\bt)\right\rangle_0^2\right].\nonumber
\end{align}
\end{lemma}

\begin{proof}
Recall that $\{\psi_\nu(t)\}_{\nu\geq 1}$ is the  trigonometrical basis on $L_2(\XX)$ and $\phi_{\vec{\bnu}}(\cdot)$ is defined in 
(\ref{eqn:phivecbnudef}). Write $\psi_{\vec{\bnu}}(\bt)=\psi_{\nu_1}(t_1)\psi_{\nu_2}(t_2)\cdots\psi_{\nu_d}(t_d)$. Note that
\begin{equation*}
\sum_{{\vec{\bnu}}\in\N^d}\left(1+\frac{\rho_{\vec{\bnu}}}{\|\phi_{\vec{\bnu}}\|_{L_2}^2}\right)^a\|\phi_{\vec{\bnu}}\|_{L_2}^2\langle f, \phi_{\vec{\bnu}}\rangle_0^2 = \sum_{{\vec{\bnu}}\in\N^d}\left(1+\frac{\rho_{\vec{\bnu}}}{\|\phi_{\vec{\bnu}}\|_{L_2}^2}\right)^a\left(\int_{\XX^d}f\psi_{\vec{\bnu}}\right)^2.
\end{equation*}
By Theorem A.2.2 and Corollary A.2.1 in \cite{lin1998tensor}, if $a>1/2m$, then for any $f,g\in\otimes^d\HH_1$,  
\begin{align}
&  \sum_{{\vec{\bnu}}\in\N^d}\left(1+\frac{\rho_{\vec{\bnu}}}{\|\phi_{\vec{\bnu}}\|_{L_2}^2}\right)^a\left(\int_{\XX^d}fg \psi_{\vec{\bnu}}\right)^2 \nonumber\\
& \lesssim  \left[\sum_{{\vec{\bnu}}\in\N^d}\left(1+\frac{\rho_{\vec{\bnu}}}{\|\phi_{\vec{\bnu}}\|_{L_2}^2}\right)^a\left(\int_{\XX^d}f \psi_{\vec{\bnu}}\right)^2\right]\left[\sum_{{\vec{\bnu}}\in\N^d}\left(1+\frac{\rho_{\vec{\bnu}}}{\|\phi_{\vec{\bnu}}\|_{L_2}^2}\right)^a\left(\int_{\XX^d}g \psi_{\vec{\bnu}}\right)^2\right].\nonumber
\end{align}
Thus,
\begin{align}
&\sum_{{\vec{\bnu}}\in\N^d}\left(1+\frac{\rho_{\vec{\bnu}}}{\|\phi_{\vec{\bnu}}\|_{L_2}^2}\right)^a\|\phi_{\vec{\bnu}}\|_{L_2}^2\left\langle \frac{\partial f(\bt)}{\partial t_j}\frac{\partial g(\bt)}{\partial t_j}, \phi_{\vec{\bnu}}(\bt)\right\rangle_0^2\nonumber\\
&  = \sum_{{\vec{\bnu}}\in\N^d}\left(1+\frac{\rho_{\vec{\bnu}}}{\|\phi_{\vec{\bnu}}\|_{L_2}^2}\right)^a\left(\int_{\XX^d}\frac{\partial f(\bt)}{\partial t_j}\frac{\partial g(\bt)}{\partial t_j} \psi_{\vec{\bnu}}(\bt)\right)^2 \nonumber\\
& \lesssim  \left[\sum_{{\vec{\bnu}}\in\N^d}\nu_j^2\left(1+\prod_{k=1}^d\nu_k^{2m}\right)^{a}\left(\int_{\XX^d}f(\bt)\psi_{\vec{\bnu}}(\bt)\right)^{2}\right] \left[\sum_{{\vec{\bnu}}\in\N^d}\left(1+\frac{\rho_{\vec{\bnu}}}{\|\phi_{\vec{\bnu}}\|_{L_2}^2}\right)^a\left(\int_{\XX^d}\frac{\partial g(\bt)}{\partial t_j} \psi_{\vec{\bnu}}(\bt)\right)^2\right]\nonumber\\
& \leq \left[\sum_{{\vec{\bnu}}\in\N^d}\left(1+\prod_{k=1}^d\nu_k^{2m}\right)^{a+\frac{1}{m}}\left(\int_{\XX^d}f(\bt)\psi_{\vec{\bnu}}(\bt)\right)^{2}\right]  \left[\sum_{{\vec{\bnu}}\in\N^d}\left(1+\frac{\rho_{\vec{\bnu}}}{\|\phi_{\vec{\bnu}}\|_{L_2}^2}\right)^a\left(\int_{\XX^d}\frac{\partial g(\bt)}{\partial t_j} \psi_{\vec{\bnu}}(\bt)\right)^2\right]\nonumber\\
& \asymp \|f\|^2_{L_2(a+1/m)} \left[\sum_{{\vec{\bnu}}\in\N^d}\left(1+\frac{\rho_{\vec{\bnu}}}{\|\phi_{\vec{\bnu}}\|_{L_2}^2}\right)^a\left(\int_{\XX^d}\frac{\partial g(\bt)}{\partial t_j} \psi_{\vec{\bnu}}(\bt)\right)^2\right].\nonumber
\end{align}
This completes the proof.
\end{proof}

\begin{lemma}[A variant of Young's inequality] 
\label{lem:aversioofyoungsineq}
For any $a,b\geq 0$ and $0<\tau<1$, we have
\begin{equation}
\label{eqn:ineqab-2leqtau}
(a+b)^{-2}\leq \frac{(1-\tau)^{1-\tau}(1+\tau)^{1+\tau}}{4}a^{-(1+\tau)} b^{-(1-\tau)}.
\end{equation}
When $\tau$ is small, the coefficient $(1-\tau)^{1-\tau}(1+\tau)^{1+\tau}/4$ is close to $1/4$.
\end{lemma}
\begin{proof}
To prove (\ref{eqn:ineqab-2leqtau}), it is sufficient to show
\begin{equation*}
a+b\geq 2(1-\tau)^{-(1-\tau)/2}(1+\tau)^{-(1+\tau)/2}a^{(1+\tau)/2}b^{(1-\tau)/2}.
\end{equation*}
Letting $p=2/(1+\tau)$, $a'=a^{1/p}$, $b'=[b/(p-1)]^{(p-1)/p}$, the above formula is equivalent to 
\begin{equation*}
\frac{a'}{p}+\frac{\left(b'\right)^{p/(p-1)}}{p/(p-1)}\geq a'b',
\end{equation*}
which holds by Young's inequality. This completes the proof.
\end{proof}

\begin{lemma}
\label{lem:x1xrgtrxi}
Suppose that $\beta\leq -1$ and $\alpha>0$. Then, as $\Xi\to \infty$,
\begin{align*}
& \int_{x_1\cdots x_r\geq \Xi, x_k\geq 1}\prod_{k=1}^rx_k^\beta (x_1^\alpha+x_2^\alpha+\cdots +x_r^\alpha)^{-1}dx_1\cdots dx_r\asymp \Xi^{\beta+1-\alpha/r}. 
\end{align*}
\end{lemma}
\begin{proof}
The proof is similar to the proof for Lemma \ref{lemma:intx1xrzxk1z2}. We omit the details here.
\end{proof}
\section{Proofs for Section \ref{sec:minmaxriskregularlat}}
\label{subsec:reglattproof}
\noindent
For brevity,  we  consider the regular lattice  $l_1=\cdots =l_d = l$ and $n=l^d$.  Other regular lattices can be shown similarly.
Write 
\begin{equation}
\label{eqn:trigbasisexp}
\psi_1(t)=1, \quad \psi_{2\nu}(t) = \sqrt{2}\cos 2\pi\nu t, \quad \psi_{2\nu+1}(t) = \sqrt{2}\sin 2\pi\nu t,
\end{equation}
for $\nu\geq 1$. 
As discussed in Section \ref{sec:minmaxriskregularlat}, it is without loss of generality to assume that $f_0: \XX^d \mapsto \R$ has a periodic boundaries on $\XX^d$. Hence $\{\psi_\nu(t)\}_{\nu\geq 1}$ forms an orthonormal system in $L_2(\XX)$ and an eigenfunction system for $K$.
For a $d$-dimensional vector ${\vec{\bnu}}=(\nu_1,\ldots,\nu_d)\in\N^d$, write 
\begin{equation}
\label{eqn:gammabnuprodgamma}
\psi_{{\vec{\bnu}}}(\bt) = \psi_{\nu_1}(t_1)\cdots\psi_{\nu_d}(t_d) \quad \mbox{ and }\quad
\lambda_{\vec{\bnu}} = \lambda_{\nu_1}\lambda_{\nu_2}\cdots\lambda_{\nu_d},
\end{equation}
where $\lambda_{\nu_j}$s and $\psi_{\nu_j}(t_j)$s are defined according to the spectral theorem, $j=1,\ldots,d$.  
Then, any function $f(\cdot)$ in $\HH$ admits the Fourier expansion $f(\bt) = \sum_{{\vec{\bnu}}\in\N^d}\theta_{\vec{\bnu}} \psi_{\vec{\bnu}}(\bt)$, where $\theta_{\vec{\bnu}} = \langle f(\bt),\psi_{\vec{\bnu}}(\bt)\rangle_{L_2}$,
and $J(f) = \sum_{\vec{\bnu}\in\N^d}\lambda_{\vec{\bnu}}^{-1}\theta_{\vec{\bnu}}^2$. 
We also write $f_0(\bt) = \sum_{{\vec{\bnu}}\in\N^d}\theta_{\vec{\bnu}}^0\psi_{\vec{\bnu}}(\bt)$. 

By Page 23 of  \cite{wahba1990}, it is known that 
\begin{equation*}
l^{-1}\sum_{i=1}^{l} \psi_{\mu}(i/l)\psi_\nu(i/l) = 
\begin{cases}
1,\quad & \mbox{if } \mu=\nu= 1,\ldots,l,\\
0,\quad & \mbox{if } \mu\neq \nu, \mu, \nu= 1,\ldots,l.
\end{cases}
\end{equation*}
Define
\begin{equation*}
\vec{\psi}_{{\vec{\bnu}}} = (\psi_{{\vec{\bnu}}}(\bt_1), \ldots, \psi_{{\vec{\bnu}}}(\bt_n))^\top,
\end{equation*} 
where $\{\bt_1,\ldots,\bt_n\}$ are the regular lattice design points. Thus, we have
\begin{equation*}
\langle \vec{\psi}_{{\vec{\bnu}}}, \vec{\psi}_{{\vec{\bmu}}} \rangle_n = 
\begin{cases}
1, \quad &\mbox{ if } \nu_j=\mu_j=1,\ldots, l; \ j=1,\ldots, d,\\
0, \quad &\mbox{ if there exists some } j \mbox{ such that } \nu_{j}\neq \mu_{j},
\end{cases}
\end{equation*}
where $\langle\cdot,\cdot\rangle_n$ is the empirical inner product in $\R^n$. 
This implies that $\{\vec{\psi}_{{\vec{\bnu}}} \ | \ \nu_j=1,\ldots, l; j=1,\ldots, d\}$ form an orthogonal basis in $\R^n$ with respect to the empirical norm $\|\cdot\|_n$. 
Denote the observed data vectors by  $\by^{(0)} = (y_1^{(0)},\ldots,y_n^{(0)})^\top$ and $\by^{(j)} = (y_1^{(j)},\ldots,y_n^{(j)})^\top$, and write
\begin{equation}
\label{eqn:distransdataz}
\begin{cases}
z_{{\vec{\bnu}}}^{(0)} & = \langle\by^{(0)}, \vec{\psi}_{{\vec{\bnu}}}\rangle_n,\\
z_{\nu_1,\ldots, 2\nu_j-1,\ldots,\nu_d}^{(j)} & =  (2\pi)^{-1}\langle \by^{(j)},  \vec{\psi}_{\nu_1,\ldots, 2\nu_j,\ldots, \nu_d}\rangle_n,\\
z_{\nu_1,\ldots, 2\nu_j,\ldots,\nu_d}^{(j)} & = - (2\pi)^{-1}\langle \by^{(j)}, \vec{\psi}_{\nu_1,\ldots, 2\nu_j-1,\ldots, \nu_d}\rangle_n,
\end{cases}
\end{equation}
for $\nu_j=1,\ldots, l$ and $j=1,\ldots,d$. Then, $z_{{\vec{\bnu}}}^{(0)} = \tilde{\theta}_{\vec{\bnu}}^0 + \delta^{(0)}_{\vec{\bnu}}$ and $z_{{\vec{\bnu}}}^{(j)} = \nu_j\tilde{\theta}_{\vec{\bnu}}^0 + \delta^{(j)}_{\vec{\bnu}}$, where $\tilde{\theta}_{\vec{\bnu}}^0 = \theta_{\vec{\bnu}}^0 + \sum_{\mu_j\geq l+1, j=1,\ldots,d}\theta_{\vec{\bmu}}^0\langle \vec{\psi}_{\vec{\bnu}}, \vec{\psi}_{\vec{\bmu}}\rangle_n$. The errors $\delta^{(0)}_{\vec{\bnu}}$ satisfy
\begin{equation*}
\begin{aligned}
\E[\delta^{(0)}_{\vec{\bnu}}] & = \frac{1}{n}\sum_{i=1}^n\E[\epsilon_i^{(0)}] \vec{\psi}_{{\vec{\bnu}}}(i)\leq \frac{1}{n}\sqrt{\sum_{i=1}^n\{\E[\epsilon_i^{(0)}]\}^2}\sqrt{\sum_{i=1}^n\vec{\psi}^2_{{\vec{\bnu}}}(i)} = o(n^{-1/2}),\\
\text{Var}[\delta^{(0)}_{\vec{\bnu}}] & = \frac{1}{n^2}\sum_{i=1}^n\text{Var}[\epsilon_i^{(0)}]\vec{\psi}_{{\vec{\bnu}}}^2(i) + \frac{1}{n^2}\sum_{i\neq i'}\text{Cov}[\epsilon_i^{(0)},\epsilon_{i'}^{(0)}]\vec{\psi}_{{\vec{\bnu}}}(i)\vec{\psi}_{{\vec{\bnu}}}(i')\\
& \leq \frac{\sigma_0^2}{n}\cdot\frac{1}{n}\sum_{i=1}^n\vec{\psi}_{{\vec{\bnu}}}^2(i) + \frac{2}{n^2}\sum_{i\neq i'}\text{Cov}[\epsilon_i^{(0)},\epsilon_{i'}^{(0)}]\\
& = O(n^{-1}) + \frac{2}{n^2}\sum_{i\neq i'}{o(|i-i'|^{-\Upsilon})}  = O(n^{-1}) + o(n^{-1}) = O(n^{-1}).
\end{aligned}
\end{equation*}
Similarly for any $j$, $\delta^{(j)}_{\vec{\bnu}}$s have mean $o(n^{-1/2})$ and covariances $O(n^{-1})$ . 

\subsection{Proof of Theorem \ref{theorem:lowerbdfNlambdareg}}
\label{sec:prooflowbdreg}

\noindent
We now prove the lower bound under the regular lattices. By the data transformation (\ref{eqn:distransdataz}), it suffices to show the optimal rate in a special case
\begin{equation}
\label{eqn:ynuejwhitenoise}
\begin{cases}
z_{\vec{\bnu}}^{(0)} & = \theta^0_{\vec{\bnu}} + \delta^{(0)}_{\vec{\bnu}}, \\
z_{\vec{\bnu}}^{(j)} & = \nu_j\theta^0_{\vec{\bnu}} + \delta^{(j)}_{\vec{\bnu}},\quad \mbox{ for } 1\leq j\leq p,
\end{cases}
\end{equation}
where $\delta^{(j)}_{\vec{\bnu}}\sim \NN(0,\sigma_{j}^2/n)$ are independent. For any $\vec{\bnu}\in\N^d$, if we have the prior that $| \tilde{\theta}^0_{\vec{\bnu}} |\leq \pi_{\vec{\bnu}}$, then the minimax linear estimator is 
\begin{equation*}
\widehat{\theta}_{\vec{\bnu}}^L = \frac{\sigma_0^{-2}z^{(0)}_{\vec{\bnu}}+\sum_{j=1}^p\sigma_j^{-2}\nu_jz^{(j)}_{\vec{\bnu}}}{n^{-1}\pi_{\vec{\bnu}}^{-2} + \sigma_0^{-2}+\sum_{j=1}^p\sigma_j^{-2}\nu_j^2},
\end{equation*}
and the minimax linear risk is
\begin{equation*}
n^{-1}\left[n^{-1}\pi_{\vec{\bnu}}^{-2} + \sigma_0^{-2}+\sum_{j=1}^p\sigma_j^{-2}\nu_j^2\right]^{-1}.
\end{equation*}
By Lemma 6 and Theorem 7 in  \cite{donoho1990minimax}, if $\sigma_j^2$s are known, 
the minimax risk of estimating $\theta_{\vec{\bnu}}^0$ under the model (\ref{eqn:ynuejwhitenoise}) is larger than $80\%$ of the minimax linear risk of the hardest rectangle subproblem, and the latter linear risk is
\begin{equation}
\label{eqn:linearriskRLsumvecv}
R^{L} = n^{-1}\max_{\sum_{\vec{\bnu}\in \bV}(1+\lambda_{\vec{\bnu}})\pi_{\vec{\bnu}}^2=1} \sum_{\vec{\bnu}\in \bV} \left[n^{-1}\pi_{\vec{\bnu}}^{-2} + \sigma_0^{-2}+\sum_{j=1}^p\sigma_j^{-2}\nu_j^2\right]^{-1},
\end{equation}
where $\lambda_{\vec{\bnu}}$ is the product of eigenvalues in (\ref{eqn:gammabnuprodgamma}) and recall that the set  $V$ is defined in (\ref{def:Vbnu}).

We use the Lagrange multiplier method to find $\pi_{\vec{\bnu}}^2$ for solving (\ref{eqn:linearriskRLsumvecv}). Let $a$ be the scalar multiplier and  define
\begin{equation*}
L(\pi_{\vec{\bnu}}^2,a) = \sum_{\vec{\bnu}\in \bV} \left[n^{-1}\pi_{\vec{\bnu}}^{-2} + \sigma_0^{-2}+\sum_{j=1}^p\sigma_j^{-2}\nu_j^2\right]^{-1}  - a (1+\lambda_{\vec{\bnu}})\pi_{\vec{\bnu}}^2.
\end{equation*}
Taking partial derivative with respect to $\pi_{\vec{\bnu}}^2$ gives
\begin{equation*}
\frac{\partial L}{\partial \pi_{\vec{\bnu}}^2} = n^{-1}\left[n^{-1}+\left(\sigma_0^{-2}+\sum_{j=1}^p\sigma_j^{-2}\nu_j^2\right)\pi_{\vec{\bnu}}^2\right]^{-2} - a(1+\lambda_{\vec{\bnu}}) = 0.
\end{equation*}
This implies
\begin{equation*}
\widehat{\pi}_{\vec{\bnu}}^2 = \left(\sigma_0^{-2}+\sum_{j=1}^p\sigma_j^{-2}\nu_j^2\right)^{-1}\left[b(1+\lambda_{\vec{\bnu}})^{-1/2}- n^{-1}\right]_+,
\end{equation*}
where $b=(na)^{-1/2}$. On one hand, plugging the above formula into the constraint $\sum_{\vec{\bnu}\in \bV}(1+\lambda_{\vec{\bnu}})\pi_{\vec{\bnu}}^2=1$ gives,
\begin{equation*}
\sum_{\vec{\bnu}\in \bV} \prod_{k=1}^d\nu_k^{2m}\left(\sigma_0^{-2}+\sum_{j=1}^p\sigma_j^{-2}\nu_j^2\right)^{-1}\left[b\prod_{k=1}^d\nu_k^{-m}- n^{-1}\right]_+ \asymp 1.
\end{equation*}
By restricting $\prod_{j=1}^d\nu_j\leq (nb)^{1/m}$, this becomes
\begin{equation}
\label{eqn:sumbnuinVprodk1d}
\begin{aligned}
& \sum_{\vec{\bnu}\in \bV, \prod_{k=1}^d\nu_k\leq (nb)^{1/m}}\left(\sigma_0^{-2}+\sum_{j=1}^p\sigma_j^{-2}\nu_j^2\right)^{-1} \left(b\prod_{k=1}^d\nu_k^{m}- n^{-1}\prod_{k=1}^d\nu_k^{2m}\right) \asymp 1.
\end{aligned}
\end{equation}
On the other hand, the linear risk in (\ref{eqn:linearriskRLsumvecv}) can be written as
\begin{equation}
\label{eqn:explicitexpofRL}
\begin{aligned}
R^L \asymp n^{-1}& \sum_{\vec{\bnu}\in \bV,\prod_{k=1}^d\nu_k\leq (nb)^{1/m}}\left(1-\frac{1}{nb}\prod_{k=1}^d\nu_k^m\right)\times \left(\sigma_0^{-2}+\sum_{j=1}^p\sigma_j^{-2}\nu_j^2\right)^{-1}.
\end{aligned}
\end{equation}
We discuss for $R^L$ in the above (\ref{eqn:explicitexpofRL}) under the condition (\ref{eqn:sumbnuinVprodk1d}) for three cases with $0\leq p\leq d-r$, $d-r<p<d$ and $p=d$.

If $0\leq p\leq d-r$, since $\vec{\bnu}\in \bV$, there are at most $r$ of $\nu_1,\ldots,\nu_d$ not equal to 1, which implies that the number of combinations of non-1 indices being summed in (\ref{eqn:sumbnuinVprodk1d}) is no greater than $C_d^1+C_d^2+\cdots+C_d^r<\infty$. Due to the term $(\sigma_0^{-2}+\sum_{j=1}^p\sigma_j^{-2}\nu_j^2)^{-1}$, the largest terms of the summation (\ref{eqn:sumbnuinVprodk1d}) over ${\vec{\bnu}}\in \bV$ correspond to the combinations of indices whereas fewer $\nu_1,\ldots,\nu_p$ being summed as possible, for example, $v_k\equiv 1$ for $k\leq p$ and $k> p+r$, and $(\nu_{p+1},\ldots,\nu_{p+r})\in\N^r$ are non-1. Thus, (\ref{eqn:sumbnuinVprodk1d}) is equivalent to 
\begin{equation*}
\sum_{\prod_{k=1}^r\nu_{p+k}\leq (nb)^{1/m}}\left(b\prod_{k=1}^r\nu_{p+k}^{m}- n^{-1}\prod_{k=1}^r\nu_{p+k}^{2m}\right) \asymp 1.
\end{equation*}
Using the integral approximation, we have
\begin{equation*}
\int_{\prod_{k=1}^rx_{p+k}\leq (nb)^{1/m},x_{p+k}\geq1} \left(b\prod_{k=1}^rx_{p+k}^m - \frac{1}{n}\prod_{k=1}^rx_{p+k}^{2m}\right)dx_{p+1}\cdots dx_{p+r}\asymp 1.
\end{equation*}
By letting $z_{j} = \prod_{1\leq k\leq j}x_{p+k}$, $j=1,2,\ldots,r$, we have
\begin{equation*}
\int_1^{(nb)^{1/m}}\left[\int_1^{z_r}\cdots\int_{1}^{z_2} \left(bz_r^m - \frac{1}{n}z_r^{2m}\right)z_1^{-1}\cdots z_{r-1}^{-1}dz_1\cdots dz_{r-1}\right]dz_r\asymp 1,
\end{equation*}
where the left-hand side term is the order of $n^{(m+1)/m}b^{(2m+1)/m}[\log(nb)]^{r-1}$ and hence
\begin{equation}
\label{eqn:p0d-rbn-m12m1log}
b\asymp n^{-(m+1)/(2m+1)}(\log n)^{-m(r-1)/(2m+1)}.
\end{equation}
The linear risk in (\ref{eqn:explicitexpofRL}) becomes
\begin{align}
R^L & \asymp n^{-1}\int_{\prod_{k=1}^rx_{p+k}\leq (nb)^{1/m},x_{p+k}\geq 1}\left(1-\frac{1}{nb}\prod_{k=1}^rx_{p+k}^m\right)\nonumber\\
& \asymp [\log(nb)]^{r-1}n^{-1+1/m}b^{1/m} \asymp  [n(\log n)^{1-r}]^{-2m/(2m+1)},\nonumber
\end{align}
where the last step is by (\ref{eqn:p0d-rbn-m12m1log}).

If $d-r< p< d$, as discussed in the previous case, 
the number of combinations of non-1 indices being summed is finite, and 
the largest terms of the summation (\ref{eqn:sumbnuinVprodk1d}) over ${\vec{\bnu}}\in \bV$ correspond to the combinations of indices whereas fewer than $\nu_1,\ldots,\nu_p$ being summed as possible, for example, $v_k\equiv 1$ for $k\leq d-r$, and $(\nu_{d-r+1},\ldots, \nu_d)\in\N^r$ are non-1. Thus, (\ref{eqn:sumbnuinVprodk1d}) is equivalent to 
\begin{equation*}
\begin{aligned}
 \sum_{\prod_{k=1}^r\nu_{d-r+k}\leq (nb)^{1/m}}\left(b\prod_{k=1}^r\nu_{d-r+k}^{m}- n^{-1}\prod_{k=1}^r\nu_{d-r+k}^{2m}\right) \left(1+\sum_{j=d-r+1}^p\nu_j^2\right)^{-1} \asymp 1.
\end{aligned}
\end{equation*}
Using the integral approximation, we have
\begin{align}
& \int_{\prod_{k=1}^rx_{d-r+k}\leq (nb)^{1/m},x_{d-r+k}\geq1}  \left(b\prod_{k=1}^rx_{d-r+k}^m - n^{-1}\prod_{k=1}^rx_{d-r+k}^{2m}\right)\nonumber\\
& \quad\quad\quad\quad \quad\quad\quad\quad\quad \times \left(1+\sum_{j=d-r+1}^px_j^2\right)^{-1}dx_{d-r+1}\cdots dx_{d}\asymp 1.\nonumber
\end{align}
By letting $z_{j} =x_{p+1}x_{p+2}\cdots x_j$, $j=p+1,\ldots, d$, we get
\begin{align*}
1& \asymp \int_{x_{d-r+1}\cdots x_pz_d\leq (nb)^{1/m}} \left[\int_1^{z_d}\cdots\int_{1}^{z_{p+2}} \right.\\
& \quad\quad \left(bx_{d-r+1}^m\cdots x_p^mz_d^m - \frac{1}{n}x_{d-r+1}^{2m}\cdots x_p^{2m}z_{d}^{2m}\right)z_{p+1}^{-1}\cdots z_{d-1}^{-1}\nonumber\\
& \quad\quad\left.\times\left(1+x_{d-r+1}^2+\cdots +x_p^2\right)^{-1}dz_{p+1}\cdots dz_{d-1}\vphantom{\int_1^{z_d}}\right]dx_{d-r+1}\cdots dx_pdz_d\nonumber\\
& =\int_{x_{d-r+1}\cdots x_pz_d\leq (nb)^{1/m}}bx_{d-r+1}^m\cdots x_p^mz_d^m\left(1 - \frac{1}{nb}x_{d-r+1}^{m}\cdots x_p^{m}z_{d}^{m}\right)\nonumber\\
& \quad\quad\quad\times (\log z_d)^{d-p-1} \left(1+x_{d-r+1}^2+\cdots +x_p^2\right)^{-1}dx_{d-r+1}\cdots dx_pdz_d\nonumber\\
& \asymp [\log (nb)]^{d-p-1} n^{1+1/m}b^{2+1/m}.\nonumber
\end{align*}
The last step is by Lemma \ref{lemma:intx1xrzxk1z1}.
Hence,
\begin{equation}
\label{eqn:p0d-rbn-m12m1log2}
b\asymp n^{-(m+1)/(2m+1)}(\log n)^{-m(d-p-1)/(2m+1)}.
\end{equation}
The linear risk in (\ref{eqn:explicitexpofRL}) becomes
\begin{align}
R^L & \asymp n^{-1}\int_{\prod_{k=d-r+1}^dx_{k}\leq (nb)^{1/m},x_{k}\geq 1}\left(1-\frac{1}{nb}x_{d-r+1}^m\cdots x_d^m\right)\nonumber\\
& \quad\quad\quad\quad\quad\quad\quad\quad\quad\cdot (1+x^2_{d-r+1}+\cdots+x_p^2)^{-1}dx_{d-r+1}\cdots dx_d \nonumber\\
& \asymp n^{-1} \int_{x_{d-r+1}\cdots x_pz_d\leq (nb)^{1/m}}\left(1-\frac{1}{nb}x_{d-r+1}^m\cdots x_p^mz_d^m\right)(\log z_d)^{d-p-1}\nonumber\\
&\quad\quad\quad\quad\quad\quad\quad \quad\cdot(1+x_{d-r+1}^2+\cdots+x_p^2)^{-1}dx_{d-r+1}\cdots dx_pdz_d\nonumber\\
& \asymp  [\log (nb)]^{d-p-1}n^{-1+1/m}b^{1/m},\nonumber
\end{align}
where the second step uses the same change of variables by letting $z_{j} =x_{p+1}x_{p+2}\cdots x_j$, $j=p+1,\ldots, d$, and the last step is by Lemma \ref{lemma:intx1xrzxk1z1}.
By (\ref{eqn:p0d-rbn-m12m1log2}), we have
\begin{align}
R^L &  \asymp  [n(\log n)^{1+p-d}]^{-2m/(2m+1)}.\nonumber
\end{align}

If $p=d$, as discussed in the previous two cases, the number of combinations of non-1 indices being summed is finite, and the largest terms of the summation (\ref{eqn:sumbnuinVprodk1d}) over ${\vec{\bnu}}\in \bV$ correspond to any combinations of $r$ non-1 indices, for example, $\nu_k\equiv 1$ for $k\geq r+1$, and $(\nu_1,\ldots,\nu_r)\in\N^r$.  Thus, (\ref{eqn:sumbnuinVprodk1d}) is equivalent to 
\begin{equation*}
\sum_{\prod_{k=1}^r\nu_{k}\leq (nb)^{1/m}}\left(b\prod_{k=1}^r\nu_{k}^{m}- n^{-1}\prod_{k=1}^r\nu_{k}^{2m}\right)\left(1+\sum_{k=1}^r\nu_k^2\right)^{-1} \asymp 1.
\end{equation*}
Using the integral approximation, we have
\begin{align}
1& \asymp \int_{\prod_{k=1}^rx_{k}\leq (nb)^{1/m},x_{k}\geq1}  \left(b\prod_{k=1}^rx_{k}^m - n^{-1}\prod_{k=1}^rx_{k}^{2m}\right) \left(1+\sum_{k=1}^rx_k^2\right)^{-1}dx_{1}\cdots dx_{r}\nonumber\\
& \asymp \int_{\prod_{k=1}^rx_{k}\leq (nb)^{1/m},x_{k}\geq1}  b\prod_{k=1}^rx_{k}^m \left(1+\sum_{k=1}^rx_k^2\right)^{-1}dx_{1}\cdots dx_{r}\nonumber
\end{align}
By letting $\beta=m>1$ and $\alpha=2$ in Lemma \ref{lemma:intx1xrzxk1z2}, we have for any $r\geq 1$, 
\begin{equation}
\label{eqn:p0d-rbn-m12m1log3}
b\asymp  n^{-(mr+r-2)/(2mr+r-2)}.
\end{equation}
The linear risk in (\ref{eqn:explicitexpofRL}) becomes
\begin{align}
R^L & \asymp n^{-1}\int_{\prod_{k=1}^rx_{k}\leq (nb)^{1/m},x_{k}\geq 1}\left(1-\frac{1}{nb}x_{1}^m\cdots x_r^m\right)\nonumber\\
& \quad\quad\quad\quad\quad\quad\quad\quad\quad\quad\quad\quad\quad\quad\quad\cdot (1+x^2_{1}+\cdots+x_r^2)^{-1}dx_{1}\cdots dx_r\nonumber\\
& \asymp n^{-1}\int_{\prod_{k=1}^rx_{k}\leq (nb)^{1/m},x_{k}\geq 1}(1+x^2_{1}+\cdots+x_r^2)^{-1}dx_{1}\cdots dx_r\nonumber\\
& \asymp \left[n^{-1}(nb)^{(r-2)/(mr)}\right] \mathbbm{1}_{r\geq 3} + \left[n^{-1}\log(nb)\right]\mathbbm{1}_{r=2}+\left(n^{-1}\right)\mathbbm{1}_{r=1},\nonumber
\end{align}
where the last step uses Lemma \ref{lemma:intx1xrzxk1z2} with $\beta=0$ and $\alpha=2$. 
By (\ref{eqn:p0d-rbn-m12m1log3}), we have
\begin{equation*}
R^L \asymp\left[n^{-(2mr)/[(2m+1)r-2]}\right] \mathbbm{1}_{r\geq 3} + \left[n^{-1}\log(n)\right]\mathbbm{1}_{r=2}+n^{-1}\mathbbm{1}_{r=1},
\end{equation*}
where the constant factor does not depend on $n$. This completes the proof.

\subsection{Proof of  Theorem \ref{thm;deterdesgnupperbdgeneralpdreg}}
\label{sec:proofupperbdreg}
\noindent
We now prove the theorem for only $r=d$ and $p=d-1$. Other settings can be shown similarly.
Using the discrete transformed data (\ref{eqn:distransdataz}), the estimator $\widehat{f}_{n}$ in (\ref{scheme1}) can be obtained through 
\begin{align}
\widehat{\theta}_{\vec{\bnu}}=\underset{\tilde{\theta}_{\vec{\bnu}}\in\R}{\arg\min} &\left\{ \frac{1}{n(p+1)}\left[\frac{1}{\sigma_0^2}\sum_{\vec{\bnu}\in V, \|\vec{\bnu}\|_{\min}\leq l}\left(z_{\vec{\bnu}}^{(0)} - \theta_{\vec{\bnu}}\right)^2\right.\right.\nonumber\\
&\left.\left.+\sum_{j=1}^p\frac{1}{\sigma_j^2}\sum_{\vec{\bnu}\in V, \|\vec{\bnu}\|_{\min}\leq l }\left(z^{(j)}_{\vec{\bnu}}- \nu_j\theta_{\vec{\bnu}}\right)^2\right]+\lambda\sum_{\vec{\bnu}\in V, \|\vec{\bnu}\|_{\min}\leq l}\lambda_{\vec{\bnu}}\theta_{\vec{\bnu}}^2
\right\}\nonumber
\end{align}
and $\widehat{f}_{n}(\bt) = \underset{\vec{\bnu}\in \bV, \|\vec{\bnu}\|_{\min}\leq l}{\sum}\widehat{\theta}_{\vec{\bnu}}\psi_{\vec{\bnu}}(\bt)$, where $\bV$ is defined in (\ref{def:Vbnu}).
Direct calculations give
\begin{equation*}
\widehat{\theta}_{\vec{\bnu}} =  \frac{\sigma_0^{-2}z^{(0)}_{\vec{\bnu}}+\sum_{j=1}^p\sigma_j^{-2}\nu_jz^{(j)}_{\vec{\bnu}}}{ \sigma_0^{-2}+\sum_{j=1}^p\sigma_j^{-2}\nu_j^2 + \lambda\lambda^{-1}_{\vec{\bnu}}}. 
\end{equation*}

The deterministic error of $\widehat{f}_{n}$ can be analyzed in two parts. On one hand, since $f_0\in\HH$ and $\lambda_\nu\asymp\nu^{-2m}$, we know $\sum_{\vec{\bnu}\in \bV, \|\vec{\bnu}\|_{\min}\geq l+1}(\theta_{\vec{\bnu}}^0)^2\asymp n^{-2m}$. This is the truncation error due to $\widehat{\theta}_{\vec{\bnu}} =0$ for $\nu_k\geq l+1$, $1\leq k\leq d$.
On the other hand, note that $\langle\vec{\psi}_{\vec{\bnu}},\vec{\psi}_{\vec{\bmu}}\rangle_n^2\leq 1$ and then 
\begin{equation*}
\left(\sum_{\vec{\bmu}\in \bV, \|\vec{\bmu}\|_{\min}\geq l+1}\theta_{\vec{\bmu}}^0\langle\vec{\psi}_{\vec{\bnu}},\vec{\psi}_{\vec{\bmu}}\rangle_n\right)^2\leq \sum_{\vec{\bmu}\in \bV, \|\vec{\bmu}\|_{\min}\geq l+1}(\theta^0_{\vec{\bmu}})^2 \asymp n^{-2m}.
\end{equation*}
Thus,
\begin{equation*}
\begin{aligned}
& \sum_{\vec{\bnu}\in \bV, \|\vec{\bnu}\|_{\min}\leq l}\left(\E\widehat{\theta}_{\vec{\bnu}} - \theta_{\vec{\bnu}}^0\right)^2 \\
&  {\lesssim  \sum_{\vec{\bnu}\in \bV, \|\vec{\bnu}\|_{\min}\leq l} \frac{(\lambda\lambda^{-1}_{\vec{\bnu}}\theta_{\vec{\bnu}}^0)^2 +  [\E\delta^{(0)}_{\vec{\bnu}}]^2 + \sum_{j=1}^p\nu_j^2[\E\delta^{(j)}_{\vec{\bnu}}]^2}{(\sigma_0^{-2}+\sum_{j=1}^p\sigma_j^{-2}\nu_j^2+\lambda\lambda^{-1}_{{\vec{\bnu}}})^2}} + n^{-2m+1}\\
& \leq \lambda^2\sup_{{\vec{\bnu}}\in \bV}\frac{\lambda^{-1}_{\vec{\bnu}}}{\left(\sigma_0^{-2}+\sum_{j=1}^p\sigma_j^{-2}\nu_j^2+\lambda\lambda^{-1}_{{\vec{\bnu}}}\right)^2} \sum_{{\vec{\bnu}}\in \bV}\lambda_{\vec{\bnu}}^{-1}(\theta_{\vec{\bnu}}^0)^2\\
& \quad+o(n^{-1})\sum_{{\vec{\bnu}}\in \bV,\|\vec{\bnu}\|_{\min}\leq l}\frac{1+\sum_{j=1}^p\nu_j^2}{(1+\sum_{j=1}^p\nu_j^2+\lambda\nu_1^{2m}\cdots\nu_d^{2m})^2}+n^{-2m+1} \\
&  \asymp \lambda^2 J(f_0)\sup_{{\vec{\bnu}}\in \bV} \frac{\nu_1^{2m}\cdots\nu_d^{2m}}{(1+\sum_{j=1}^p\nu_j^2+\lambda\nu_1^{2m}\cdots\nu_d^{2m})^2} +o\{n^{-1}\lambda^{-1/2m}\} + n^{-2m+1},
\end{aligned}
\end{equation*}
where the last step uses Lemma \ref{lemma:varthetamultivariatetensorrank} with $a=0$ and $p=d-1$.
Define that
\begin{equation*}
B_\lambda({\vec{\bnu}}) =  \frac{\nu_1^{2m}\cdots\nu_d^{2m}}{(1+\sum_{j=1}^p\nu_j^2+\lambda\nu_1^{2m}\cdots\nu_d^{2m})^2}.
\end{equation*}
For the $\sup_{{\vec{\bnu}}\in \bV}B_\lambda({\vec{\bnu}})$ term above, suppose that $\prod_{j=1}^d\nu_j^{2m}>0$ is fixed and denoted by $x^{-1}$, then $B_\lambda({\vec{\bnu}})$ is maximized by letting $\sum_{j=1}^p\nu_j^2$ be as small as possible, where $p=d-1$. This suggests $\nu_1=\nu_2=\cdots=\nu_p=1$, and 
\begin{align}
\underset{\vec{\bnu}\in \bV}{\sup}B_{\lambda}({\vec{\bnu}}) 
& \asymp \sup_{x>0}\frac{x^{-1}}{(1+\lambda x^{-1})^2} \asymp \lambda^{-1}, \nonumber
\end{align}
where the last step is achieved when $x \asymp \lambda$. Combining all parts of bias gives
\begin{equation}
\label{eqn:detererrordeterdesign}
 \sum_{{\vec{\bnu}}\in \bV}\left(\E\widehat{\theta}_{\vec{\bnu}} - \theta_{\vec{\bnu}}^0\right)^2\\
 =O\left\{\lambda J(f_0) + n^{-2m+1}\right\} +o\{n^{-1}\lambda^{-1/2m}\}.
\end{equation}
The constant factor on the upper bound does not depend on $n$.

The stochastic error is bounded as follows:
\begin{align}
\sum_{\vec{\bnu}\in \bV}\E\left(\widehat{\theta}_{\vec{\bnu}} - \E\widehat{\theta}_{\vec{\bnu}}\right)^2 & = \sum_{\vec{\bnu}\in \bV,\|\vec{\bnu}\|_{\min}\leq l}\frac{n^{-1}(\sigma_0^{-2}+\sum_{j=1}^p\sigma_j^{-2}\nu_j^2)}{(\sigma_0^{-2}+\sum_{j=1}^p\sigma_j^{-2}\nu_j^2+\lambda\lambda_{\vec{\bnu}}^{-1})^2}\nonumber\\
& \lesssim\sum_{{\vec{\bnu}}\in \bV,\|\vec{\bnu}\|_{\min}\leq l}\frac{1+\sum_{j=1}^p\nu_j^2}{n(1+\sum_{j=1}^p\nu_j^2+\lambda\nu_1^{2m}\cdots\nu_d^{2m})^2}.\nonumber
\end{align}
Using Lemma \ref{lemma:varthetamultivariatetensorrank}  with $a=0$ and $p=d-1$, we have 
\begin{equation}
\label{eqn:vardeterdesign}
\sum_{{\vec{\bnu}}\in \bV}\E\left(\widehat{\theta}_{\vec{\bnu}} - \E\widehat{\theta}_{\vec{\bnu}}\right)^2 = O\left\{n^{-1}\lambda^{-1/2m}\right\}.
\end{equation}
Combining (\ref{eqn:detererrordeterdesign}) and (\ref{eqn:vardeterdesign}) and letting $\lambda\asymp n^{-2m/(2m+1)}$ completes the proof.


\subsection{Auxiliary Lemmas for Theorems  \ref{theorem:lowerbdfNlambdareg} and  \ref{thm;deterdesgnupperbdgeneralpdreg}}

\begin{lemma} 
\label{lemma:intx1xrzxk1z1}
Suppose that $s\geq 1$, $\beta\geq 0$ and $\beta\neq 1$, and $r\geq 1$. Then as $\Xi\to\infty$,
\begin{equation*}
 \int_{x_1\cdots x_r\cdot z\leq \Xi, x_k\geq 1, z\geq 1} x_1^\beta\cdots x_r^\beta z^\beta(\log z)^s(x_1^2+\cdots +x_r^2)^{-1}dx_1\cdots dx_rdz\asymp \Xi^{\beta+1}(\log \Xi)^s.
\end{equation*}
\end{lemma}
\begin{proof}
For any $\tau\geq 1$, we have 
\begin{equation*}
\{1\leq z\leq \Xi\tau^{-r},1\leq x_k\leq \tau, k=1,\ldots,r\}\subset \{x_1\cdots x_r\cdot z\leq \Xi, z\geq 1, x_k\geq 1, k=1,\ldots,r \}.
\end{equation*} Thus, if $\Xi\to \infty$,
\begin{align}
& \int_{x_1\cdots x_r\cdot z\leq \Xi, x_k\geq 1, z\geq 1} x_1^\beta\cdots x_r^\beta z^\beta(\log z)^s(x_1^2+\cdots +x_r^2)^{-1}dx_1\cdots dx_rdz\nonumber \\
& \geq \int_1^{\Xi\tau^{-r}}\int_1^\tau\cdots\int_1^\tau z^\beta(\log z)^sx_1^{\beta-2}\cdots x_r^{\beta-2}dx_1\cdots dx_rdz\nonumber\\
& \asymp \Xi^{\beta+1}\tau^{-r(\beta+1)}(\log \Xi-r\log \tau)^s\tau^{r(\beta-1)}.\nonumber
\end{align}
Let $\tau\to 1$, we have $\int_{x_1\cdots x_r\cdot z\leq \Xi, x_k\geq 1, z\geq 1} (\log z)^s(x_1^2+\cdots +x_r^2)^{-1}dx_1\cdots dx_rdz\gtrsim \Xi^{\beta+1}(\log \Xi)^s$.

On the other hand, define $u=x_1\cdots x_r\cdot z$ and change the variable $z$ to $u$. We have that as $\Xi\to \infty$,
\begin{align}
& \int_{x_1\cdots x_r\cdot z\leq \Xi, x_k\geq 1, z\geq 1} x_1^\beta\cdots x_r^\beta z^\beta(\log z)^s(x_1^2+\cdots +x_r^2)^{-1}dx_1\cdots dx_rdz\nonumber \\
&= \int_{1}^\Xi\int_1^{u}\int_1^{u/x_r}\cdots \int_1^{u/(x_rx_{r-1}\cdots x_{2})}u^\beta(\log u-\log x_r-\cdots -\log x_1)^s\nonumber\\
& \quad\quad\quad\quad \cdot\left(x_1^2+\cdots + x_{r-1}^2 +x_r^2\right)^{-1}x_1^{-1}\cdots x_{r-1}^{-1}x_{r}^{-1}dx_1\cdots dx_{r-1}dx_{r}du\nonumber\\
&\lesssim \int_{1}^\Xi\int_1^{u}\int_1^{u/x_r}\cdots \int_1^{u/(x_rx_{r-1}\cdots x_{2})}u^\beta(\log u-\log x_r-\cdots -\log x_1)^s\nonumber\\
& \quad\quad\quad\quad \cdot x_1^{-1-2/r}\cdots x_{r-1}^{-1-2/r}x_{r}^{-1-2/r}dx_1\cdots dx_{r-1}dx_{r}du\nonumber\\
& \lesssim \int_{1}^\Xi u^\beta(\log u)^s du \asymp \Xi^{\beta+1}(\log \Xi)^s.\nonumber
\end{align}
The second step is by Lemma \ref{lem:aversioofyoungsineq}. This completes the proof.
\end{proof}

\end{document}